\documentclass[twocolumn,tighten]{aastex63}

\hypersetup{linkcolor=red,citecolor=blue,filecolor=cyan,urlcolor=magenta}

\received{}
\revised{}
\submitjournal{ApJ}

\shorttitle{Cannibalism Caught in the Act}
\shortauthors{Hsu et al.}
\graphicspath{{./}{figures/}}

\usepackage{CJK}
\begin{document}
\begin{CJK*}{UTF8}{gbsn}

\title{SDSS-IV MaNGA: Cannibalism Caught in the Act -- on the Frequency of Occurrence of Multiple Cores in Brightest Cluster Galaxies}



\correspondingauthor{Yun-Hsin Hsu}
\email{Y.Hsu@physik.lmu.de}

\author[0000-0003-0381-562X]{Yun-Hsin Hsu (徐允心)}
\affiliation{Institute of Astronomy and Astrophysics, Academia Sinica  (ASIAA),  Taipei 10617, Taiwan}
\affiliation{Institute of Astronomy, National Tsing Hua University,  Hsinchu 30013, Taiwan}
\affiliation{University Observatory, Faculty of Physics, Ludwig-Maximilians-Universit\"{a}t M\"{u}nchen, Scheinerstr. 1, 81679 Munich, Germany}
\affiliation{Max-Planck-Institut f\"{u}r extraterrestrische Physik (MPE), Giessenbachstrasse 1, D-85748 Garching bei M\"{u}nchen, Germany}

\author[0000-0001-7146-4687]{Yen-Ting Lin}
\affiliation{Institute of Astronomy and Astrophysics, Academia Sinica  (ASIAA),  Taipei 10617, Taiwan}

\author{Song Huang}
\affiliation{Department of Astrophysical Sciences, Princeton University,  Princeton, NJ 08544, USA}
\affiliation{Department of Astronomy, Tsinghua University, Beijing 100084, China}

\author{Dylan Nelson}
\affiliation{Institut f\"{u}r theoretische Astrophysik, Zentrum f\"{u}r Astronomie, Universit\"{a}t Heidelberg, D-69120 Heidelberg, Baden-W\"{u}rttemburg, Germany}

\author[0000-0002-9495-0079]{Vicente Rodriguez-Gomez}
\affiliation{Instituto de Radioastronom\'ia y Astrof\'isica,  Universidad Nacional Aut\'onoma de M\'exico, Apdo. Postal 72-3, 58089 Morelia, Mexico}

\author{Hsuan-Ting Lai}
\affiliation{Graduate Institute of Astronomy, National Central University,  Taoyuan City 32001, Taiwan}
\affiliation{Institute of Astronomy and Astrophysics, Academia Sinica  (ASIAA),  Taipei 10617, Taiwan}

\author{Jenny Greene}
\affiliation{Department of Astrophysical Sciences, Princeton University,  Princeton, NJ 08544, USA}

\author{Alexie Leauthaud}
\affiliation{University of California Observatories - Lick Observatory, University of California, Santa Cruz,  Santa Cruz, CA 95064, USA}

\author[0000-0001-8215-1256]{Alfonso Arag\'{o}n-Salamanca}
\affiliation{School of Physics and Astronomy, University of Nottingham, University Park, Nottingham, NG7 2RD, UK}

\author{Kevin Bundy}
\affiliation{University of California Observatories - Lick Observatory, University of California, Santa Cruz,  Santa Cruz, CA 95064, USA}

\author{Eric Emsellem}
\affiliation{European Southern Observatory,  85748 Garching, Germany}

\author[0000-0002-4202-4727]{Michael Merrifield}
\affiliation{School of Physics and Astronomy, University of Nottingham, University Park, Nottingham, NG7 2RD, UK}

\author{Surhud More}
\affiliation{The Inter-University Centre for Astronomy and Astrophysics, Pune, India}

\author{Nobuhiro Okabe}
\affiliation{Graduate School of Advanced Science and Engineering, Hiroshima University,
 Hiroshima,
739-8526, Japan}

\author[0000-0002-2204-6558]{Yu Rong}
\affiliation{Chinese Academy of Sciences South America Center for Astronomy, National Astronomical Observatories, Chinese}
\affiliation{Institute of Astrophysics, Pontificia Universidad Cat\'olica de Chile, Av.~Vicu\~na Mackenna 4860, 7820436 Macul, Santiago, Chile}

\author[0000-0002-8725-1069]{Joel R. Brownstein}
\affiliation{Department of Physics and Astronomy, University of Utah,  Salt Lake City, UT 84112, USA}

\author[0000-0003-1805-0316]{Richard R.~Lane}
\affiliation{Centro de Investigaci\'{o}n en Astronom\'{i}a, Universidad Bernardo O'Higgins, Avenida Viel 1497, Santiago, Chile}

\author{Kaike Pan}
\affiliation{Apache Point Observatory and New Mexico State
University,  Sunspot, NM, 88349-0059, USA}

\author[0000-0001-7240-7449]{Donald P.~Schneider}
\affiliation{Department of Astronomy and Astrophysics, The Pennsylvania State University, University Park, PA 16802}
\affiliation{Institute for Gravitation and the Cosmos, The Pennsylvania State University, University Park, PA 16802}

\begin{abstract}

Although it is generally accepted that massive galaxies form in a two-phased fashion, beginning with a rapid mass buildup through intense starburst activities,  followed by primarily dry mergers that mainly deposit stellar mass at outskirts, the late time stellar mass growth of brightest cluster galaxies (BCGs), the most massive galaxies in the universe, is still not well understood.  Several independent measurements have indicated a slower mass growth rate than predictions from theoretical models.  We attempt to resolve the discrepancy by measuring the frequency of BCGs with multiple-cores, which  serve as a proxy of the merger rates in the central region and facilitate a more direct comparison with theoretical predictions.  Using 79 BCGs at $z=0.06-0.15$ with integral field spectroscopic (IFS) data from the Mapping Nearby Galaxies at APO (MaNGA) project, we obtain a multiple-core fraction of $0.11 \pm 0.04$ at $z\approx 0.1$ within a 18\,kpc radius from the center, which is comparable to the value of $0.08 \pm 0.04$ derived from mock observations of 218 simulated BCGs from the cosmological hydrodynamical simulation {\it IllustrisTNG}.  We find that most of cores that appear close to the BCGs from imaging data turn out to be physically associated systems.  
Anchoring on the similarity in the multiple-core frequency between the MaNGA and IllustrisTNG, we discuss the mass growth rate of BCGs over the past 4.5\,Gyr.
\end{abstract}


\section{Introduction}\label{sec:intro}

In the current cosmological paradigm, the mass content of the universe is dominated by cold dark matter (CDM) and the expansion is governed by the so-called dark energy (which could take the form of a cosmological constant $\Lambda$). Structure formation proceeds in a bottom-up fashion; small dark matter halos form first, then grow by merging and accreting smaller halos (e.g., \citealt{peebles82}, also see \citealt{baugh06} for a review). In modern theories of galaxy formation, galaxies are believed to form within dark matter halos \citep[e.g.,][]{rees77,swhite78,  swhite91}. The dominant galaxy in a halo is often referred to  as the central galaxy, and all other galaxies as satellites. As halos grow by mergers, their galaxy population grows correspondingly. Particularly, because of  dynamical friction, massive galaxies from an infalling halo would typically sink quickly to the center of the larger halo and merge with the central galaxy, creating an even more massive galaxy, a process once called ``galactic cannibalism'' \citep{ostriker75}. At the present time, the culmination of the hierarchical structure formation is clusters of galaxies, whose central galaxies are often known as ``brightest cluster galaxies'' (BCGs).

The growth paths BCGs experience potentially contains important constraints on cluster and galaxy formation. BCGs are usually found  at or near the center in their host cluster (e.g., \citealt{lin04b}).
It is observed that $K_s$-band luminosity or stellar mass of BCG shows a correlation with the mass and velocity dispersion of its host cluster \citep[e.g.,][]{lin04b, whiley08, lidman12, kravtsov18,gorden-marx21}. 
Furthermore, the extended stellar envelop of BCGs could potentially serve as a better proxy of cluster halo mass than richness \citep{huang21}.

A variety of evidence points to the special status of BCGs among  cluster member galaxies. 
Because of its central location within the host cluster, galactic cannibalism inevitably takes place.
The tidal debris stripped from  cluster galaxies contributes to the light of central galaxies \citep{richstone76}. 
In addition, BCGs are found to form a separate population, distinctive with respect to the extreme of the cluster galaxy luminosity/stellar mass function  
\citep{tremaine77,lin10, rong18,dalal21}. 
Moreover, the major axis of BCGs are found to often align with the cluster orientation \citep{sastry68, niederste10}.

Recent numerical simulations and semi-analytical models (SAMs) suggest that massive galaxies, BCGs included, form in a two-phase scenario. Stars form intensely in the progenitors at high redshifts, and  late time ($z<1$) assembly is dominated by dissipationless mergers \citep[e.g.,][]{delucia07,oser10,laporte13,rodriguez-gomez16,rogone18,jing21}.
However, 
there appears to be a discrepancy in  BCG stellar mass growth between model predictions and observations. 
\citet{lin13} find
a good agreement in the mass growth history between observations and model prediction at
$z=0.5 -1.5$; however, there seems to be a halt in the growth of real BCGs at $z<0.5$, while the model BCGs continue to grow.
\citet{inagaki15}
investigate the mass growth in BCG at $z<0.5$, using the so-called ``top-$N$'' method (that is, selecting the top $N$ most massive clusters in a given comoving volume over different cosmic epochs), and conclude from observations that the mass growth is less than 14$\%$ between $z=0.4$ and $0.2$, while the SAM of \citet{guo11} predicts at least 30$\%$. 
Similarly,
\citet{lidman12} find a factor of 1.5 times smaller mass growth at $z=0.3-1$ compared to the simulation prediction (see also \citealt{zhang16}).
A recent work by \citet{lin17}, using deep photometry from the Subaru Hyper Suprime-Cam Survey \citep{aihara18}, shows that BCGs typically grow by about 35\% between $z=1$ and $z=0.3$ (again using the top-$N$ approach), while the SAM of \citet{guo13} suggests a factor of two larger growth rate.

Such a discrepancy could be explained if, for mergers occurring at late times, BCGs mainly accrete mass into their extended outskirts,  beyond the observational photometry apertures \citep{whiley08, inagaki15}. \citet{rogone18} analyze hydrodynamical simulations and obtain a smaller stellar mass growth factor that is consistent with observations by using an aperture similar to that of observations (30 \& 50\,kpc).
This result suggests that a more direct comparison between observation and simulation is required to solve this discrepancy. However, it is difficult to measure the total luminosity of BCGs, which often have extended surface brightness profiles in the crowded cluster regions. 
It requires not only deep imaging data with a flattened sky and very careful treatments of background subtraction and source masking, but also sophisticated modeling techniques
\citep[e.g.,][]{Huang2013TheGalaxies, Huang2016TheGalaxies, meert13, meert16, huang18,fischer19,wang21}.

Another approach to  this problem is to measure the merger rate of BCGs close to their centers. The $N$-body simulations of \citet{gao04} suggest that BCGs have gone through many merging events that bring material to the  innermost region of $\sim 10\,$kpc, even at $z<1$. 
This implies that these mergers, corresponding to the ``second phase'' in the two-phase scenario mentioned above, not only affect the outskirts of the BCGs, but also have strongly observable effects in the central region. 

One can define the merger rate
as the probability of a BCG with two or more closely-separated cores to be observed per unit time:\\
\begin{equation}\label{form:merger}
\mathcal{R} = \frac{ N_{\rm multiple-core} }{ N_{\rm BCG} }  \frac{ 1 }{ t_{\rm vis} },
\end{equation}
which is the combination of the ``multiple-core frequency'' 
with a merger timescale, which we term the ``visibility time'' here. Very close pairs are also called multiple-nuclei or multiple-cores,  because the secondary/satellite galaxies, during the merger process with a BCG, often appear as an additional core of the BCG 
\citep{schneider83a,lauer88}. We use the term ``multiple-core frequency'' $f_{\rm mc}$ for the fraction of BCGs that appear as multiple-cored in a volume-limited sample. 
The visibility time, defined to be the duration for a satellite to remain ``visible'' (i.e., identifiable from imaging or spectroscopy) during the course of galactic cannibalism, has to be derived from numerical simulations, or estimated from theory. On the other hand, the multiple-core frequency is an observable that provides the opportunity for a direct comparison between observations and models. The same quantity for pairs with larger separation (e.g., when the two galaxies are clearly seem as separate entities) is often named ``pair fraction'' in the literature \citep[e.g.,][]{liu09, mcintosh08, groenewald17}.

The pair fraction as a critical step toward deriving merger rates of central galaxies in massive halos such as groups and clusters has been widely used \citep[e.g.,][]{edwards12, burke13, lidman13}.
While morphological distortions of galaxies in a pair can be an indication of interaction, thus serving as an (indirect) indicator of physical association of the pair \citep{lauer88, mcintosh08, liu09, liu15}, the most reliable way to identify pairs is through spectroscopy \citep[e.g.,][]{groenewald17}.
\citet{brough11} and \citet{jimmy13} conduct the first targeted integral field spectroscopy (IFS) observation of BCGs with close companions.

In this work, we present a measurement of multiple-core frequency of the largest sample of BCGs to date, using IFS data from the Mapping Nearby Galaxies at APO (MaNGA; \citealt{bundy15, drory15, law15, law16, yan16, yan16b, law21}) project, which is part of the
fourth generation of Sloan Digital Sky Survey (SDSS-IV; \citealt{blanton17,gunn06,smee13}).
We further compare our measurement with results from the cosmological hydrodynamical simulation IllustrisTNG \citep{weinberger17,springel18,pillepich18,pillepich18b,naiman18,marinacci18,nelson18,nelson19} to examine the consistency between observations and models.

This paper is structured as follows.  In Section~\ref{sec:basics} we present the essential ingredients of our analysis, including the cluster sample, the imaging and IFS data, and the simulation.  In Section~\ref{sec:IFU} we describe in detail our method for extracting the multiple-core frequency: from core detection in images to confirmation of physical association using MaNGA velocity maps.  We carry out a similar analysis on mock images of simulated BCGs in Section~\ref{sec:TNG}.  We compare our results with some of the findings from the literature in Section~\ref{sec:dis}, where we also show that the BCG samples used in our analysis are unbiased with respect to the general BCG population.
We conclude in Section~\ref{sec:conclusion}.
In Appendix~\ref{appendix:Re_const} we present a comparison  of several kinds of photometric measurements used in our analysis, showing a consistency among them.  In Appendix~\ref{appendix:note} we describe  BCGs that either require special treatment for their photometry, or have to be excluded due to various reasons.
We list our BCG sample and the detected cores in Appendix~\ref{appendix:tab}.

We adopt a cosmology with a Hubble constant of ${H_0}=100\, h\, \mathrm{km\, s}^{-1}\mathrm{Mpc}^{-1}$, with $h=0.73$, $\Omega_M=0.27$, and $\Omega_\Lambda=0.73$ throughout this paper. We use halo mass defined as $M_{180m}$ in observations and  $M_{200m}$ in the simulation. These are  the mass contained within a radius $R_{180m}$ ($R_{200m}$) within  which the mean  density is 180 (200) times the mean density of the universe. The difference between $M_{180m}$ and $M_{200m}$ is within 2\% so the two can be approximated as the same quantity.

\section{Elements of Analysis}
\label{sec:basics}

\subsection{The MaNGA BCG Sample}
\label{sec:mangabcg}

MaNGA has obtained spatially resolved spectroscopy for 
 about 10,000 galaxies out to $z=0.15$. The  data is obtained by integral field units (IFUs) built with  fiber bundles, which have diameters ranging from $12''$ to 32$''$, providing a spatial sampling $1-2$\,kpc (at the typical redshift of MaNGA galaxies, $z\approx 0.03$). The MaNGA sample is constructed to have a flat stellar mass distribution, and  consists of the primary, secondary, color-enhanced, and ancillary samples \citep{wake17}. The primary sample has their IFU coverage to 1.5 times the effective radius ($R_e$), and the secondary to 2.5\,$R_e$. The ancillary programs focus on special types of galaxies such as massive galaxies, merger candidates, active galaxies, etc. 
 In particular, the ``BCG'' ancillary program has enabled comprehensive studies of the kinematic morphology--density relation and the angular momentum content of massive, central galaxies \citep{greene17,greene18}.

Our parent BCG sample is taken from the group and cluster catalog of \citet[][hereafter Y07]{yang07}, updated to the version based on the SDSS data release 7  \citep[DR7;][]{sdssdr7}.
Among the 3 versions of catalogs  provided,
we adopt the one that is constructed using the SDSS model magnitude\footnote{Using the version based on the Petrosian magnitude would affect the BCG selection at less than 1\% level (X.~Yang, 2022, private communication).} and includes additional redshifts from the literature, in order to 
have the largest number of clusters.
We apply a cut in the cluster mass $M_{180m}\geq 10^{14} \, h^{-1}M_{\odot}$,
which results in 4033 clusters. We note in passing that the halo mass provided by Y07 is estimated by the ranking of total stellar mass of a cluster/group. 

 The details of BCG selection are described in \citet[][see Section 3.2 therein]{yang05c}.  Note that the BCG is the most luminous galaxy among the members and may not necessarily be close to the cluster center (e.g., \citealt{skibba11}), which is the geometric and luminosity-weighted center of member galaxies.
Matching the 4033 BCGs with the 8113 galaxies released from MaNGA Product Launch-9 (MPL-9), 
we obtain 128 BCGs. These galaxies belong to the MaNGA primary,  secondary, and color-enhanced sample, as well as the ``BCG'' and ``MASSIVE'' ancillary programs \citep{wake17}. The 128 clusters lie at  $z =0.02-0.15$, and are all detected in the X-rays by \citet{wang14}. Hereafter we shall refer to this sample as  ``MPL-9 BCGs'' (see Tables~\ref{tab:defn} and \ref{tab:BCG}).

\begin{deluxetable*}{ccc}
\tablecaption{Definition of BCG samples\label{tab:defn}}

\tablehead{
\colhead{Name} & \colhead{Number} & \colhead{Definition} 
} 

\startdata
All & 4033 & All of Y07 clusters  with $M_{180m}\geq 10^{14} \, h^{-1}M_{\odot}$ (\S~\ref{sec:mangabcg}) \\
MPL-9 & 128 & ``All'' matched to MaNGA MPL-9 (\S~\ref{sec:mangabcg}) \\
Main & 79 & ``MPL-9'' with problematic BCGs removed and having IFU coverage to $\ge 18$\,kpc  (\S~\ref{subsubsec:IFU cov}) \\
Volume-limited & 73 & Same as ``Main'', but with  stellar mass above stellar completeness limit (Eqn.~\ref{eqn:c_lim}) (\S~\ref{subsubsec:vlim})\\
Parent & 1359 & Same as ``All'', but  above the completeness limit  and at $z=0.02-0.15$ (\S~\ref{subsubsec:vlim})\\
TNG-Comparison & 225 & Similar to  Parent, but  at $z=0.07-0.11$ and within a volume of (300\,Mpc)$^3$ (\S~\ref{sec:tngsample})\\
Not-in-MaNGA & 1237 & Same as ``Parent'', but excluding the MPL-9 sample (\S~\ref{appendix:unbias})\\
\enddata



\end{deluxetable*}

The algorithm used in the cluster finder of \citet{yang05b} selects the BCG solely based on the luminosity. However, sometimes the brightest galaxy in a cluster has a spiral morphology.
Given that it is unlikely for a  central galaxy of a virialized, matured cluster to be a spiral \citep{coziol09}, we decide to visually inspect all MPL-9 BCGs using images from SDSS,
with the aid of the $g-r$ {\it vs.}~$i$ color-magnitude diagrams of cluster members (see e.g., Fig.~\ref{fig:changed_cmd} for examples). 
In this paper we regard BCGs to be of early type morphology\footnote{It is found by \citet{zhao15} that about 4\% of the 625 BCGs studied by \citet{vonderlinden07} are spiral galaxies.}, and also the most luminous galaxies in each cluster.
If the BCG candidates show a spiral morphology (which makes it quite difficult for the detection of multiple-cores given our methodology as described below; in total  5 spiral BCG candidates are discarded), or are not the brightest galaxy on the red sequence, 
we search for other possible candidates. If there is no better candidate, or the better candidate is not observed by MaNGA, we remove the cluster from the sample. Six clusters are removed due to the above reasons ( please see Appendices \ref{appendix:changed} and \ref{appendix:BCG_id} for more details). 
The Coma cluster is also removed because it does not have the same type of data products as other BCGs in our sample.  
Therefore, we obtain 121 visually confirmed BCGs.
 We emphasize that our BCG selection is primarily that of Y07; only 5 of the 128 galaxies initially defined as BCGS were redefined through visual inspection (see Appendices \ref{appendix:changed} and \ref{appendix:BCG_id}). Our main conclusion is not affected by redefining these 5 BCGs.
%

 In addition, there are 6 BCGs with  nearby bright stars, or are severely affected by the edges of mosaic frames that would make their photometry unreliable (see Section~\ref{subsubsec:Phot}). Since these events are independent of the multiple-core frequency, these BCGs are also removed from our analysis (see Appendix \ref{appendix:bad_im}). There are 115 BCGs left after these cuts.

\subsection{The BCG Sample from IllustrisTNG}
\label{sec:tngbcgsample}

IllustrisTNG is a series of cosmological hydrodynamical simulations which has three simulation volumes, TNG50, TNG100, and TNG300. We use the simulation TNG300-1 (hereafter simply TNG300), which has a box size of 300\,Mpc on a side, to maximize our sample size of simulated BCGs.  TNG300-1 has $2\times 2500^3$ resolution elements, and  a mass resolution of $1.1\times10^7 M_{\odot}$ for baryons and $5.9\times10^7 M_{\odot}$ for dark matter.  The gravitational softening length (for stars and dark matter) of 1.5\,kpc at $z=0$.

The average redshift of our MPL-9 sample of 128 BCGs is $z=0.1$.
Therefore, we select BCGs from snapshot 91 of TNG300, which corresponds to $z=0.0994$. There are 225 halos with mass $M_{200m} \geq 10^{14} h^{-1}M_{\odot}$ identified using the friends-of-friends algorithm \citep{davis85}. 
The main subhalos of these halos are identified as the BCGs (via the SUBFIND algorithm; \citealt{springel01b}).
In Section~\ref{sec:TNG} we will describe our methods to mimic the selection of BCGs as closely as possible to our MaNGA sample, and how we derive multiple-core frequency from synthetic images of the resulting BCG sample.

\section{An IFU Survey of Multiple-Core Frequency of Brightest Cluster Galaxies}\label{sec:IFU}

Our analysis consists of 4 steps: (1) modeling the light distribution of a BCG using SDSS imaging data; (2) subtracting the best model of the BCG from the image and measuring the position and fluxes of the core(s), if present; (3) finding the counterpart(s) of the core(s) in the MaNGA stellar velocity map, and determining whether the core(s) are physically associated with the BCG; and (4) estimating the core-to-BCG flux ratio.  
Furthermore, we need to examine the completeness of our BCG selection, and apply  correction factors where needed.
We describe each of these steps in detail in the following.

Although in Eqn.~\ref{form:merger} it is implied that the multiple-core frequency is simply the number of BCGs with multiple cores ($N_{\rm BCG, mc}$) divided by the total number of BCGs in a complete sample ($N_{\rm BCG}$), in reality, there are often cases where more than one satellite is merging with a BCG (i.e., there will be $>1$ cores).   Given that each merger event should be independent, we thus define formally the multiple-core frequency $f_{\rm mc}$ to be 
\begin{equation}
f_{\rm mc} \equiv N_{\rm multiple-core}  / N_{\rm BCG},
\label{eq:mc}
\end{equation}
where the numerator on the right denotes the total number of cores, instead of the number of BCGs with multiple cores.

\subsection{Photometry of BCGs}
\label{subsubsec:Phot}

The BCG photometry is somewhat  ill-defined for several reasons. First, 
the surface brightness profiles of elliptical galaxies can usually be described by a S\'{e}rsic \citep{sersic63} profile with an index $n\gtrsim 3$ or so (e.g., \citealt{kormendy09}), and such profiles, as well as actual observations, do not exhibit a well-defined/sharp edge.
Moreover, since BCGs are very spatially extended, with a substantial fraction of their flux below the sky  level, we can only extrapolate the profile we assumed to obtain the flux in this unconstrained region.
Second, BCGs have much more complex profiles than common ellipticals, and it may require $>2$ S\'{e}rsic components to describe their surface brightness profiles. The properties such as position angle or color of the inner to outer region of BCGs can be quite different \citep{Huang2013TheGalaxies, Huang2016TheGalaxies}.
Third, BCGs are often located in crowded regions. Cluster members surround, touch, or merge with BCGs, making it  difficult to mask them out or deblend them from BCGs without affecting the photometric measurement.
These all add to the uncertainty in the photometry of BCGs. Below we discuss how we obtain  BCG photometry that is reliable enough for our needs.

We have two ways to obtain the photometric measurements, such as $R_e$ and total magnitude.
Our primary resources are the photometric catalogs of \citet[][hereafter M16]{meert16}  and \citet[][hereafter F19]{fischer19}. These catalogs are generated by the 2D fitting pipeline {\tt{PyMorph}} \citep{meert13,meert15} that uses {\tt GALFIT} \citep{peng02} as the engine for galaxy morphology modeling, and have a better estimation of the brightness than the SDSS pipeline for the most luminous galaxies, because of better sky subtraction, as well as  more flexible modeling (2 S\'{e}rsic components;
\citealt{bernardi17}). We use the ``Best model'' table of M16 and remove the galaxies flagged as bad (flag\,$=20$). 
For the BCGs that do not have a good fit in M16, we use the F19 catalog. F19 mark the preferred model for each galaxy with the ``FLAG\underline{ }FIT'' flag; if there is no preference, we use the S\'{e}rsic$+$Exponential model. 74 out of our 115 BCGs have reliable magnitude and $R_e$ measurements from these 2 catalogs.

For  BCGs not included in either of the catalogs of M16 or F19, we obtain their total magnitudes by running the code {\tt Ellipse} on SDSS mosaic images (see below). {\tt Ellipse}  is a fully automated Python package for fitting ellipses to isophotal contours of galaxies, developed  by Dr.~G.~Torrealba\footnote{\url{https://github.com/Grillard/GalfitPyWrap}}. 
As a consistency check, we
 fit single S\'{e}rsic profile to the surface brightness measured by {\tt Ellipse} and find good agreement in the total flux  with M16 and F19 catalogs (please see Appendix \ref{appendix:Re_const} for more details).

The images of BCGs are taken from SDSS DR12  \citep{sdssdr12}.
Our BCG sample has a typical $R_e$ of $10''$ at its mean redshift of 0.1. 
Given the size of the BCGs, we need to have a large enough area to capture the extended profile and successfully perform the sky subtraction. 
Since the BCGs often do not lie within one single ``corrected frame'' of SDSS, we need to construct mosaic'd images, which are obtained from
the SDSS DR12 Science Archive Server (SAS) as well as through the URL tool of DESI Legacy Imaging Survey\footnote{\url{https://www.legacysurvey.org/dr9/description/}} \citep{dey19}. 
We have confirmed that the images obtained from the two methods are identical. 
In practice, we use the URL tool of the DESI Legacy Imaging Survey because the SDSS SAS does not support bulk downloads. 
We use the $i$-band images for modeling the BCGs, because they show the multiple-core features most clearly, and some cores are only distinguishable from the BCG in the $i$-band.

\subsection{Maximum Projected Distance and IFU Coverage for Core Detection}
\label{subsubsec:IFU cov}

As we want to focus on mergers taking place in the central parts of BCGs, we need to define a maximum distance (from the BCG center) for our search of secondary cores.  There are two factors in our consideration for the maximum distance.  The first one is the aperture size of the IFUs, as it directly limits the maximum separation of multiple-cores practically.  The second one is whether to have the distance defined to be a certain fraction of $R_e$.  We choose to use a fixed metric distance, so that a direct comparison can be made when applying our procedures to mock images of simulated BCGs (see Section~\ref{subsubsec:TNG_MC}).

The median $R_e$ of the 74 BCGs with photometric measurements from M16 and F19 is 17.5\,kpc.  Balancing between the IFU coverage and the maximum projected distance, we decide to select the BCGs that are covered by their IFU to at least 18\,kpc, in order to have the largest sample size (which effectively also sets a lower redshift limit in our sample at $z\approx 0.06$).
Our final sample consists of 79 BCGs, which shall be referred to as the ``Main'' sample (Table \ref{tab:defn}).

\subsection{Identifying Multiple-Cores in SDSS Images}
\label{subsubsec:multi}

\begin{figure*}[]
\epsscale{1.2}
\plotone{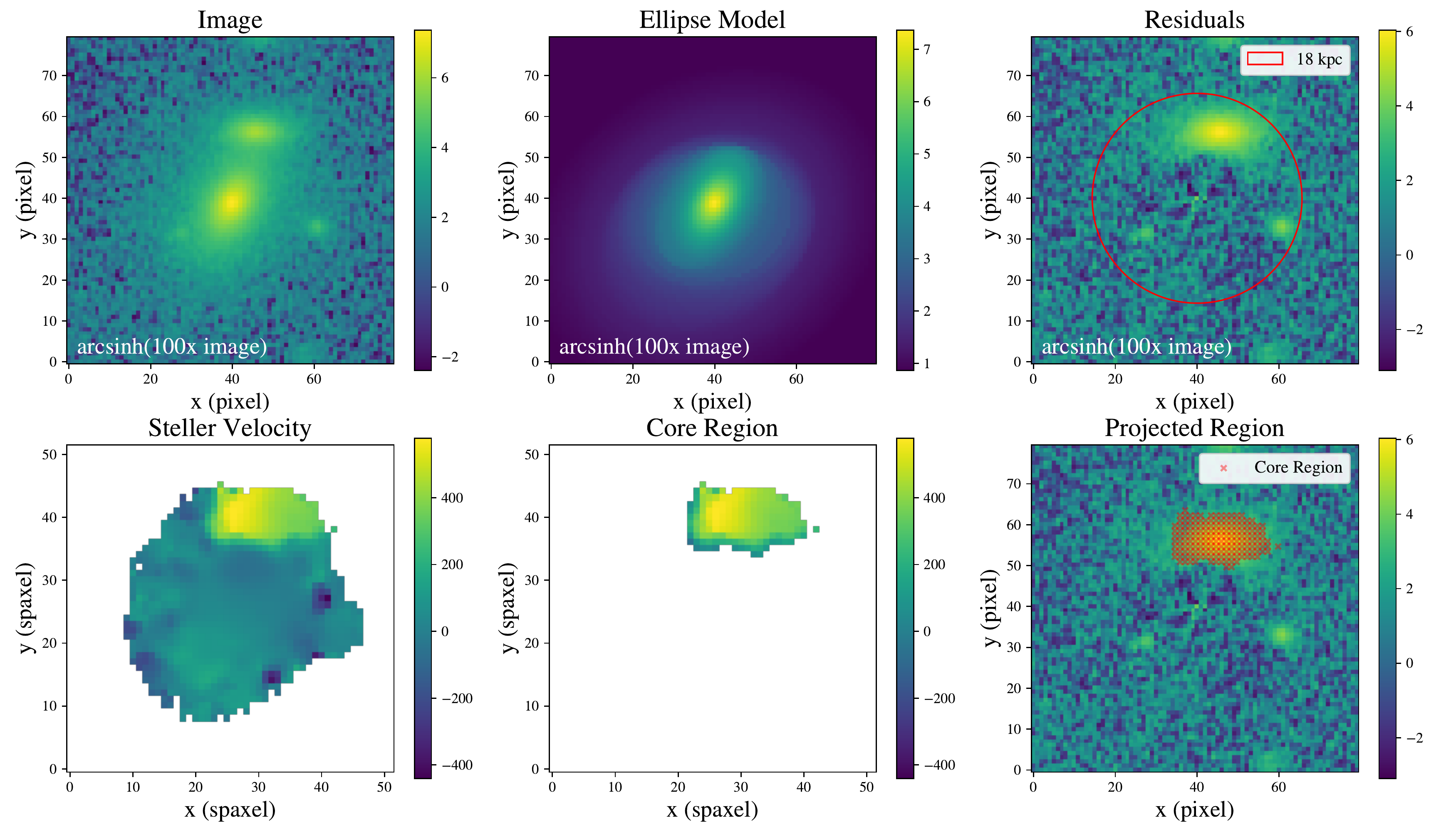}
\caption{
Demonstration of the procedure of  core detection, using BCG No.~6 in our sample as an example. The upper row shows the SDSS $i$-band image of BCG No.~6, the model produced by {\tt Ellipse}, and the residual map. These three images are displayed in the stretching of arcsinh(100$\times$), and zoomed in to the central 80 pixels. {\it The values shown on the color bars correspond to  arcsinh(100$\times$ flux/nanomaggy), a convention used in all such SDSS (or mock SDSS) images throughout the paper.}
The red circle of 18\,kpc from the BCG center is plotted on the residual map. The pixel scale of the SDSS images is $0.396''$/pixel.
The lower row shows the stellar velocity map with systemic velocity corrected (if needed), the detected core segmentation region, and the position of the core segmentation spaxels over the residual map. Note that the two rows are not plotted on the exact same scale.\label{fig:core_step}}
\end{figure*}

After downloading the SDSS mosaics,
the images are cropped to sizes between 500$\times$500  pixels and 1818$\times$1818 pixels ($6'\times 6'$) for further analyses. 
We focus on the profile within 150\,kpc, which corresponds to an image size of 682 pixels for the most nearby BCG in our sample. We also generate  axisymmetric  galaxy models with {\tt{GALFIT}} to examine the effect of limited image size on the recovery of $R_e$ and total flux. For galaxy models with $R_e$ in the range of  $10-40$ pixels and S\'{e}rsic index between $1-8$, 
results from our tests suggest that
images of 800$\times$800 pixels can result in $78-100\%$ of the true $R_e$. Hence image sizes larger than 800 pixels on a side would serve our goals well. 

We feed these images cutouts to the source extraction software {\tt SExtractor} \citep{bertin96} to obtain their segmentation maps. 
By varying parameters such as BACK$\_$SIZE (controlling the size of the grid of background measurement), the way weight maps are obtained (either supplied by SDSS or generated by {\tt SExtractor}), and the sizes of input images, we find that the resulting maps do not sensitively depend on these choices.
Small differences occur occasionally on some images with very bright stars or very crowded regions. We mainly use the 800$\times$800 pixel images and 
set BACK$\_$SIZE\,$=160$ (that is, 1/5 of the image size).
In 4 cases (out of 89), we need to resort to 1000$\times$1000 pixel images in order to obtain a reasonable segmentation map.

To detect the cores in the images, we need to remove the light of the  main body of the BCGs. We subtract the {\tt SExtractor} measured background from the images, masked out the segmentation region of the sources touching the BCGs, and substitute the masked regions that are not connected to the BCG with a Gaussian noise that has the same standard deviation as the sky 
 measured by {\tt SExtractor}. 
These images, now with the BCG left as the only source, are then fed to {\tt Ellipse}, which outputs empirical surface brightness models and profiles (in a similar fashion to the ``ellipse'' task in {\tt IRAF}; \citealt{jedrzejewski87}). We subtract the empirical models from the image  to obtain the ``BCG-free'' residual maps. An example of an image, the model, and a residual map is shown in the upper row of Fig.~\ref{fig:core_step}.

On the residual maps, we run the Python implementation of {\tt SExtractor}, {\tt SEP} \citep{barbary16}, with strong deblending parameters\footnote{Detection threshold\,$=0.2$, minimum area\,$=1$, deblend threshold\,$=64$, deblend contrast\,$=0.0001$, clean parameter\,$=1.0$.} and a low detection threshold to detect any possible core with $1.8-18$ kpc separation from the BCG center. 
The lower limit, 1.8\,kpc, is set to avoid identifying residuals of the BCG main body due to imperfections in the model as spurious cores; such cases, if present, will be safe-guarded by our next step (kinematic confirmation via MaNGA velocity maps) as well as our final visual inspection.  In addition, such a lower limit can avoid the blurring of images due to seeing.
The upper-right panel of Fig.~\ref{fig:core_step} shows the 18 kpc circle and a detected core.

\subsection{Identifying True Merging Systems with MaNGA Velocity Maps}\label{subsubsec:merg}

To distinguish the merging systems from chance projections, we make use of the Python package {\tt Marvin} \citep{cherinka19}, specifically designed to display and conduct calculations with various IFU maps produced by the MaNGA Data Analysis Pipeline \citep[DAP;][]{westfall19,belfiore19}. 
We apply the DAP ``DONOTUSE'' mask to the maps to avoid spaxels that are not suitable for scientific analyses.

Subsequently, we need to remove any contribution from the systemic velocity of the galaxy. The MaNGA stellar velocity maps are corrected to the redshift from the NASA-Sloan Atlas\footnote{\url{http://nsatlas.org/}} (NSA) catalog if available, otherwise the redshift is estimated by the  DAP. However, sometimes, especially for complex galaxies that have multiple cores (or a fiber bundle containing foreground/background objects), the redshift can be inaccurate, or is not corrected to the object we identify as the  main body of the BCG. We deal with this issue through the following steps. We first take the minimum absolute value between the value of central spaxel and the   median value of the spaxels with Signal-to-Noise Ratio (SNR) larger than 10. If this absolute value is $\leq 50\,$km/s,  we regard this object to have a reasonable redshift measurement. 
If the absolute value is $> 50\,$km/s, the redshift may be problematic and requires correction. We set the new reference point at the median velocity of spaxels with SNR\,$>10$. This definition avoids contamination from the cores in the central region. We apply the equation in Sec.~7.1.4 in \citet{westfall19} to correct the velocity map relative to the new reference point. These corrected maps are used to calculate the velocity offset between the cores and the BCG.
For  BCGs with rotation features, these features might be detected as a large core by our 
extraction pipeline, however. Therefore, we manually select the velocity maps with strong rotation features, fit a 3D plane to it, and subtract the velocity structure  of that plane. We only use these subtracted maps in the extraction process, and do not use them when calculating the velocity offsets of the cores. The BCGs with strong rotation features are nos.~6, 25, 26, 39, 51, 102, 107, 116, 117, 124 (Table~\ref{tab:BCG}).

Once the velocity maps are systemic velocity-corrected and rotation-subtracted (if needed), we extract sources by running {\tt SEP}. The spaxels that have an SNR\,$< 3$ are masked out during the extraction process. 
If a core is detected in both the residual image and the velocity map within a tolerance separation, it is regarded as a robust detection.
The tolerance separation is set to be 3 times the 
geometric mean of the major and minor axes of the isophotal image output by {\tt SEP}, 
since this size appears to best resemble the core region  identified by visual inspection, and it generally well represents the isophotal limits of a detected object \citep{barbary16}. If there is more than one region on the velocity map that satisfies the criteria, the nearest one is considered as the (kinematic) counterpart. 
Given that {\tt SEP} only detected positive peaks, while the cores could have both positive or negative velocity offsets,  both the original map and its negative are source extracted.
If more than one secondary core associated with one BCG is confirmed, we record them separately. 

We have explored the  SNR threshold for the exclusion of spaxels.  
By varying the lower limit in SNR between $2-5$, we find that the effect is to
slightly change the sizes of the core segmentation area on the velocity maps, and a few faintest cores would not be detected if the SNR limit is high. They do not affect the relatively bright cores (see Section~\ref{subsubsec:f_ratio}) that are used in our main results. 

The bottom row of Fig.~\ref{fig:core_step} demonstrates the results of the core confirmation process. Afterwards, we remove stars that are not masked by MaNGA masks using the following procedure. We match the confirmed cores with objects in the Gaia early data release 3 catalog \citep{gaiaedr3, gaiaedr3b, seabroke21, lindegren21} and see if they have significant parallax or proper motion. We also match the cores with SDSS objects and see if they are classified as ``STAR''. If they do, we flag the cores as ``star'' and remove them. Finally, we remove false confirmations by visual inspection, which are caused by masked stars and the spaxels around the masked region. 

There is one  core having a SDSS spectrum showing it is a galaxy at a  redshift  ($z=0.23584$) different from  BCG no.~99 ($z=0.12935$), so it is removed.  It is curious that the galaxy does not show dramatic velocity difference in the MaNGA DAP velocity map (see below), which again shows the importance of visual inspection (for this particular case, the background galaxy is starbursting, hence its color is quite distinct from the typical red colors that cores associated with BCGs have).

Finally, we note that the MaNGA DAP assumes all objects (spaxels) in a given datacube belong to one single galaxy, and all spectra are fit with a range of $\pm 2000\,$km/s from the NSA redshift of the primary target \citep{westfall19}.  We have thus paid special attention to check whether the velocity offsets from the DAP of all cores are reliable, by examining the model fits to the spectra of the cores.

\begin{figure}[]
\epsscale{1.2}
\plotone{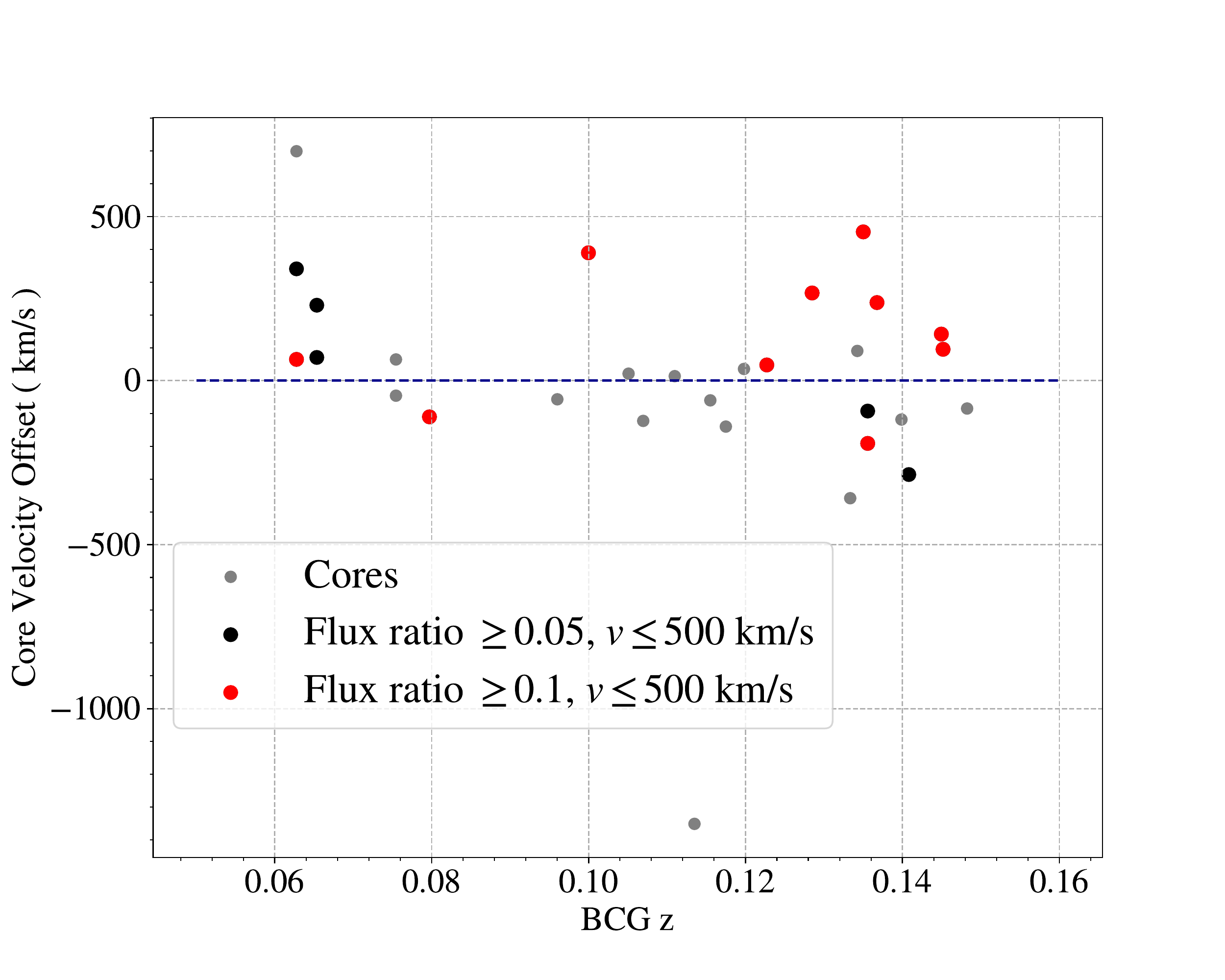}
\caption{
The velocity offset of the cores confirmed in velocity maps to be potentially associated with their BCGs. Excluding the 2 extreme points with velocity offset larger than 500\,km/s, there are 28 cores with a high possibility to merge with their BCGs.
Among these 28, we show the 5 cores with flux ratio $\mathcal{F}_{\rm core}=0.05-0.1$ as black points, while those 10 with $\mathcal{F}_{\rm core}\ge 0.1$ as red points.
Here a positive velocity offset means the core has a peculiar velocity moving away from us (relative to the BCG).
\label{fig:v_off}}
\end{figure}

The 30 confirmed cores and the velocity offsets from the main body of the median of their spaxels are shown in Fig.~\ref{fig:v_off} (all points). 
To select cores that have a high probability to merge with their BCGs, we limit ourselves to those with a maximum velocity offset of 500\,km/s;\footnote{We caution that this choice depends on the cluster velocity dispersion; for the clusters in our sample, which have mass about $10^{14}\,M_\odot$, this is adequate.} 28 out of 30 cores satisfy this cut.

\subsection{Flux Ratio}
\label{subsubsec:f_ratio}

\begin{figure*}[ht!]
\centering
\includegraphics[width=\textwidth]{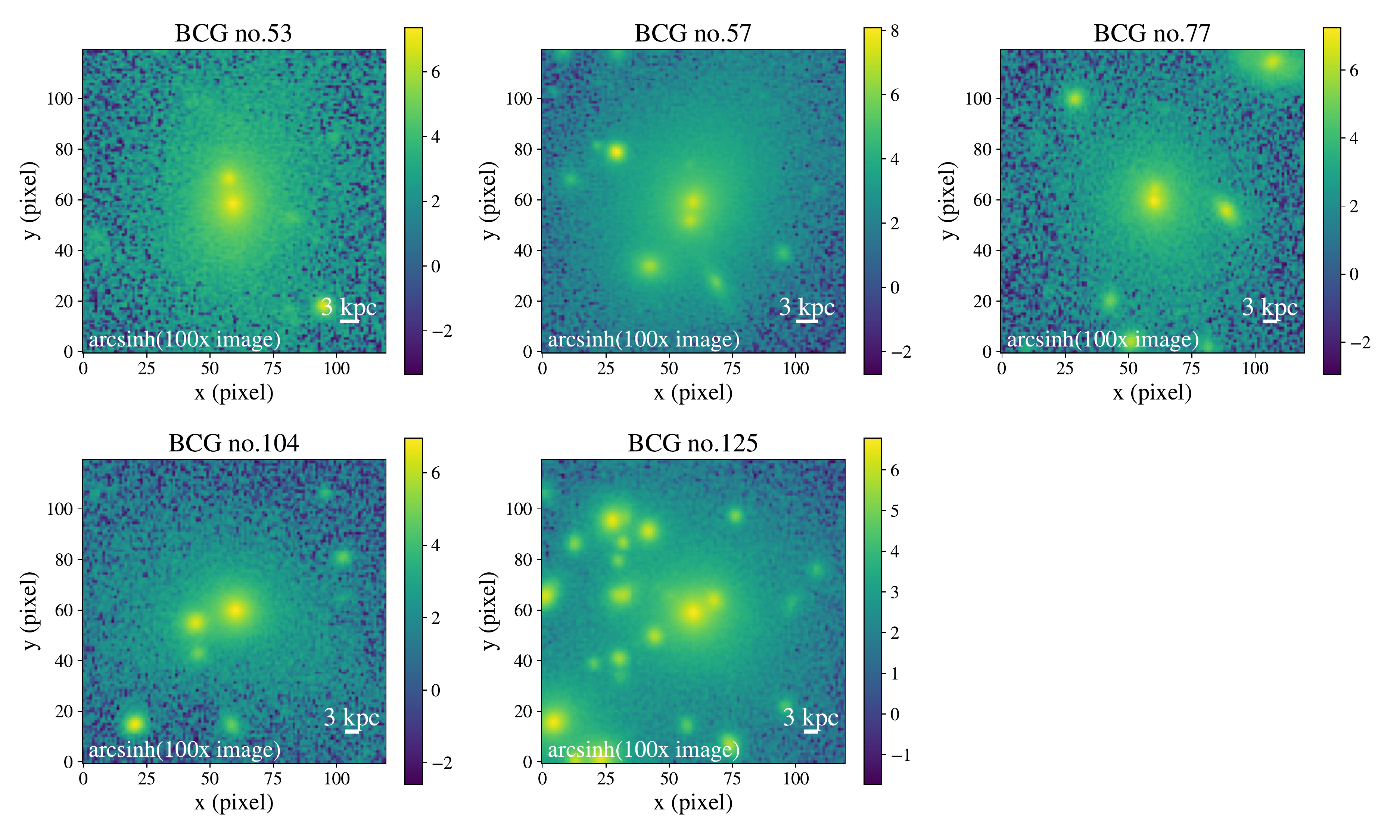}
\caption{\label{fig:maj}
The 5 additional major mergers identified by visual inspection. These images are displayed in the stretching of arcsinh(100$\times$), and zoomed in to the central 120 pixels (see caption of Fig.~\ref{fig:core_step} for more details).  The white horizontal bar indicates a scale of 3\,kpc.
}
\end{figure*}

Our next task is to determine the flux ratio between the detected cores and the main body of the BCG, $\mathcal{F}_{\rm core}$, which then allows us to estimate whether the merger is major (e.g., mass ratio $>4$) or minor.  However, given the following practical considerations, we have to set a lower limit in the flux ratio that we can measure.
First, for the cores with flux ratio $\mathcal{F}_{\rm core}=0.01-0.05$, the contamination rate from star grows quickly. Second, the tidal plumes, masked stars, uncleaned residuals start to cause false detections in this flux ratio range. Third, the systematic uncertainty of BCG photometry is at few percent level. 

Therefore, in this paper we present the multiple-core fraction with minimum flux ratios of 0.1 and 0.05, respectively. 
The distribution of flux ratio of our sample shows a trend that quickly decreases towards high values of $\mathcal{F}_{\rm core}$.  However, taking a closer look at the distribution, there appears to be a gap above $\mathcal{F}_{\rm core}>0.1$, justifying our choice of $\mathcal{F}_{\rm core,min}=0.1$.
For the flux estimates of the cores, we consider the maximum value among the following 7 kinds of measurements: 
 ({\it i}) the sum of pixels in the {\tt SEP} segmentation region defined on the image, ({\it ii}) the sum of the positive pixels of the {\tt SEP} segmentation region defined on the velocity map, and ({\it iii-vii}) the sum of the pixels within a radius of 1, 2,... to 5 kpc. 

Method ({\it i}) works the best for the large or non-circular cores, while method ({\it ii}) is best suited for  cores that are very close to the BCG center and thus often suffered from over-subtraction in the residual maps. Except for the one defined on the velocity map, the rest work well for the cores with different sizes or on the edge of the IFUs. There are 5 and 11 cores having flux ratios greater than 0.1 and 0.05 through the above procedures, respectively.

In close major mergers (i.e., when two cores of comparable brightness are very close in projection, $\lesssim 2$\,kpc), deblending and, in turn, getting good photometry of the secondary galaxies become exceedingly difficult, so we visually select the cases that are certain to have flux ratios larger than 0.1, in parallel to the automatic measurements mentioned above. There are 10 visually selected major mergers, including 5 that are detected by our pipeline. The extra 5 cases added by visual selection are shown in Fig.~\ref{fig:maj}. In short, there are 10 and 15 cores with flux ratio greater than 0.1 and 0.05, respectively (see Table~\ref{tab:ob_core}).   The velocity offsets of these cores are shown in Fig.~\ref{fig:v_off} (as black and red points).

\subsection{Completeness Correction and the Multiple-Core Frequency}
\label{subsubsec:vlim}

\begin{deluxetable*}{cccc}
\tablecaption{Statistics of Core Detection\label{tab:vol_lim}}

\tablehead{
\colhead{redshift range} & \colhead{$0.02-0.1025$} & \colhead{$0.1025-0.13$} & \colhead{$0.13-0.149$}
} 

\startdata
Parent sample & 590 & 409 & 360\\
Volume-limited (VL) sample & 22 & 26 & 25\\
Core number with $\mathcal{F}_{\rm core}\ge 0.1\ (0.05)$ in VL & 3 (6) & 2 (2)& 3 (5) \\
Core number expected in Parent & 80.5 & 31.5 & 43.2\\
Multiple-core frequency ($\mathcal{F}_{\rm core}\ge 0.1$) & 0.140 & 0.077 & 0.120\\
Multiple-core frequency ($\mathcal{F}_{\rm core}\ge 0.05$) & 0.273 & 0.077 & 0.200\\
\enddata



\end{deluxetable*}

The multiple-core frequency in Eqn.~\ref{form:merger} is defined for a volume-limited sample. So far we have been presenting the multiple-core measurements among the 79 BCGs of our Main sample, which does not constitute a volume-limited sample (see below). 
It is also not yet clear whether our BCG sample, constructed somewhat heterogeneously from the MaNGA primary, secondary, color-enhanced, and two ancillary programs, are a representative sub-sample of all BCGs at $z\le 0.15$.
In this Section,  
we describe our way of applying a completeness correction factor to the multiple-core frequency (also see a more detailed discussion in Section~\ref{sec:samplesel}).

\begin{figure}[]
\epsscale{1.2}
\plotone{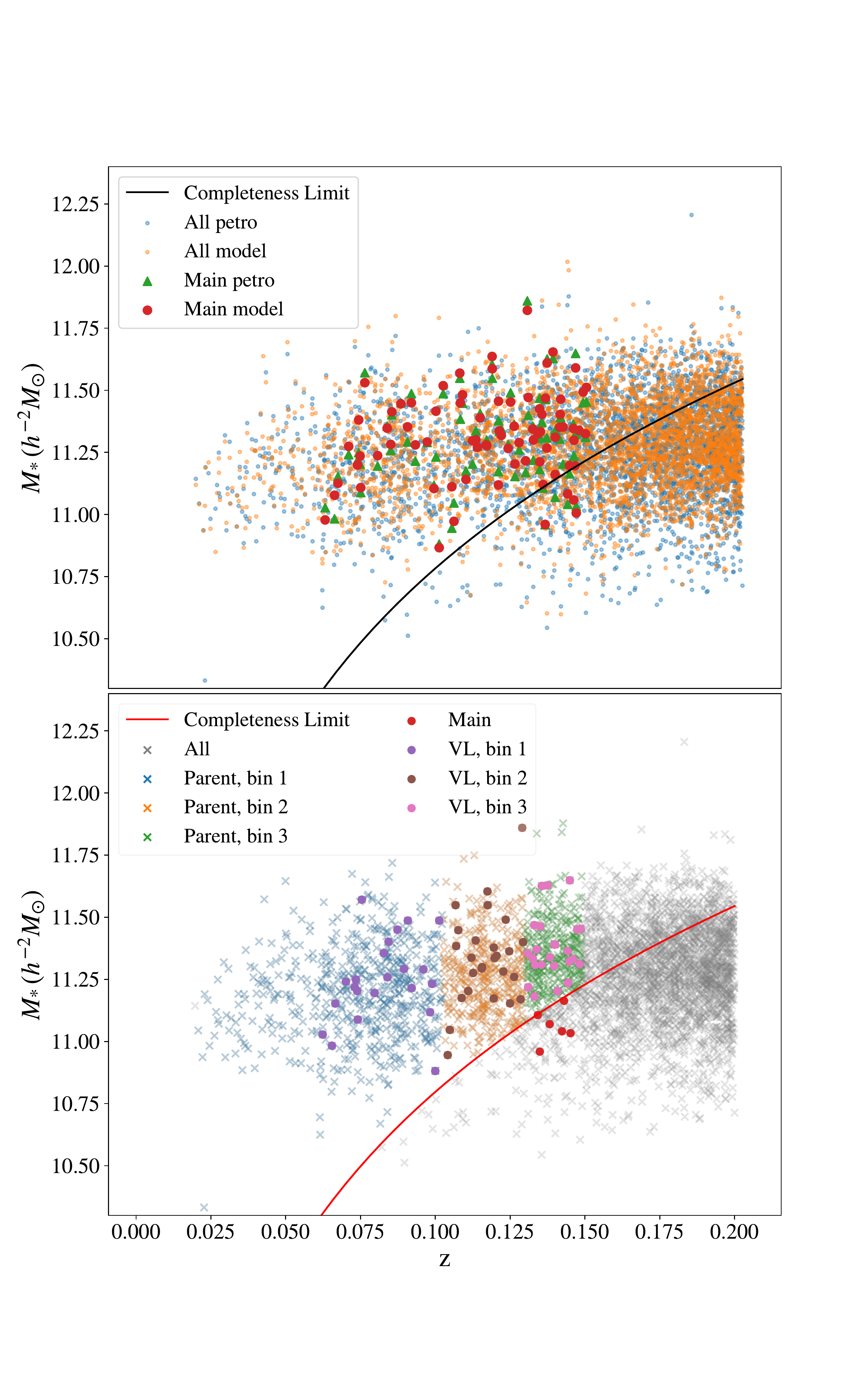}
\caption{
{\it Top:} The distribution of stellar mass  based on both the model and petrosian magnitudes  of 
our Main sample (large red and green symbols) and
All sample (orange and blue dots). 
The black curve is Eqn.~\ref{eqn:c_lim}.
{\it Bottom:}
The grey crosses show the distribution of BCGs in our ``All'' sample (Table \ref{tab:defn}).
They are further split 
into 3 redshift bins that have about the same comoving volume; for the BCGs above the completeness limit (red line), they belong to our Parent sample (color coded for ease of distinction of the 3 redshift bins). Our volume-limited sample  consists of the large circles. 
\label{fig:vlim}}
\end{figure}

Since  the SDSS main galaxy sample \citep{strauss02} is $r$-band limited, \citet{vandenbosch08} derive a corresponding completeness limit in stellar mass as a function of redshift, after considering the uncertainties in $K$-corrections in converting flux to luminosity, as well as the spread in mass-to-light ratios of red galaxies, appropriate for our BCGs:
\begin{eqnarray}
\label{eqn:c_lim}
\log{\left(\frac{M_{*, lim}}{h^{-2}M_{\odot}}\right)} & =& [ 4.852+2.246\log{D_L(z)} -1.186z \nonumber \\
 & & +1.123\log{(1+z)}  ] / (1-0.067z),
\end{eqnarray}
where $D_L$ is the luminosity distance.
We show the distribution of our BCGs in the stellar mass {\it vs.}~redshift plane (as the large green and red symbols), along with those in the All sample (as orange points) in Fig.~\ref{fig:vlim} (top panel). 
We note that 6 of the BCGs in the Main sample fall short of the completeness limit, and we shall refer to the rest, 73 BCGs, as the ``volume-limited'' sample (Table \ref{tab:defn}).
The green and red symbols represent stellar mass derived based on the SDSS Petrosian \citep{petrosian76} and model magnitudes.
It is clear that the difference is small whether the model  or Petrosian magnitudes are used for the BCG selection (note that the former is used in the Y07 catalog).

\begin{figure}[]
\epsscale{1.2}
\plotone{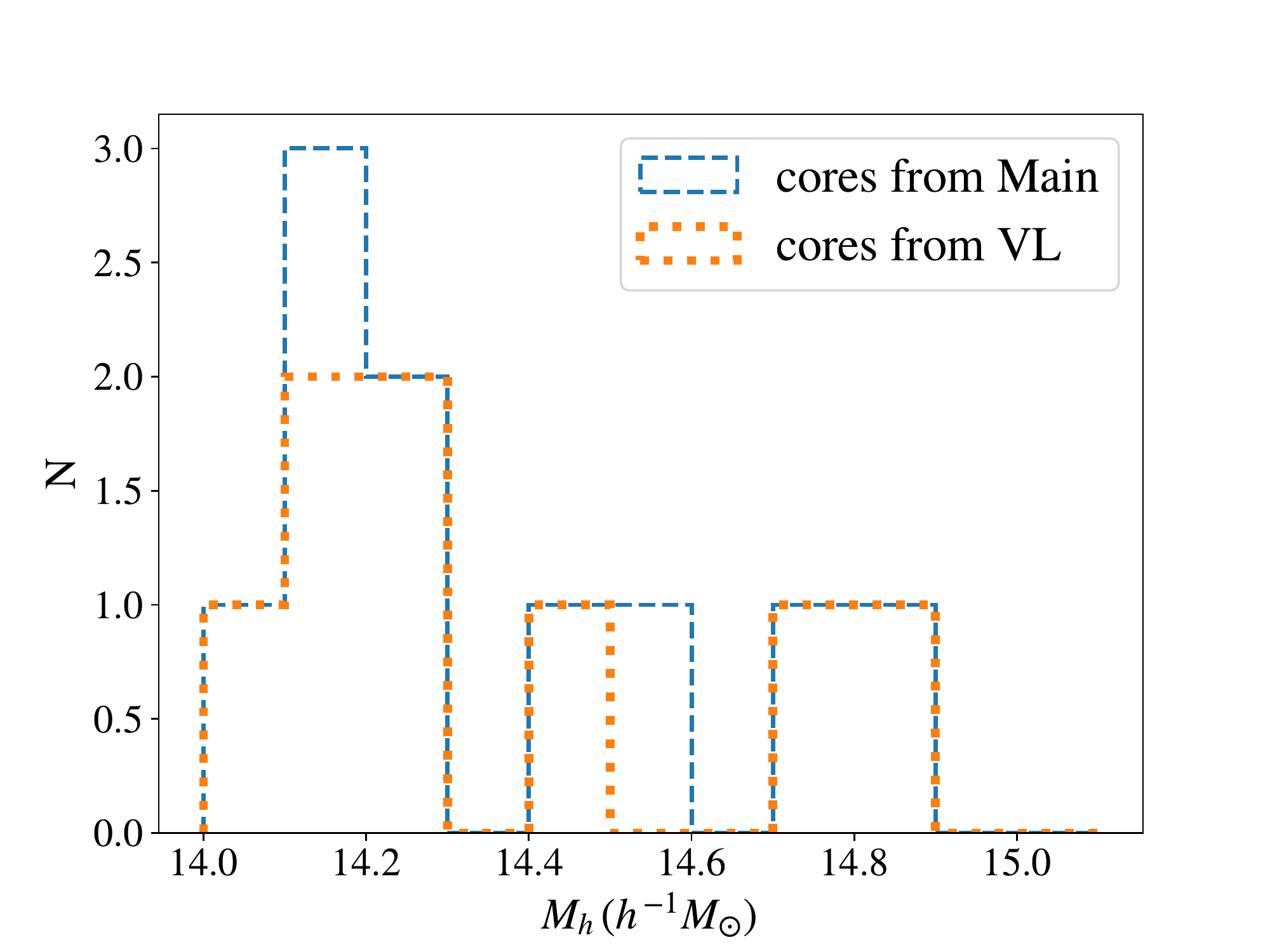}
\caption{
The cluster mass distribution of BCGs with $\mathcal{F}_{\rm core}\ge 0.1$ in the Main (volume-limited) sample, as shown by the blue (orange) histogram.  Note that with this flux ratio cut, all the BCGs host only one additional core.
\label{fig:79_core_Mh}}
\end{figure}

To proceed, we consider all  BCGs at $z=0.02-0.15$ above the completeness limit and  hosted by clusters with   $M_{180m} \geq 10^{14} h^{-1}M_{\odot}$ as the ``parent'' BCG sample (Table \ref{tab:defn}).
We split the parent sample into 3 redshift bins, $z=0.02-0.1025$, $z=0.1025-0.13$, and $z=0.13-0.149$, which are chosen to have about the same comoving volume. There are 590, 409, 360 BCGs in each bin,
among them 22, 26, 25  belonging to our Main sample (Fig.~\ref{fig:vlim}, bottom panel).
For the cores with $\mathcal{F}_{\rm core} \geq0.1$, there are 3, 2, 3 in each bin; the numbers for the case with $\mathcal{F}_{\rm core}\ge 0.05$ are 6, 2, 5, respectively 
(Table \ref{tab:vol_lim}).
We take the ratio between the number of BCGs in the volume-limited and the parent sample in each of the redshift bins as a redshift-dependent completeness correction factor. 
In this way, we obtain a multiple-core frequency of $0.114$\footnote{$f_{\rm  mc} = (80.5+31.5+43.2)/1359 = 0.114$} for the BCGs in the local universe (with $\mathcal{F}_{\rm core} \ge 0.1$; please note that in this case, whether we use $N_{\rm BCG, mc}$ or $N_{\rm multiple-core}$ in Eqn.~\ref{eq:mc} gives the same result), which is  very close to the value if we simply use the results from our volume-limited sample [$f_{\rm mc} = (3+2+3)/73 = 0.0110$].

To summarize,
including major mergers, there are 10 and 15 confirmed merging systems with flux ratios larger than 0.1 and 0.05, respectively (Fig.~\ref{fig:v_off}; Table \ref{tab:ob_core}). The corresponding ``apparent'' (i.e., not corrected for completeness) multiple-core frequencies are 
0.13 $\pm$ 0.04, 0.19 $\pm$ 0.05, assuming the error is Poissonian. 
The volume-limited multiple-core frequency  for  $\mathcal{F}_{\rm core} \ge 0.1$ is $0.11 \pm 0.04$. The corresponding value for $\mathcal{F}_{\rm core} \ge 0.05$ is $0.19 \pm 0.05$.

The halo mass distributions of the cored-BCGs in the Main  and  volume-limited samples are shown in Fig.~\ref{fig:79_core_Mh}. There are more BCGs with cores in the low halo mass end, but there are also more clusters (hence BCGs) in the low mass end. 
We measure the multiple-core frequency in two cluster mass bins [$\log M_{180m}/(h^{-1}\,M_\odot) = 14-14.55$ and $14.55-15.1$] and find values of 
$f_{\rm mc}=0.13\pm 0.05$ and $0.11\pm 0.08$,
respectively.  
Given our sample size, unfortunately we cannot meaningfully measure any cluster mass dependence.

However, if a higher $f_{\rm mc}$ is indeed found for lower mass clusters,  it could be due to the fact that, since the most massive BCGs tend to inhabit the most massive clusters, and very massive clusters must have started growing a long time ago, the growth of the most massive BCGs happened mostly in the distant past and traces of multiple cores may have now disappeared. At least some of  the less massive BCGs (living mostly in less massive clusters) could have grown more recently or be growing now, and therefore would be more likely to show multiple cores.

\section{Multiple-core Frequency of Brightest Cluster Galaxies in TNG300}
\label{sec:TNG}

Next we will measure the multiple-core frequency of BCGs from TNG300.  
As mentioned in Section~\ref{sec:tngbcgsample}, there are 225 BCGs in snapshot 91, which is the output closest to the median redshift of our MPL-9 sample.
With the pipeline that can automatically detect cores in imaging data in hand (Section~\ref{subsubsec:multi}), in principle it is straightforward to apply it to mock images of simulated BCGs.  However, we first need to construct the cluster selection function of our volume-limited sample (Table \ref{tab:defn}) and apply it to the TNG halos, such that the resulting multiple-core frequency can be better compared with the observed value.

\subsection{The Halo Sample}
\label{sec:tngsample}

\begin{figure}[]
\epsscale{1.3}
\plotone{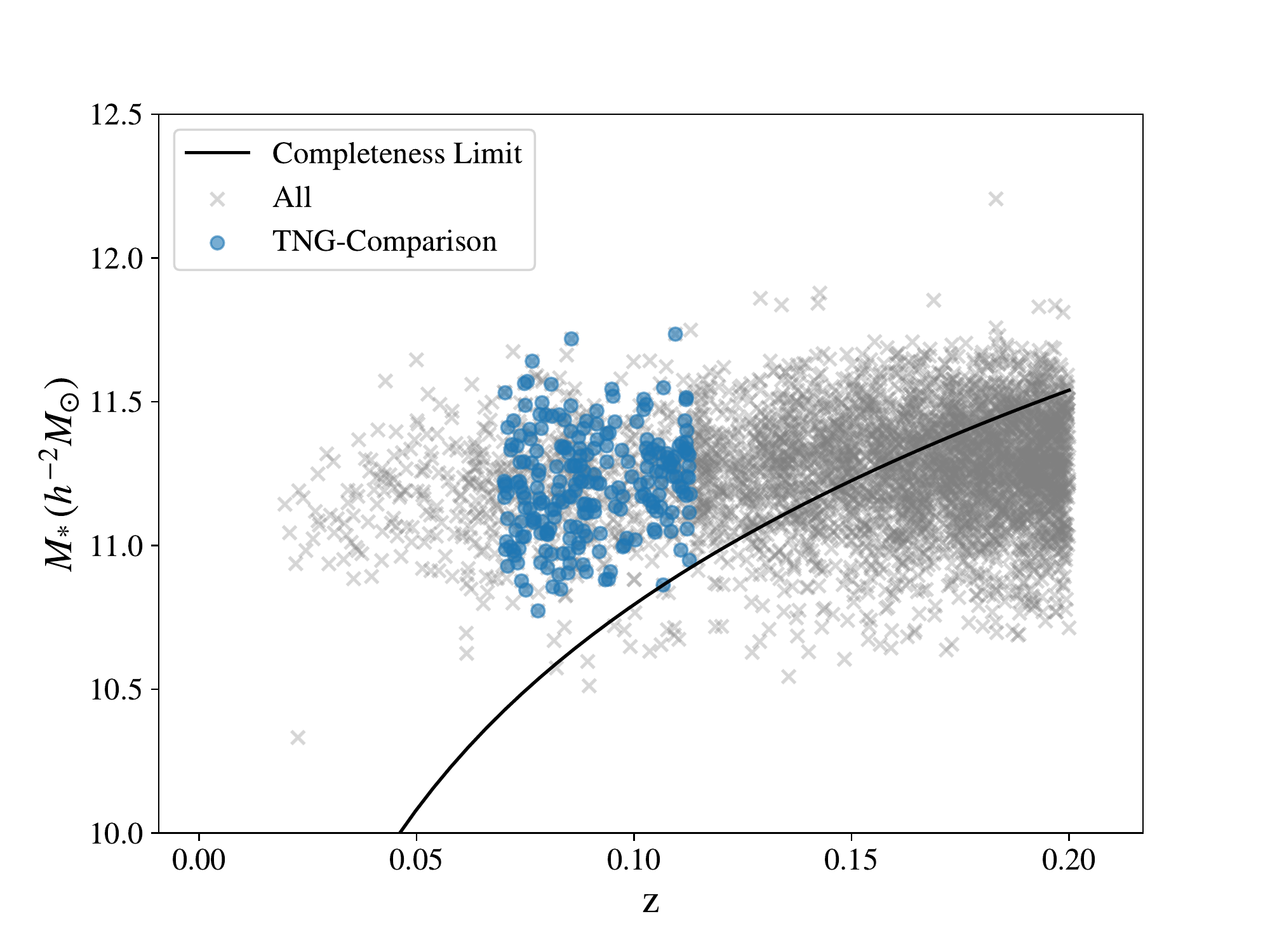}
\caption{
There are 225 BCGs with  stellar mass above the completeness limit at $z=0.07-0.11$ within a (300\,Mpc)$^3$  volume from the catalog of Y07, which are referred to as the TNG-Comparison sample (blue points).  The corresponding cluster sample is used to construct the halo mass selection function.
\label{fig:Yangbox}}
\end{figure}

\begin{figure}[]
\epsscale{1.3}
\plotone{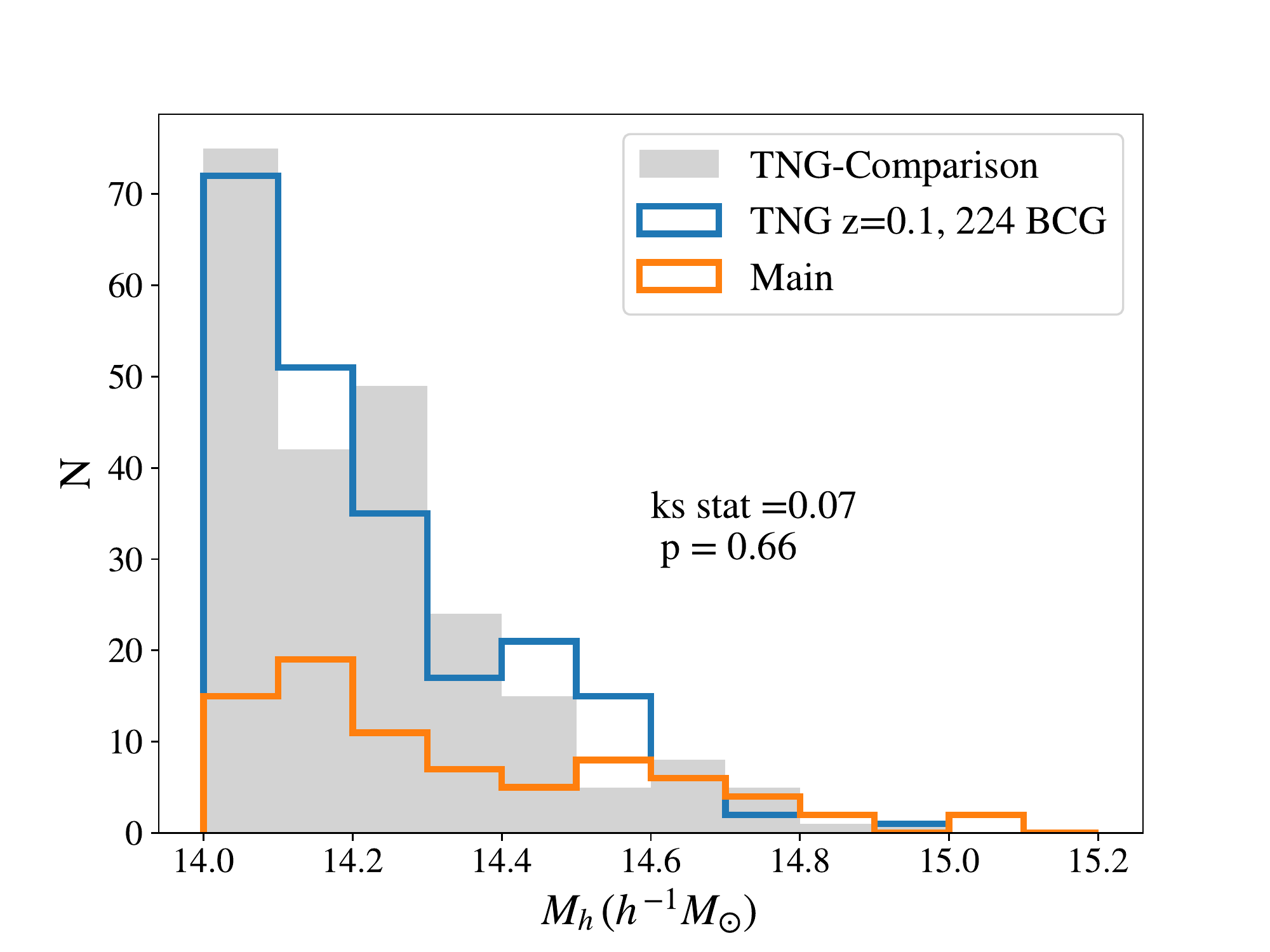}
\caption{
The halo mass distribution in 0.1 dex bins, of the BCGs in the TNG-Comparison (blue histogram), the Main samples (green histogram), and in cluster-scale TNG300 halos (with mass $M_{200m}\ge 10^{14}\,h^{-1}M_\odot$, orange histogram). 
The halo mass selection function is constructed by the ratio of the first two distributions.  The  median masses of the two observational samples are $\log M_{180m}/(h^{-1}\,M_\odot)=14.19$ and 14.25, respectively.
Performing a two-sample Kolmogorov-Smirnov test between the TNG300 and TNG-Comparison samples, we obtain a  $p$-value of 0.66, supporting the assumption that they are drawn from the same parent population.
\label{fig:Yang89}}
\end{figure}

To construct the selection function of the observed BCGs, we 
consider a subset of the Y07 cluster sample, selected to lie at $z=0.07-0.11$ within a {\it randomly} chosen area bounded by the R.A.~range of 140.27 to 229.90 deg.~and Dec.~range of 5.06 to 54.89 deg., which corresponds to a comoving volume equal to a TNG300 box.
There are, incidentally, also 225 BCGs with cluster mass $M_{180m} \geq 10^{14} h^{-1}M_{\odot}$, and stellar mass above the completeness limit; we shall refer to this sample as the ``TNG-Comparison'' sample.
Fig.~\ref{fig:Yangbox} shows the distribution of BCGs in the stellar mass {\it vs.}~redshift space of the TNG-Comparison sample (blue points), together with all the BCGs living in massive clusters
from the Y07 catalog (i.e., the ``All'' sample). The cluster mass distributions of the TNG-Comparison sample and the Main sample are shown in  Fig.~\ref{fig:Yang89}. The selection function is calculated by computing the ratio of these two distributions as a function of cluster  mass. If there are more BCGs in our Main sample than in the TNG-Comparison sample in a  mass bin,  the value of the selection function is set to 1 in that bin.

As the second step, we examine whether the halo mass distributions of the TNG300 halos and that of the TNG-Comparison sample are similar.  Before doing so, we have to remove a simulated BCG\footnote{The object has an ID 293868 (see Section~\ref{appendix:exclude_TNG_BCG}).  The reason for its removal is due to the difficulty of obtaining a good {\tt Ellipse} model for it (see Section~\ref{subsubsec:TNGphot}). 
Since $R_e$ is necessary for our further analyses, and such a failure (which is not  due to the presence of cores in the BCG) should be independent of its multiple-core frequency, removing this BCG from the sample should not affect our results.}. 
By comparing the 224 halos from TNG300 with 225 clusters from the TNG-Comparison, we show in 
Fig.~\ref{fig:Yang89}  that the mass distributions of the two  are similar. Performing a two-sample Kolmogorov-Smirnov  (KS) test, we obtain a KS statistic of $D=0.07$, and $p$-value of 0.66, 
supporting the assumption that they are drawn from the same parent population.

\begin{figure}[]
\epsscale{1.1}
\plotone{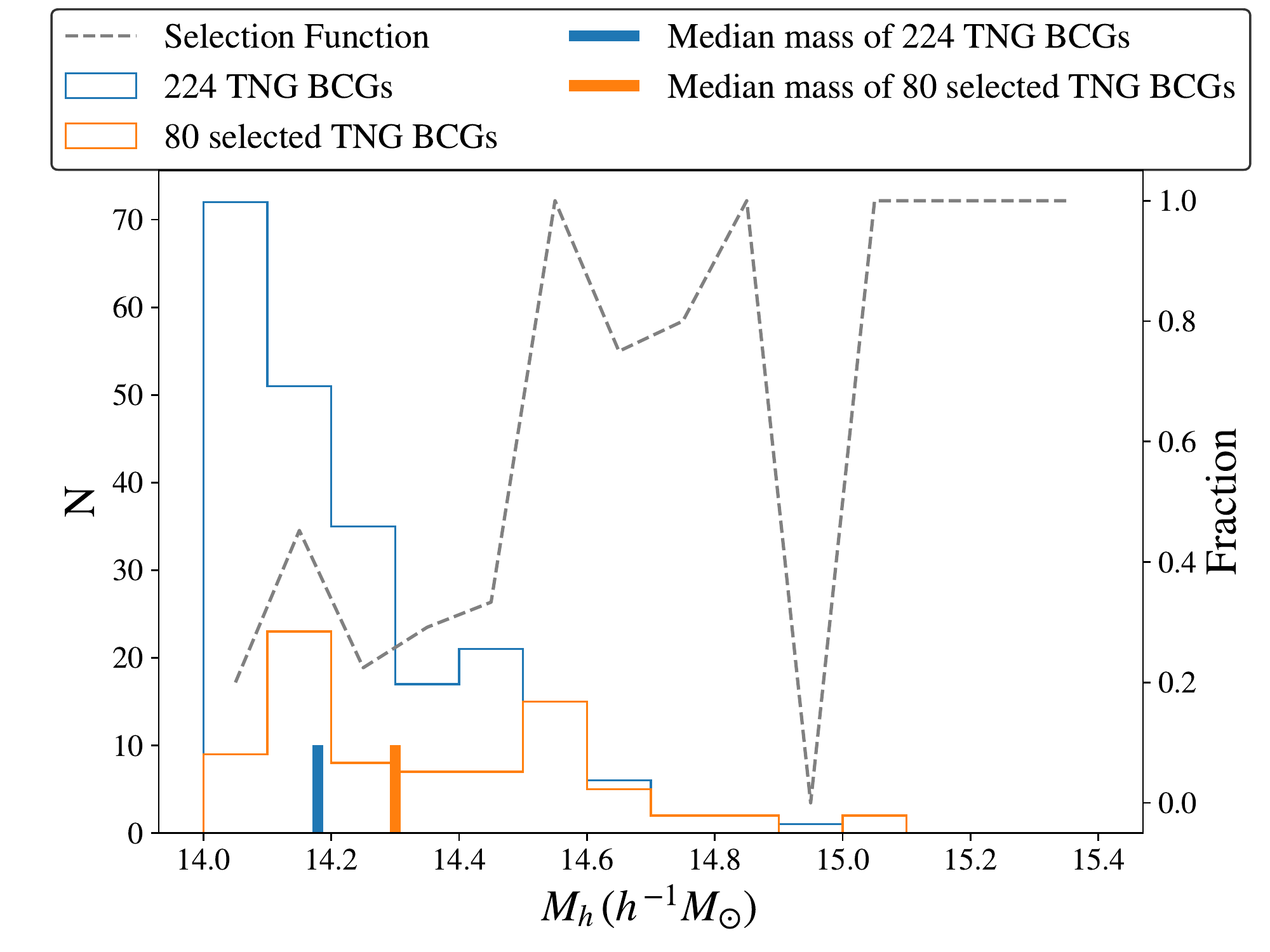}
\caption{
The halo mass distribution of one of the 50 mock TNG samples. In this example, 80 of the 224 TNG BCGs are selected, with halo mass distribution  consistent with that in our TNG-Comparison sample.
The gray dashed line is the selection function.
\label{fig:select_eg}}
\end{figure}

We now can then safely apply the selection function to the TNG sample, which is done in a Monte-Carlo fashion. 
For each TNG BCG, we draw a random number between $0-1$ and compare it with the value of the selection function corresponding to the halo mass of that BCG.  If it is smaller, we accept the BCG/halo.  Repeating this for all 224 TNG BCGs, we then have one ``mock'' BCG/halo sample, whose halo mass distribution should be similar to our volume-limited sample.
For statistical robustness, we have constructed 50 such mock samples. 
One of the mock BCG/halo samples is shown in Fig.~\ref{fig:select_eg}. We then measure the multiple-core frequency from these 50 samples in the following Sections.

\subsection{Synthetic Images}\label{sec:TNGim}

\begin{figure*}
\centering
\includegraphics[width=\textwidth]{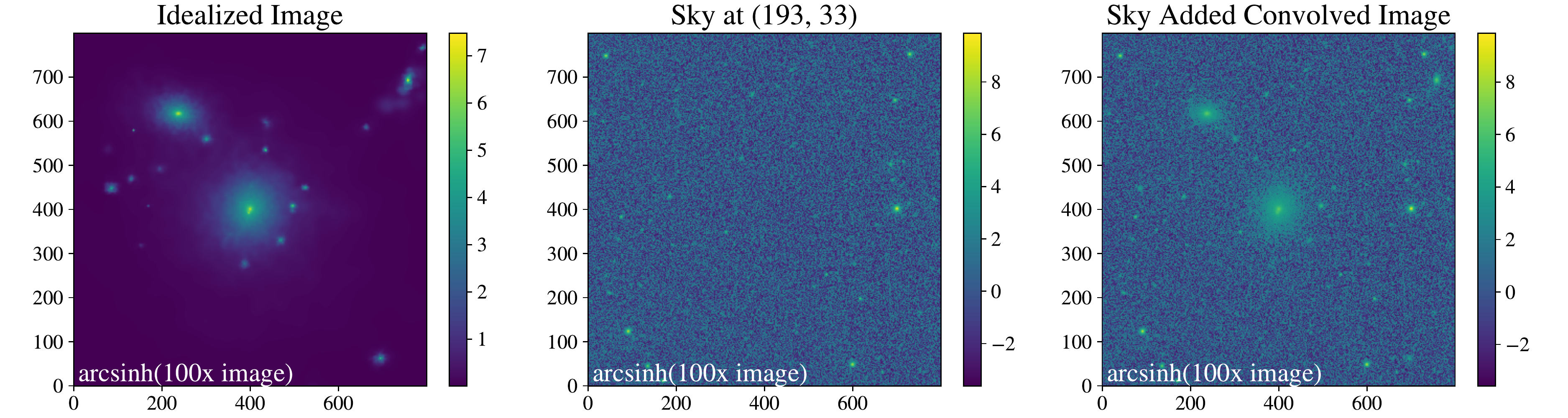}
\caption{\label{fig:synth_im}
The left panel is the idealized image of a BCG in TNG300 generated by the method described in \citet{rodriguez-gomez19}. The middle panel is a SDSS image centering at (RA, Dec) = (193, 33) degree. The right panel is the product of convolving the left panel  with a Gaussian PSF then adding the sky in the middle panel. As noted in Fig.~\ref{fig:core_step}, the values shown on the color bars correspond to  arcsinh(100$\times$ flux/nanomaggy).
}
\end{figure*}

The synthetic images are generated following the procedures described in \citet{rodriguez-gomez19}. The observing angle is perpendicular to the $xy$-plane. The pixel size is $0.396''$ as in SDSS, and the field-of-views are 1000$\times$1000 pixels and 800$\times$800 pixels, matching those of the real BCGs. Since the BCGs  mainly consist of old stellar populations with little dust \citep{vonderlinden07},  
the images are generated by the stellar population synthesis code GALAXEV \citep{bruzual03} and we have skipped the radiative transfer calculations (for justification, please see  \citealt{rodriguez-gomez19}). After the idealized $i$-band images are generated (in units of  nanomaggies, as in SDSS images), they are convolved with a Gaussian Point Spread Function (PSF) with $1.5''$ Full Width at Half Maximum. Adding the model images to a patch of real SDSS sky centered at (RA, Dec) = (193$^{\circ}$, 33$^{\circ}$)  that is void of bright galaxies and stars completes the generation of synthetic images. We show an example image in Fig.~\ref{fig:synth_im}.

\subsection{Photometry}\label{subsubsec:TNGphot}

The photometry of TNG BCGs is obtained in a similar fashion as described in Section~\ref{subsubsec:Phot}. We feed the 1000$\times$1000 pixel synthetic images to {\tt SExtractor}, with BACK$\_$SIZE=200 (1/5 of the image size), to obtain their segmentation maps. We subtract the {\tt SExtractor} measured background from the images, mask out the region of the sources touching the BCGs, and substitute the region of other sources with a Gaussian noise that has the standard deviation of the sky.  These ``BCG-only''
images are fed to {\tt Ellipse} to obtain empirical surface brightness models and profiles. We then subtract the empirical models from the synthetic images to obtain residual maps. Also, we fit a single S\'{e}rsic profile to the surface brightness profile in order to obtain the total flux and $R_e$ of the BCGs. 

Two BCGs have complex profiles that cannot be fit by a single S\'{e}rsic profile. We use the part of their curve of growth from {\tt Ellipse} within 150\,kpc and above the sky uncertainty to obtain their total flux and $R_e$ (please see Appendix \ref{appendix:Re_const} for more details). We also visually inspect all of the profiles and residuals, and find that 6 BCGs have unreliable profiles that are affected by bright neighbors in the field (please see Appendix \ref{appendix:exclude_TNG_BCG}). 
Since this fraction is small and should be independent of the multiple-core frequency, we add a warning flags to them and remove them from the further analyses.  However, one of them (ID 65561) actually has a double core structure, and we shall report the multiple-core frequency with and without these 6 BCGs in Section~\ref{subsec:TNG_results}.
In total 218 simulated BCGs have good photometry measurements.

\subsection{Identifying Multiple-Cores}
\label{subsubsec:TNG_MC}

The identification of cores for the TNG BCGs is performed in the same fashion as described in Section~\ref{subsubsec:multi}.
The only difference is the criteria of maximum separation due to the $R_e$ difference between the Main sample and the simulated sample. We compare the $R_e$ distribution of each of the 50 mock TNG samples with the Main sample using the KS test, and find some differences, which could be due to the IFU coverage criterion imposed on the observed samples (see also Section~\ref{appendix:unbias} and Fig.~\ref{fig:3bincf}). The average value of median $R_e$ for the 50 mock samples is 22.35$\,h^{-1}$\,kpc, and the median $R_e$ of the Main sample is 16.10\,$h^{-1}$kpc. We thus modify the 18\,kpc separation adopted in Section~\ref{subsubsec:IFU cov} by the ratio of $R_{e,sim}/R_{e,obs}$ and set 25\,kpc as the maximum separation for the search of cores in simulated BCGs.  To mimic what is done to the real BCGs, a minimum search radius of 2.5\,kpc is also set.

We run {\tt SEP} with the strong deblending parameters\footnote{Detection threshold\,$=0.2$, minimum area\,$=1$, deblend threshold\,$=64$, deblend contrast\,$=0.0001$, clean parameter\,$=1.0$} and a low detection threshold to detect any possible core within $2.5-25$\,kpc from the BCG center on the residual maps.

\subsection{Flux Ratio}\label{subsubsec:TNG_f_ratio}

\begin{figure}[]
\epsscale{1.2}
\plotone{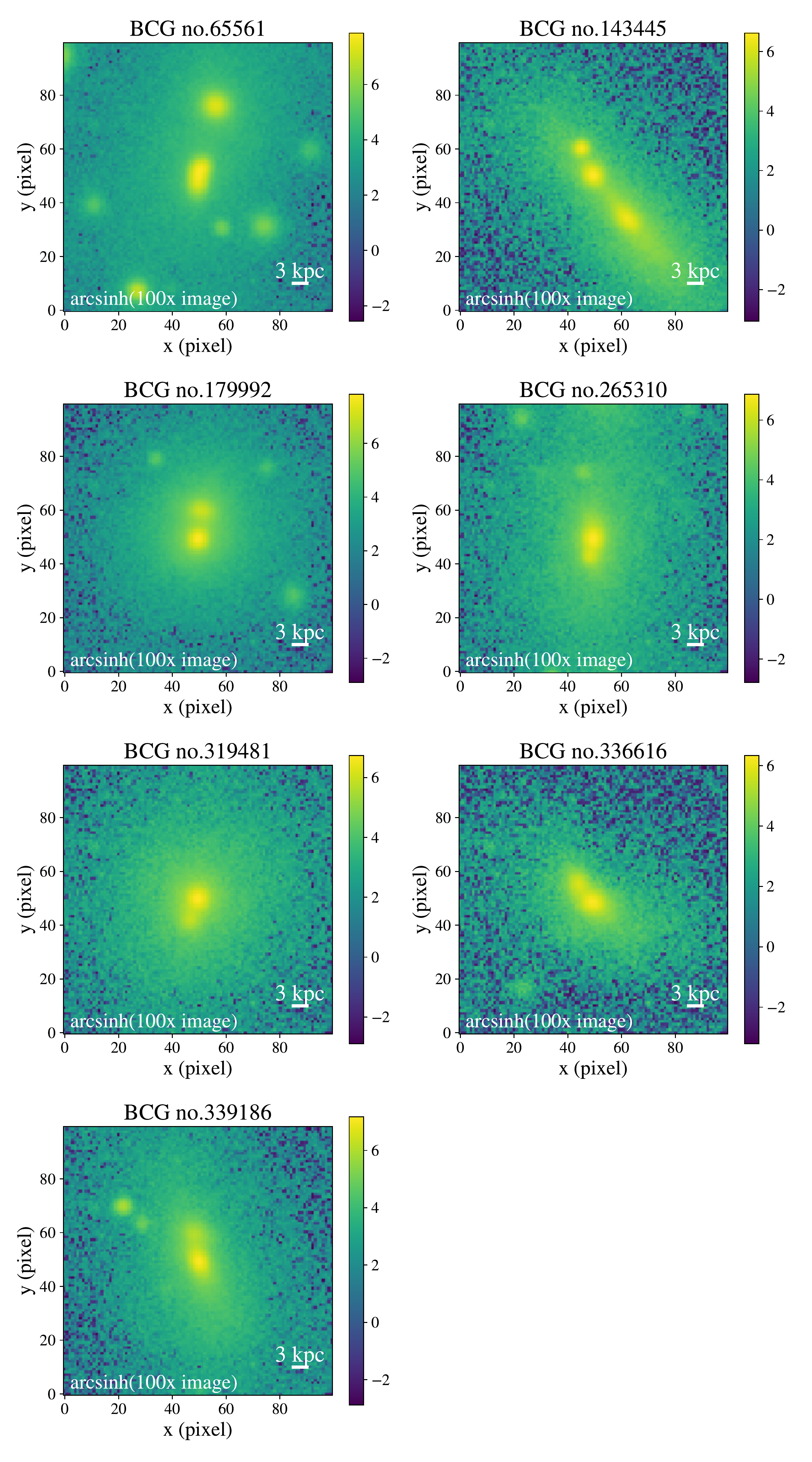}
\caption{
The zoom-in images of the extra 7 major mergers in TNG selected by visual inspection.  The white horizontal bar indicates a scale of 3\,kpc.
\label{fig:tng_maj}}
\end{figure}

The procedures are similar to that described in Section~\ref{subsubsec:f_ratio}. For the flux of the cores, we consider the maximum value among 6 types 
of measurements, including: ({\it i}) the sum of pixels in the {\tt SEP} segmentation region extracted from the images, and ({\it ii}--{\it vi}) the sum of the pixels within a radius of 1, 2... to 5\,kpc of the cores. 
Fourteen cores are found to have $\mathcal{F}_{\rm core} \ge 0.1$. 
There are additional 7 major mergers selected by visual inspection (Fig.~\ref{fig:tng_maj}). Therefore, 21 cores have $\mathcal{F}_{\rm core} \ge 0.1$. 
We note that the one BCG that is excluded due to bad photometric fit (Section~\ref{sec:tngsample}) actually has two cores.

As the synthetic images only use stellar particles that belong to the friends-of-friends (FoF) group to which a simulated BCG belongs to, there is no ``contamination'' from foreground/background objects.  Therefore, unlike in the case for MaNGA BCGs, we do not further confirm the physical association of cores with the BCGs via kinematics.

\subsection{Results}
\label{subsec:TNG_results}

\begin{figure}
\centering
\includegraphics[width=0.5\textwidth]{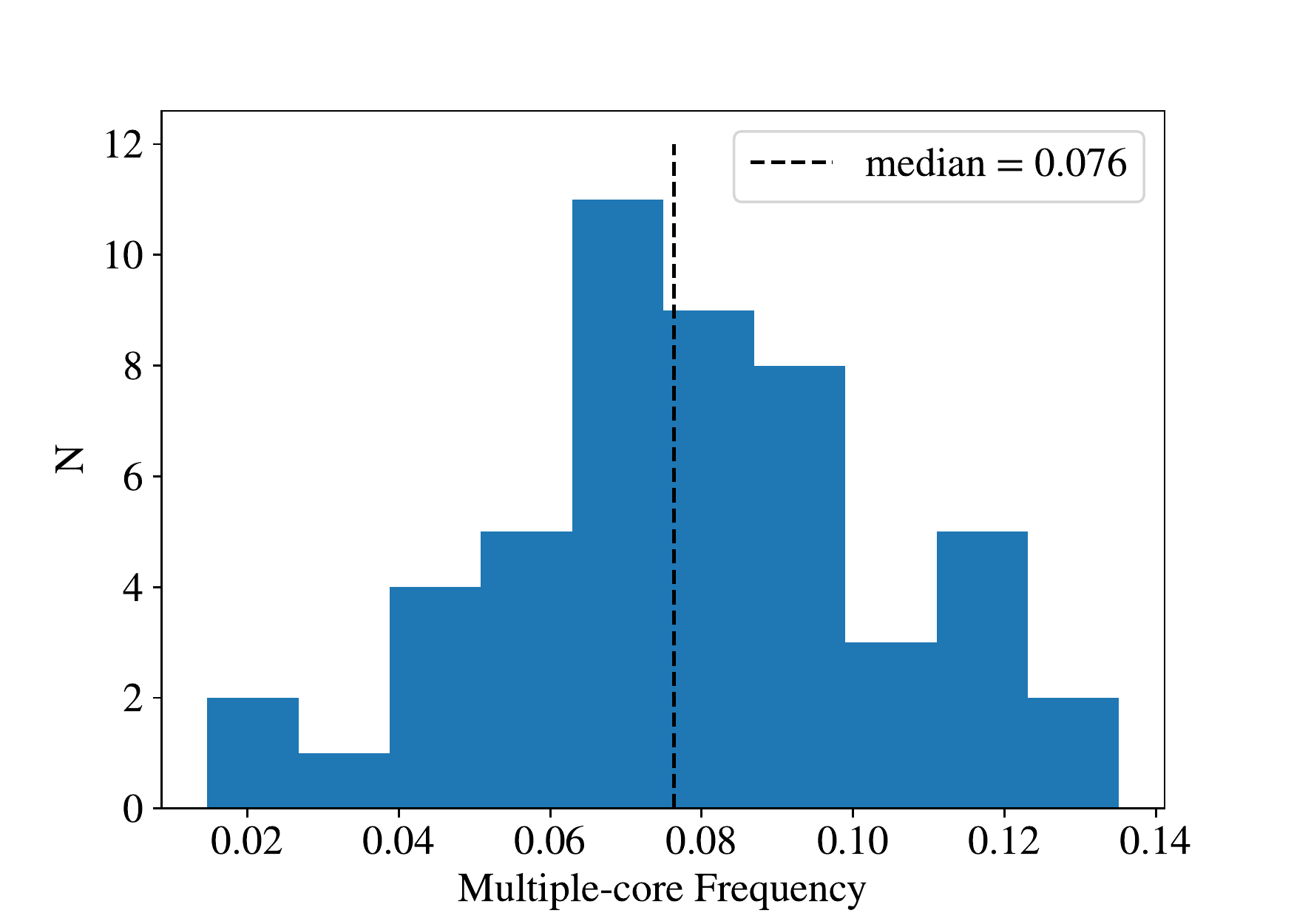}
\caption{\label{fig:mcf}
The distribution of the multiple-core frequency of the 50 mock TNG samples, for the case of $\mathcal{F}_{\rm core}\ge 0.1$. The median is 0.076, and the standard deviation is 0.027.
}
\end{figure}

\begin{figure}
\centering
\includegraphics[width=0.5\textwidth]{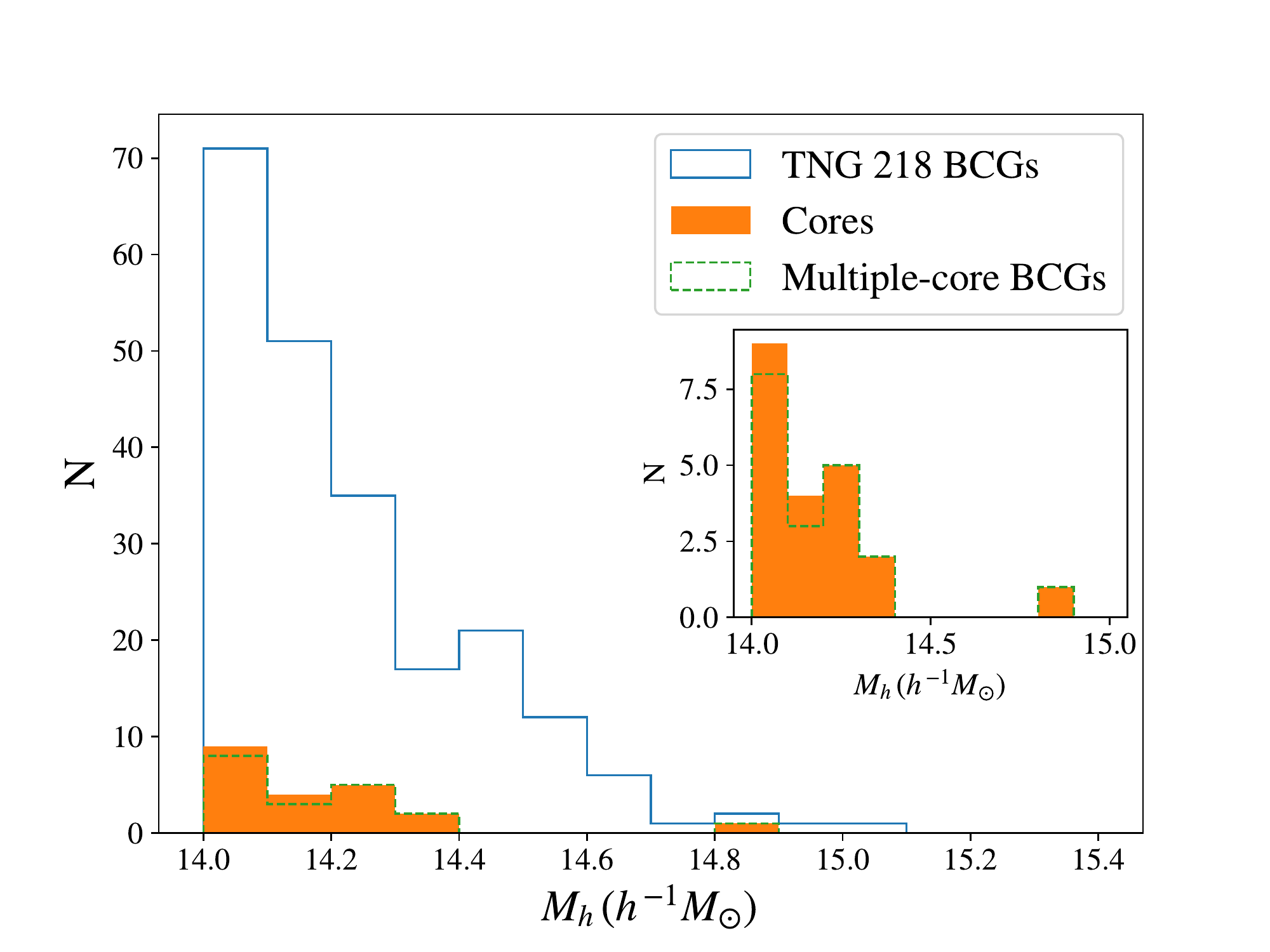}
\caption{\label{fig:TNG_core_Mh}
The halo mass distribution of simulated BCGs with $\mathcal{F}_{\rm core}\ge 0.1$ (green histogram), compared to that of the 
full TNG sample with good photometry (218 BCGs; blue histogram).  Since a BCG can  host more than one core, we also show the distribution of cores as the orange histogram, where each core contributes to the counts.  The inset shows more clearly the numbers of cores and BCGs.
}
\end{figure}

The multiple-core frequency of BCGs having $\mathcal{F}_{\rm core}\ge 0.1$ among the full TNG sample is 21 out of 218, or $f_{\rm mc} = 0.10 \pm 0.02$. It is 23 out of 225 ($f_{\rm mc}=0.10\pm 0.02$) if including BCGs without good photometry (Table \ref{tab:TNG_BCG}). 
For the case of $\mathcal{F}_{\rm core}\ge 0.05$, the numbers are 37 out of 218 ($f_{\rm mc}=0.170$) or 39 out of 225 ($f_{\rm mc}=0.173$).
Note that these are values obtained without applying the observed halo selection function and thus should not be directly compared with our observational results.

The multiple-core frequencies of the 50 mock samples for the case of $\mathcal{F}_{\rm core}\ge 0.1$ are shown in Fig.~\ref{fig:mcf}; the median value is 0.076, with a standard deviation of 0.027. The  multiple-core frequency is thus formally 
$f_{\rm mc}  = 0.08 \pm 0.02\, {\rm (Poisson)} \pm 0.03\, {\rm (Systematic)}$. 
Hereafter we shall combine the two uncertainty terms and quote $f_{\rm mc}  = 0.08 \pm 0.04$.
For the case of $\mathcal{F}_{\rm core}\ge 0.05$, we find that $f_{\rm mc}  = 0.14 \pm 0.03\, {\rm (Poisson)} \pm 0.03\, {\rm (Systematic)}$.
We also test the Monte-Carlo method by running 100 and 200 ensembles, finding that  they have nearly the same median and standard deviation as those of 50 ensembles.

The halo mass distribution of   BCGs with multiple cores is shown in Fig.~\ref{fig:TNG_core_Mh}.  It is clear that most of the cores are detected in BCGs with lower halo mass, which explains why the multiple-core frequency becomes lower after the selection function is applied, as the selection function filters out more halos at the low-mass end.

\begin{figure}[]
\epsscale{1.1}
\plotone{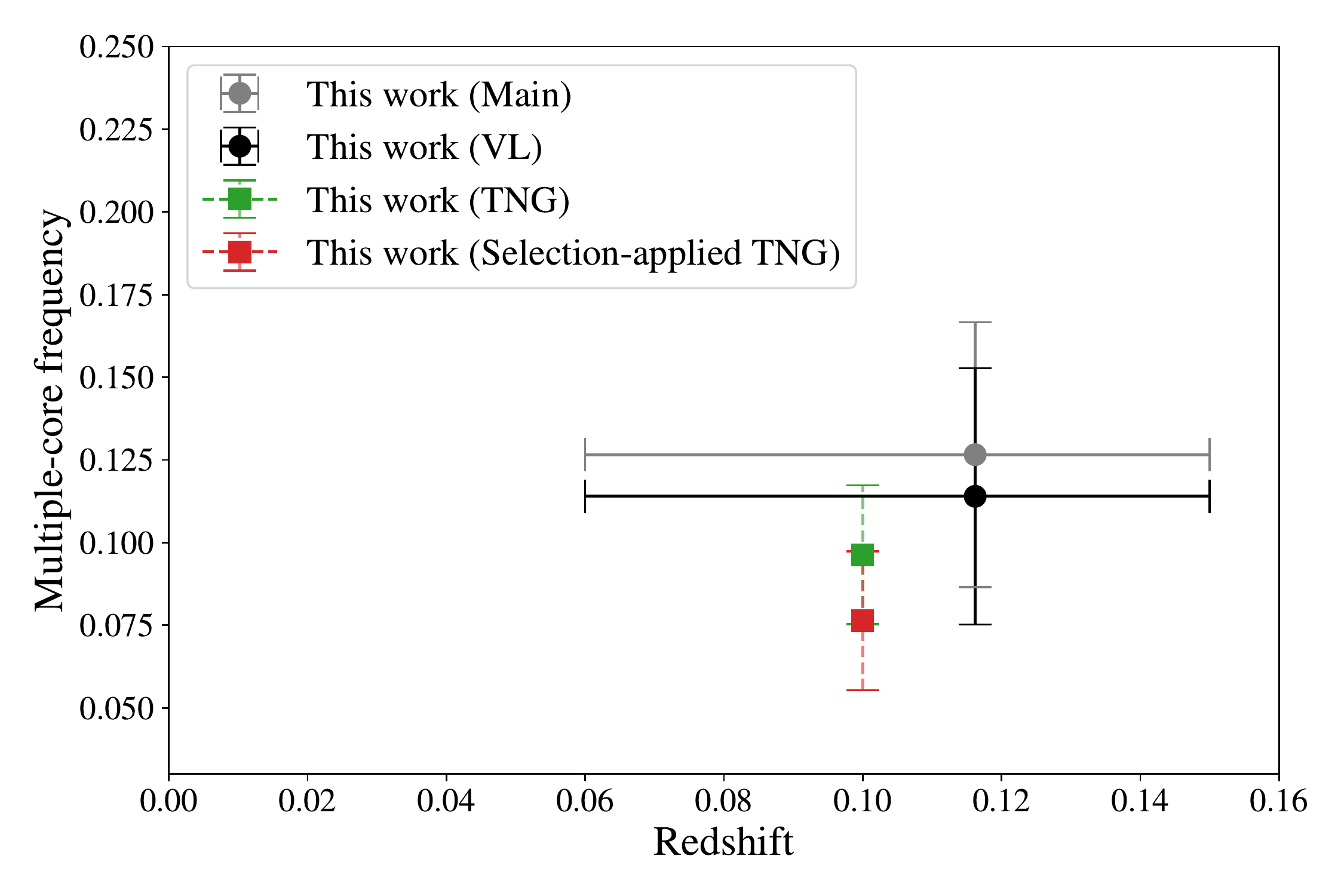}
\caption{
A comparison of our observed and simulated multiple-core frequencies $f_{\rm mc}$. 
The grey and black points are from the Main and volume-limited samples, respectively.
The error bars in redshift represent the redshift range of the samples. The error bars of the multiple-core frequencies are Poissonian. 
The green point ($f_{\rm mc}=0.0963$) is the result based on the whole TNG sample, while the red point ($f_{\rm mc}=0.0763$) is the average over the 50 mock samples.
Note that these values follow the definition of $f_{\rm mc}$ and thus a BCG would be counted $N$ times if it has $N$ cores.  If we only consider unique BCGs, the values of the green and red points will be reduced by 11\% and 7\%, respectively.
\label{fig:ltt_TNG}}
\end{figure}

The multiple-core frequency of our Main sample (the black dot in Fig.~\ref{fig:ltt_TNG}) is slightly higher than that of the TNG sample (the purple dot in Fig.~\ref{fig:ltt_TNG}), although the discrepancy is only at $1\sigma$ level.

As in Section~\ref{subsubsec:vlim}, we also measure the multiple-core frequency in two halo mass bins ($\log M_{180m}/(h^{-1}\,M_\odot) = 14-14.55$ and $14.55-15.1$).  
The values are $0.1\pm 0.2$ and $0.06\pm 0.06$, respectively.  
Again the limited halo sample size prevents us from measuring any halo mass dependence.
%

\section{Discussion}\label{sec:dis}

After having measured the multiple-core frequency from both MaNGA (Section~\ref{sec:IFU}) and IllustrisTNG (Section~\ref{sec:TNG}), here we discuss the robustness of our sample selection (Section~\ref{sec:samplesel}), showing it is representative of the local BCGs.  We compare our results with findings from the literature in Section~\ref{subsec:ltt}, measure the mass growth rate of BCGs in IllustrisTNG (Section~\ref{sec:mgh}) and, finally, discuss the effect of the presence of cores in the supermassive black hole radio activity (Section~\ref{sec:rg}).

\subsection{Velocity Offsets of the Cores and Sample Selection}
\label{sec:samplesel}

The velocity offset distribution shown in Fig.~\ref{fig:v_off}  
is slightly skewed to the positive side, and is  independent of the redshift of the BCGs. It is not clear what causes the skew. 
We have visually inspected the DAP velocity maps of the cored BCGs, and confirmed that indeed more cores show higher velocity than the main body of the BCGs, and that the spectral fits to the cores are adequate.

\begin{figure}[]
\epsscale{1.3}
\plotone{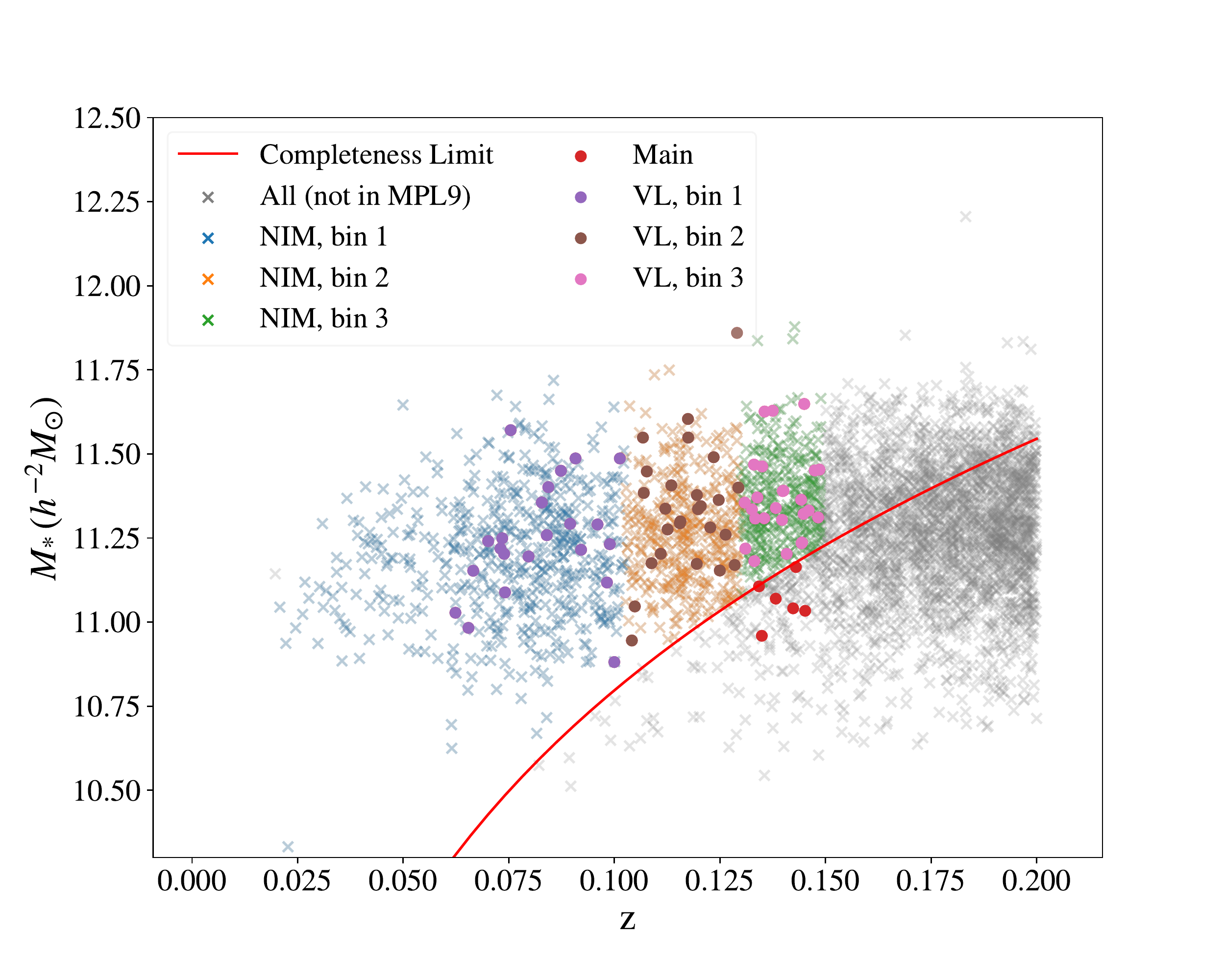}
\caption{
Illustration of the samples used in Section~\ref{appendix:unbias}: the blue, brown, and green crosses are our ``not-in-MaNGA (NIM)'' sample (split into 3 redshift bins that have about the same volume), which is constructed by excluding the MPL-9 sample from the Parent sample.   
The completeness limit is represented by the red line. Our volume-limited sample, also split into the same 3 redshift bins, are shown as large circles.
\label{fig:3bin}}
\end{figure}

One may question how representative our BCG sample (e.g., the Main or volume-limited samples) is, with respect to the overall BCG population.  This is a legitimate concern, as (1) the MaNGA sample is constructed to have a flat stellar mass distribution, thus very massive galaxies, like BCGs, could be overrepresented, compared to a volume-limited sample; (2) our BCGs are assembled from MaNGA's primary, secondary, and color-enhanced samples, as well as the BCG and MASSIVE ancillary programs, which makes the selection a bit heterogeneous.
We show in the following that {\it our sample selection criteria do not result in a biased sample of BCGs.}

\subsubsection{Unbiased Sample Selection}
\label{appendix:unbias}

\begin{figure*}
\centering
\includegraphics[width=0.7\textwidth]{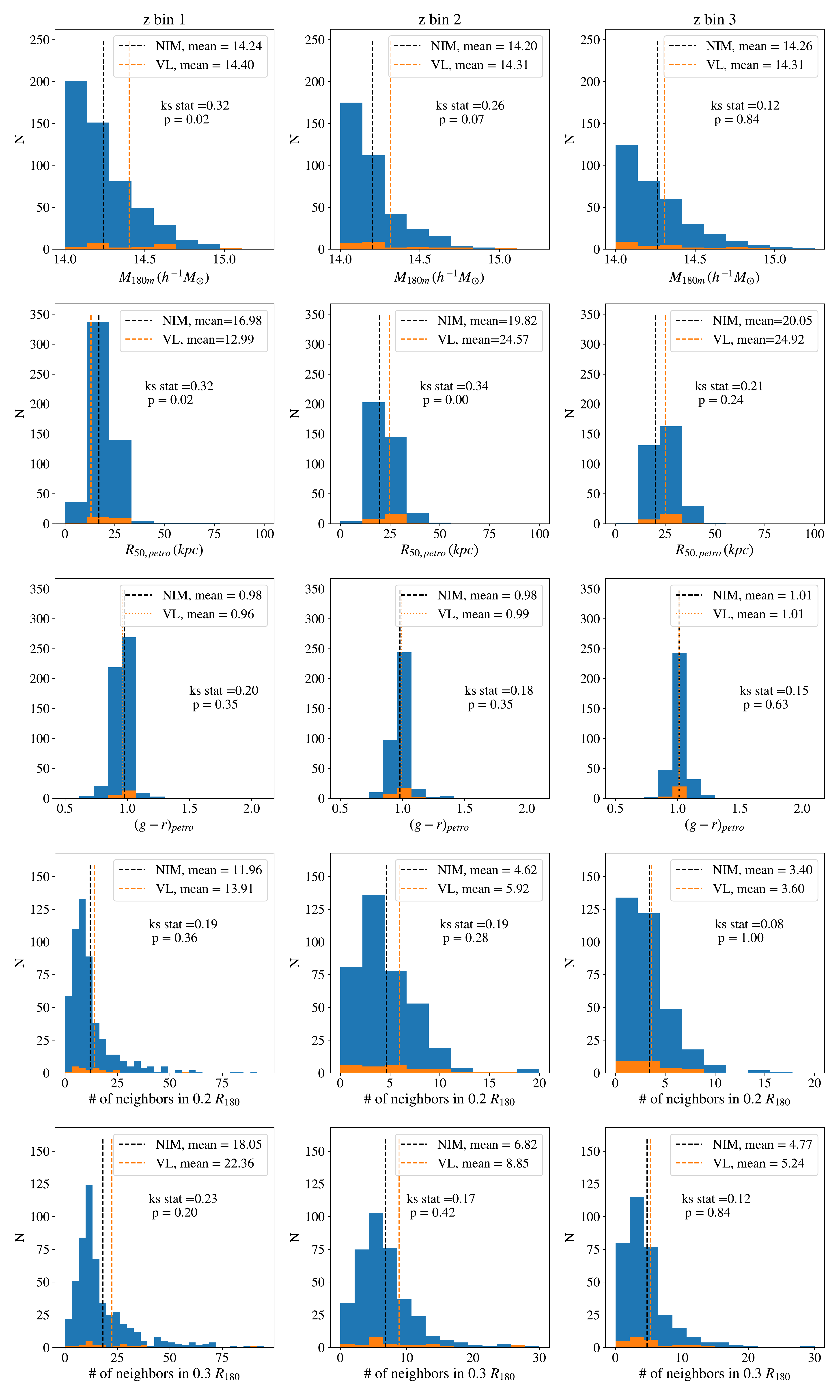}
\caption{\label{fig:3bincf}
Comparisons of various properties between our volume-limited sample (orange histograms) and the not-in-MaNGA (NIM) sample (blue histograms).  From left to right, we show the results from the 3 redshift bins ($z=0.02-0.1025, 0.1025-0.13, 0.13-0.149$); from top to bottom, the properties being considered are cluster mass, half-light radius, $g-r$ color, number of neighbors within $0.2R_{180m}$, and the number of neighbors within $0.3R_{180m}$, respectively.  Based on  KS tests, we see that only the $R_{50}$ distributions in bins 1 and  2, and the halo mass distribution in bin 1 are  different.
}
\end{figure*}

In Fig.~\ref{fig:vlim} (top panel) we 
see that the stellar mass distributions of  BCGs in our Main sample is similar to that of the All sample (Table \ref{tab:defn}).
For a more quantitative analysis, we compare various properties of our Volume-limited sample with a subset of 
clusters from Y07, which is obtained by excluding the MPL-9 BCG sample from the Parent sample, and will be referred to as the ``not-in-MaNGA (NIM)'' sample
 (Fig.~\ref{fig:3bin}; Table \ref{tab:defn}).  
Similarly to what we have done in Section~\ref{subsubsec:vlim}, we make the comparison in 3 redshift bins of comparable comoving volume ($z=0.02-0.1025, 0.1025-0.13, 0.13-0.149$; hereafter bins 1, 2 and 3).
There are 529, 376, 332 (22, 26, 25) BCGs in each bin of the NIM (volume-limited) sample.
The properties we compare are halo mass, Petrosian half-light radius, Petrosian color, and the number of neighbors, where the neighbors are defined by a certain range in projected distance and redshift (Fig.~\ref{fig:3bincf}).
These properties are obtained either directly from the Y07 catalog, or derived from the galaxy member catalog associated with the primary Y07 catalog.
 We compare these properties through their mean values and the KS test.

In Fig.~\ref{fig:3bincf}, the 3 columns represent results in each redshift bin, while the rows, from top to bottom, show comparisons in cluster mass, Petrosian half-light radius $R_{50}$, $g-r$ color, number of neighbors within $0.2R_{180m}$, and the number of neighbors within $0.3R_{180m}$, respectively. We only consider the Petrosian color within the range of $0.5-2.1$, to avoid unreasonable photometry.  In all panels, the blue (orange) histograms are for the NIM (volume-limited) sample.  
Through the two-sample KS test, we see that only the $R_{50}$ distributions in bins 1 and  2, and the halo mass distribution in bin 1 are  different.  For all other properties in all bins, we do not see obvious deviation for our volume-limited sample from the NIM sample. The differences in bin 1 are likely due to  the IFU coverage criterion we impose, which translates to a lower redshift limit at $z\approx 0.06$ for the observed samples. If we change the lower redshift limit of bin 1 for the NIM sample to $0.06$, there is then no significance difference in the halo mass distribution.

In the next Section we shall look more into bin 2 to examine the multiple-core frequency obtained from our IFU-based observations and that inferred from imaging data only.

\subsubsection{An Independent Estimate of Multiple-Core Frequency}
\label{sec:indep}

As a further test,
we run the core detection procedure as described in Section~\ref{subsubsec:multi} on the SDSS images of the 376 BCGs in  bin 2. We choose bin 2 because  (1) its redshift is  closest to that of our Main sample, and (2) there seems to be some difference in $R_{50}$ between our volume-limited sample and the NIM sample (Fig.~\ref{fig:3bincf}).
Among these, two BCGs do not have a good {\tt Ellipse} model. The core detection pipeline finds 186 cores in the remaining 374 BCGs. We fit S\'{e}sic profiles to the {\tt Ellipse} curve of growth to obtain their total flux. 18 cores have $\mathcal{F}_{\rm core}\ge 0.1$.
We also visually select additional 14 major mergers. The multiple-core frequency is thus 
$f_{\rm mc}=0.09 \pm 0.02$, 
which is shown in the right panel of Fig.~\ref{fig:ltt} as the brown point.

Given that there is minimum selection involved in this sample, ideally the multiple-core frequency based on this subsample should be consistent with that of the full TNG sample (i.e., without the selection function applied). They indeed are consistent (c.f.~the brown and red points).  
It is also interesting to note that these values are also close to what we obtain from the volume-limited sample, when similarly split into 3 redshift bins (i.e., the pink point at $z\approx 0.12$).
One should bear in mind that no spectroscopic confirmation is performed for the cores detected in the NIM sample, and thus the value quoted above should be regarded as an upper limit.  However, given our conclusion that most of the imaging-detected cores are actually physically associated with the BCGs as confirmed by kinematics (28 out of 30; see Section~\ref{subsubsec:merg}), the difference from the true value may be small.

\subsection{Comparison with the Literature}\label{subsec:ltt}

\begin{figure*}
\epsscale{1.17}
\plottwo{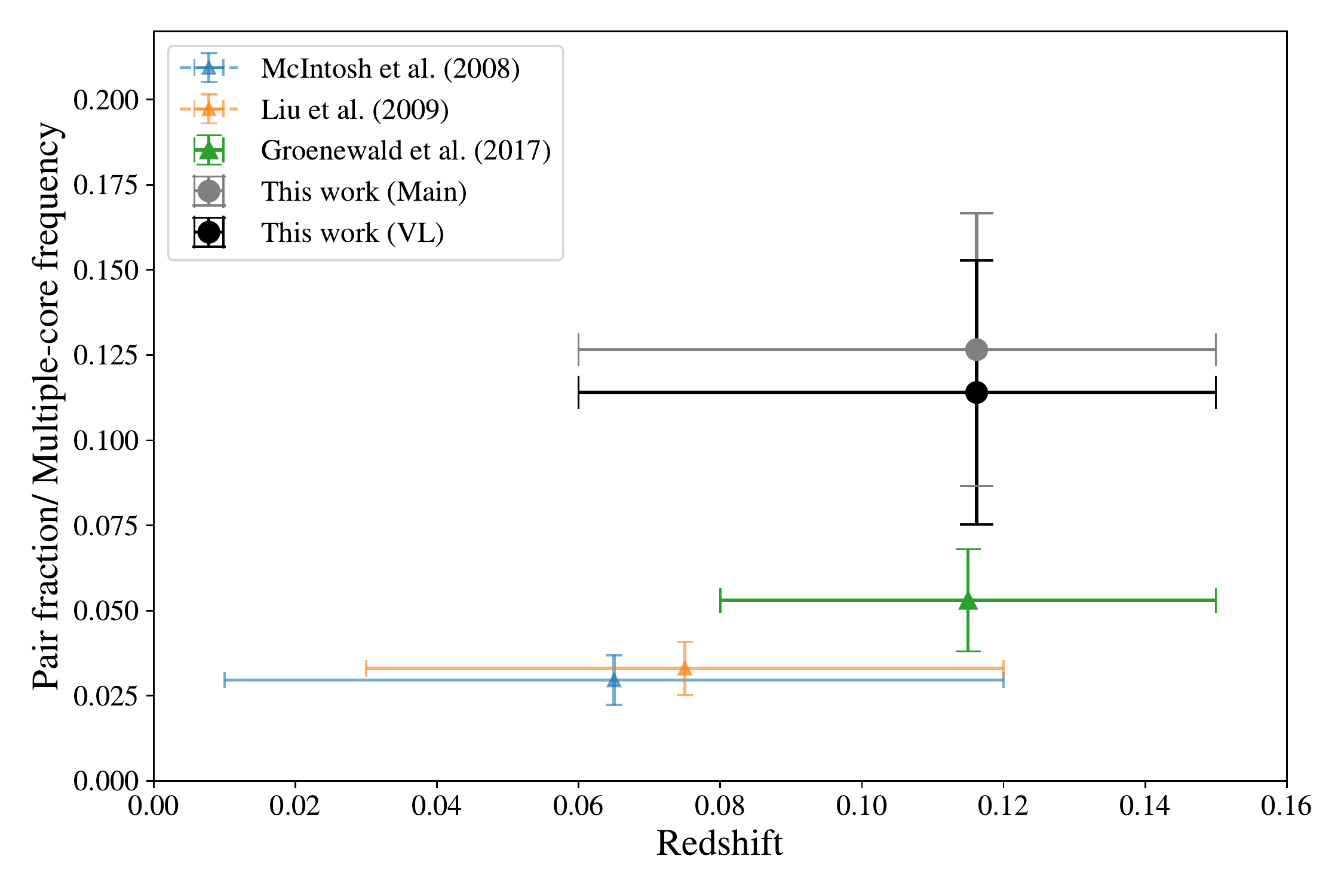}{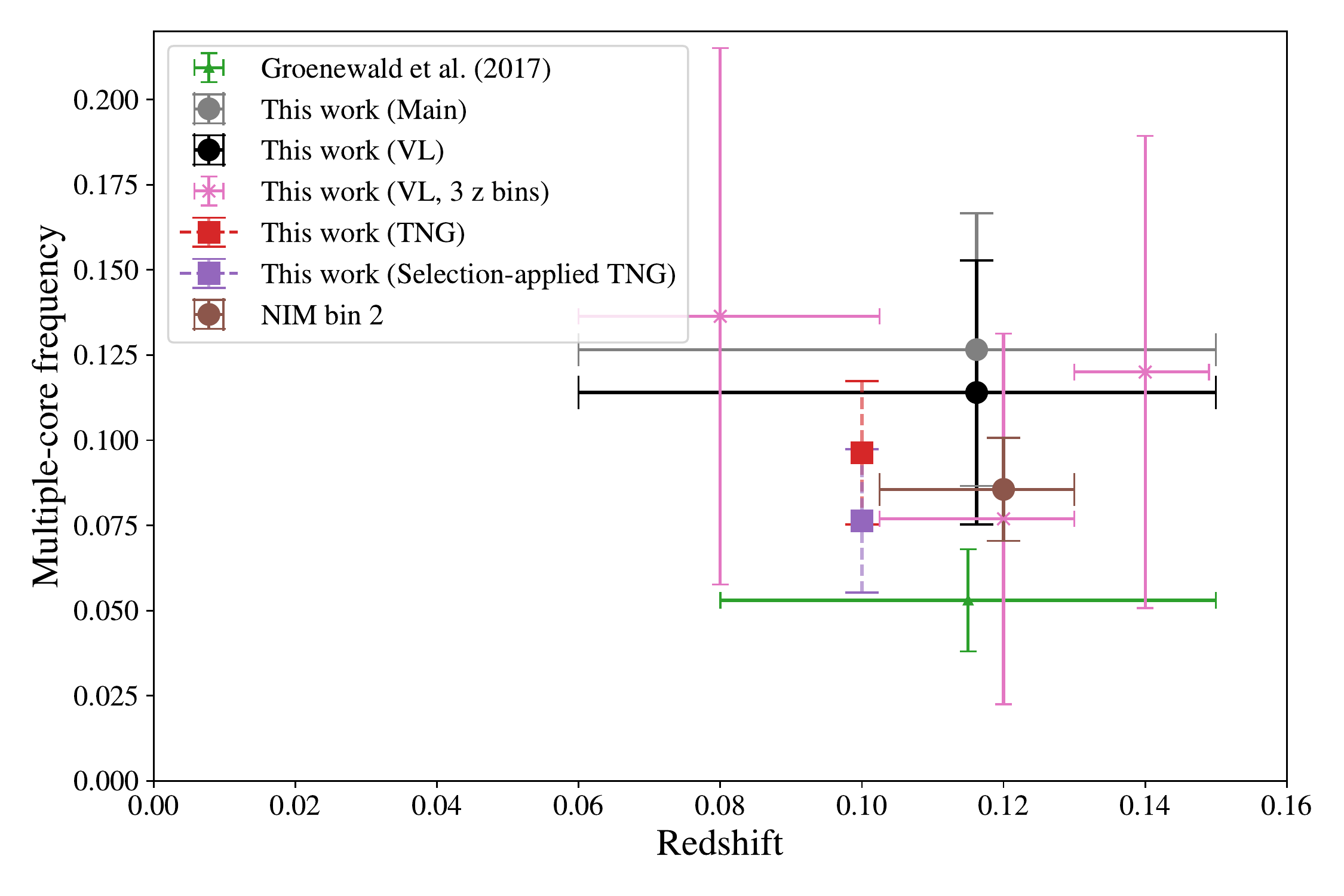}
\caption{
{\it Left:} Comparison of our multiple-core frequency with pair fraction from the literature.
The error bars in  redshift represent the redshift range of the samples. The measurements from the literature are plotted at the middle point of their redshift ranges, and the result of this work is plotted at the mean redshift of our samples.
{\it Right:} Similar to Fig.~\ref{fig:ltt_TNG}, but showing also the result from redshift bin 2 of the not-in-MaNGA (NIM) sample (brown point), as well as the pair fraction from \citet[][G17; green point]{groenewald17}.
The pink points represent the values derived from our volume-limited sample, split in 3 redshift bins. 
\label{fig:ltt}}
\end{figure*}

Fig.~\ref{fig:ltt} (left panel) shows a comparison of our measurements with some of the previous works that measure merger rates of BCGs at redshift ranges similar to ours (\citealt{mcintosh08,liu09,groenewald17}, hereafter M08, L09, and G17, respectively). It  should be noted that the BCG selection and the the merger definition are different among these works.

All of M08, L09, and G17 focus on major mergers within 30\,kpc. The sample of M08 is volume-limited and has halo mass $\geq 2.5\times 10^{13}\, M_{\odot}$;  the G17 sample  is also volume-limited, but has a much higher halo mass threshold of $\geq 2.9\times 10^{14}\, M_{\odot}$. The sample of L09 does not provide an estimate of the halo mass. M08 and G17 select pairs with mass ratios larger than 0.25, while the sample of L09 has an average luminosity ratio of 0.5. Moreover, M08 and L09 select physically related pairs by the distorted morphology. G17 select close pairs and apply a correction factor derive from a smaller spectroscopic confirmed sub-sample (12 pairs). They apply a limit in velocity difference of 300\,km/s, and conclude that a limit on 500\,km/s only increases the fraction by about 0.03 percent. 

Given the difference between our approach and that of others, it is not easy to directly compare our multiple-core frequency with the major merger pair fractions in previous works.  There are differences in sample selection, maximum separation and, most importantly, the method of finding physically related pairs. G17 note that the morphology distortion is more obvious in the late stage of the mergers \citep{lotz11}, hence {\it  it is possible to obtain a lower pair fraction if relying on the distortion in galactic shape only.}

\citet{brough11} conduct the first targeted IFS observation of BCGs with close companions, with the goal of determining the merger rates, using VIMOS on the VLT. They select 3 BCG with companions and 1 without companions within $10''$ (18\,kpc at $z\sim 0.1$). These BCGs are from the sample of the 625 BCGs from \citet{vonderlinden07} selected from the C4 catalog \citep{miller05}.  20\% of these BCGs have visually identified massive companions. They find that 2 out of 3 companions are likely bound with their BCGs.  \citet{jimmy13} apply the same method to 10 BCGs, of which 7 with companions and the rest without. They use the ``G-M20'' merger selection criteria \citep{lotz08} and conclude that 4 out of 10 BCGs have gone through mergers within the past 0.2\,Gyr, although their sample selection might be biased toward BCGs that have companions. 


\subsection{Mass Growth Rate of BCGs}
\label{sec:mgh}

The mass growth rate of massive galaxies (for which  BCGs stand at the extremal end) has been an important topic in galaxy formation.  Traditionally this is usually done through comparisons of luminosity or stellar mass functions (SMF) measured at different cosmic epochs \citep[e.g.,][]{scarlata07,bernardi13,bundy17}.  For example, \citet{bundy17} have used a large sample of massive galaxies extracted from the SDSS stripe 82 and concluded that there is very  little evolution of the massive end since $z\sim 0.7$.  They suggest that any galaxy growth would have occurred at the outskirts beyond the observational aperture, which corroborates  one critical aspect of all SMF measurements at the massive end, namely proper measurements of the ``total'' luminosity of the galaxies, a challenge that we also face in this study.  For example, using a careful sky subtraction and sophisticated modeling technique, \citet{bernardi13} show that the abundance of massive galaxies could have been underestimated by a dex in previous studies (see also \citealt{huang18}).  Another approach is to use the Hubble diagram of BCGs \citep{aragon98,whiley08}.

A different approach is employed by  \citet{masjedi08}, who use the cross-correlation function between the spectroscopic sample of luminous red galaxies (LRGs) and photometric galaxies to infer the very small scale clustering of LRGs, from which they are able to infer a growth  rate of $\sim 2\%$ per Gyr measured at $z\approx 0.25$.

Given  the consistency in multiple-core frequency between our volume-limited sample and the TNG300, in principle we can  infer the mass growth rate of BCGs, using directly the information derived from the simulation.  For each of the 225 BCGs in TNG300, we trace the stellar mass content within a 25\,kpc radius out to $z=0.6$, 
and compute the average growth rate, which is the mass difference between 
$z=0.6$ and $0.1$ divided by the time lapse between these two epochs (4.5\,Gyr).  
The median (mean) of the mass growth rate is found to be 1.3\%/Gyr (4.1\%/Gyr).

\subsection{Nuclear Radio Activity}\label{sec:rg}

By matching our Main sample to the radio galaxy catalog presented in \citet{lin18}\footnote{Available at \url{https://vizier.u-strasbg.fr/viz-bin/VizieR?-source=J/AJ/155/188}}, we find that 35\% of the BCGs have 1.4\,GHz radio power $P_{1.4}>10^{23}\,$W/Hz (a threshold typically used to separate star formation-powered and nuclear-powered radio activity), which is similar to the results shown in \citet[][Table 5 therein]{lin07}.  For the 30 BCGs with cores (irrespective of their $\mathcal{F}_{\rm core}$ values), the fraction is similar (33\%).  However, if we focus on 10 BCGs with $\mathcal{F}_{\rm core}\ge 0.1$, the fraction increases to 50\%.  
It is tempting to attribute to the elevated radio activity to the mergers with massive satellites, but given the small number of the BCGs, we do not attempt to further interpret the finding.
%
We note, however, if we increase the radio power threshold (e.g., to $P_{1.4}>10^{24}\,$W/Hz), the presence of cores only makes a small enhancement in the radio-loud fraction compared  to the BCGs  without multiple cores (20\% {\it  v.s.} 18\%).

\section{Conclusion and Prospects}\label{sec:conclusion}

The motivation of this work is to solve the discrepancy of stellar mass growth of BCGs at $z<0.5$ between models and observations, which may be caused by the use of fixed aperture photometry adopted in observations. To tackle this problem, deep photometry of BCGs and careful work on sky subtraction are required.

On the other hand, studying the merger rate in the inner regions of BCGs can be a good alternative  to solving this discrepancy. However, studying merger rates requires  the combination of the frequency of multiple-cores and merger timescale, and the latter needs to come from simulations. Hence, in this work we focus on the multiple-core frequency, which is a direct observable.

We have used the largest sample of BCGs with IFS data -- about 7 times larger than previous attempts -- to study the BCG multiple-core frequency, defined to be the fraction of BCGs that host one or more physically associated dense cores (with core-to-total flux ratio $\ge 0.1$ and velocity offset $\le 500$\,km/s) in a volume-limited sample.
Our observational result, $f_{\rm mc}  = 0.11 \pm 0.04$, appears to be consistent with the state-of-the-art cosmological hydrodynamical simulation IllustrisTNG ($f_{\rm mc}  = 0.08 \pm 0.04$), which is small compared to the discrepancy in the stellar mass growth revealed in some of the earlier works.   Our results are not very sensitive to sample selection, as long as it is volume limited.

Thus, we may have obtained a better understanding of the stellar mass assembly of BCGs: while the discrepancy in the growth of ``total'' mass may be due to the different apertures used in observations and simulations \citep[e.g.,][]{rogone18}, the multiple-core frequency in the innermost part of the BCGs appears to be comparable in observations and theory.
 Given such a reasonable agreement, we further trace the formation history of simulated BCGs back to $z=0.6$ and obtain a mean growth rate of 4.1\% per Gyr within the central 25\,kpc radius.

Our main conclusions are:
\begin{enumerate}
\item Cores detected based on images often are indeed associated with their BCGs (about 93\% of the time), although stars need to be carefully removed.
\item It is important to have realistic simulated images for the observation {\it vs.}~simulation comparisons. Applying stellar population synthesis modeling, effects of PSF, and sky noise are all critical.
\item Cores are mostly detected in BCGs of low mass clusters (around $10^{14}\,h^{-1}M_\odot$), which may be mainly because of the higher abundance of such clusters (although it might also be due to different evolutionary stages of low mass clusters compared to more massive ones; please see the discussion at the end of Section~\ref{subsubsec:vlim}). 

\item We obtain a multiple-core fraction of $0.11 \pm 0.04$ at $z \approx 0.1$ within a 18\,kpc radius from the center, which is comparable to the value of $0.08 \pm 0.04$ derived from mock observations of 218 simulated BCGs in IllustrisTNG300 at $z=0.1$.

\end{enumerate}

We have established that cores seen in BCGs are most likely to be physically associated, and therefore one can obtain a rough estimate of multiple-core frequency purely based on imaging data (e.g., Section~\ref{sec:indep}). However, in principle, the IFS data could further allow a  detailed investigation of the properties of the cores, such as the stellar populations of the satellites that are in the process of merging with the BCG.  
We shall leave such an analysis applying to the full MaNGA BCG sample to a future study.
In addition, we are not able to determine whether there is a dependence on halo mass of $f_{\rm mc}$, both in observations and simulations.  For the latter,
this could be somewhat mitigated by taking 3 projection directions per simulated halo (currently we only consider the projection along the simulation $z$-axis), and also considering more snapshots between $z=0.06$ and 0.15, and by considering lower mass halos (e.g., down to the group regime), so that the simulation statistics could be greatly boosted, potentially enough to search for such trends.

The fully automatic pipeline we developed can be readily applied to the whole MaNGA sample (MPL-11; already fully puclic). However, it is still critical to conduct visual inspection of BCGs, as these systems often are too challenging even for the most sophisticated software. Nevertheless, the pipeline can be readily applied to deep images from Hyper Suprime-Cam \citep{aihara18}  
or data from the upcoming Rubin Observatory's Legacy Survey of Space and Time\footnote{\url{https://www.lsst.org/}}, and slitless spectroscopy from Roman Space Telescope\footnote{\url{https://roman.gsfc.nasa.gov/}} or Euclid\footnote{\url{https://sci.esa.int/web/euclid}}  to study the multiple core frequency at higher redshifts, where mergers are expected to take place more often.

This kind of study can also be extend to lower mass clusters and groups \citep{banks21}, which may provide more stringent constraints, given the much higher abundance of groups.

\acknowledgments
We thank Gabriel Torrealba for developing the {\tt Ellipse} package used in this work.  
We are grateful to Wei-Hao Wang, Kyle Westfall, James Lottes, Andrew Cooper, David Wake,   Michael Blanton,  Xiaohu Yang, and Ting-Wen Lan for helpful comments, and Taira Oogi and Abdurro'uf for help with handling of simulated and MaNGA data, respectively.
We thank an anonymous referee whose comments have improved the clarity of the paper.
YHH and YTL are grateful for supports from the Ministry of Science and Technology of Taiwan under grants MOST 110-2112-M-001-004 and MOST 109-2112-M-001-005, and a Career Development Award from Academia Sinica (AS-CDA-106-M01).
DN acknowledges funding from the Deutsche Forschungsgemeinschaft (DFG) through an Emmy Noether Research Group (grant number NE 2441/1-1).
YTL thanks IH, LYL and ALL for constant encouragement and inspiration.

Funding for the Sloan Digital Sky Survey IV has been provided by the Alfred P.~Sloan Foundation, the U.S.~Department of Energy Office of Science, and the Participating Institutions. SDSS acknowledges support and resources from the Center for High-Performance Computing at the University of Utah. The SDSS web site is \url{www.sdss.org}.

SDSS is managed by the Astrophysical Research Consortium for the Participating Institutions of the SDSS Collaboration including the Brazilian Participation Group, the Carnegie Institution for Science, Carnegie Mellon University, Center for Astrophysics $\vert$ Harvard \& Smithsonian (CfA), the Chilean Participation Group, the French Participation Group, Instituto de Astrof\'{i}sica de Canarias, The Johns Hopkins University, Kavli Institute for the Physics and Mathematics of the Universe (IPMU)/University of Tokyo, the Korean Participation Group, Lawrence Berkeley National Laboratory, Leibniz Institut f\"{u}r Astrophysik Potsdam (AIP), Max-Planck-Institut f\"{u}r Astronomie (MPIA Heidelberg), Max-Planck-Institut f\"{u}r Astrophysik (MPA Garching), Max-Planck-Institut f\"{u}r Extraterrestrische Physik (MPE), National Astronomical Observatories of China, New Mexico State University, New York University, University of Notre Dame, Observat\'{o}rio Nacional/MCTI, The Ohio State University, Pennsylvania State University, Shanghai Astronomical Observatory, United Kingdom Participation Group, Universidad Nacional Aut\'{o}noma de M\'{e}xico, University of Arizona, University of Colorado Boulder, University of Oxford, University of Portsmouth, University of Utah, University of Virginia, University of Washington, University of Wisconsin, Vanderbilt University, and Yale University.

The IllustrisTNG simulations were undertaken with compute time awarded by the Gauss Centre for Supercomputing (GCS) under GCS Large-Scale Projects GCS-ILLU and GCS-DWAR on the GCS share of the supercomputer Hazel Hen at the High Performance Computing Center Stuttgart (HLRS), as well as on the machines of the Max Planck Computing and Data Facility (MPCDF) in Garching, Germany.

This work has made use of data from the European Space Agency (ESA) mission
{\it Gaia} (\url{https://www.cosmos.esa.int/gaia}), processed by the {\it Gaia}
Data Processing and Analysis Consortium (DPAC,
\url{https://www.cosmos.esa.int/web/gaia/dpac/consortium}). Funding for the DPAC
has been provided by national institutions, in particular the institutions
participating in the {\it Gaia} Multilateral Agreement.

\appendix

\section{Consistency of Photometry between {\tt Ellipse} and {\tt PyMorph}}\label{appendix:Re_const}

In this Section, we compare all kinds of the photometry we have used, and show that they are consistent (to the degree of our needs). In Sec.~\ref{subsubsec:Phot} we have described the two sources of photometry of our BCGs, namely that based on {\tt PyMorph} and from {\tt Ellipse}. The former (hereafter {\it{PyMorph}}) is from the photometric catalogs of M16 and F19  (please refer to that Section for the procedures of combining the best models in these 2 catalogs for the photometry of 74 BCGs, which shall be referred to as  the UPenn sample). The latter ({\tt Ellipse}-based) also provides two kinds of measurements: one is the S\'{e}rsic fitting mentioned in Section~\ref{subsubsec:Phot} (hereafter {\it{Sersic}}), and the other is the photometry within 150\,kpc mentioned in Sec.~\ref{subsubsec:TNGphot} (hereafter {\it{150\,kpc}}). The priority order in our usage is {\it{PyMorph}}, {\it{Sersic}}, then {\it{150\,kpc}}. Since {\it{PyMorph}} and {\it{Sersic}} are parametric models that can be integrated  to infinity, while the {\it{150\,kpc}} one is an empirical model that sums up the flux within a finite radius,  the total flux from {\it{150\,kpc}} should be systematically smaller than the parametric models. To understand the differences among the 3 photometric measurements and to check whether they are consistent, here we compare the photometry based on the UPenn sample.

\begin{figure}
\centering
\includegraphics[width=0.8\textwidth]{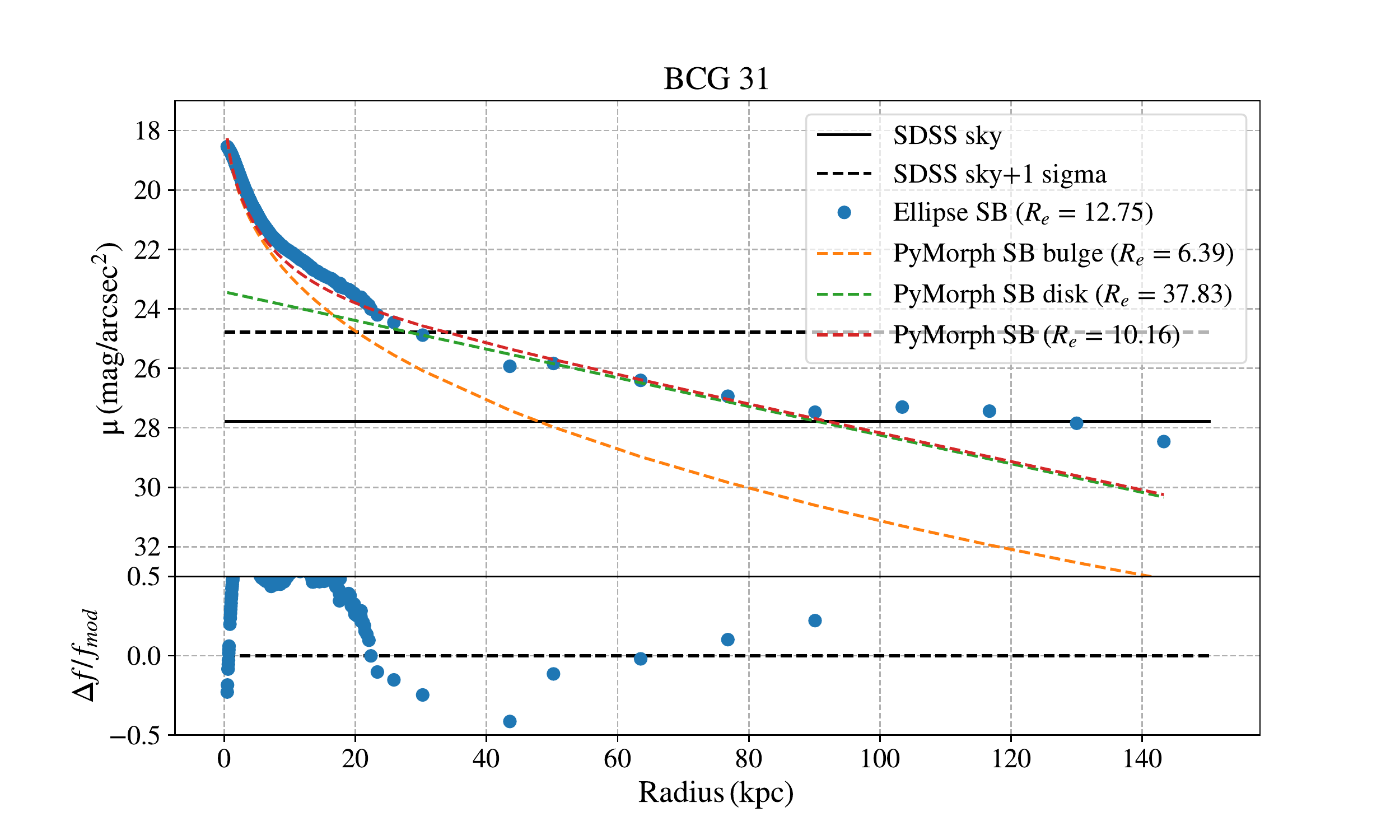}
\caption{\label{fig:31sb_eg.png}
An example (BCG no.~31) for the surface brightness profiles of {\tt Ellipse} and the 2 component model of PyMorph. The upper panel shows the surface brightness profile from {\tt Ellipse} within 150\,kpc (blue dots), the bulge component from {\tt PyMorph} (orange dashed line), the disk component from {\tt PyMorph} (green dashed line), and the combination of the two (red dashed line). The surface brightness profiles, as well as the half-light radius, are presented as a function of the radius of the generalized ellipse profile in unit of kpc. The sky and sky uncertainty measured by SExtractor is also shown. The lower panel shows the difference between two profiles as $ ({\rm flux}_{\tt Ellipse}-{\rm flux}_{\tt Pymorph} ) / {\rm flux}_{\tt Pymorph}$.
}
\end{figure}

\begin{figure}
\centering
\includegraphics[width=0.8\textwidth]{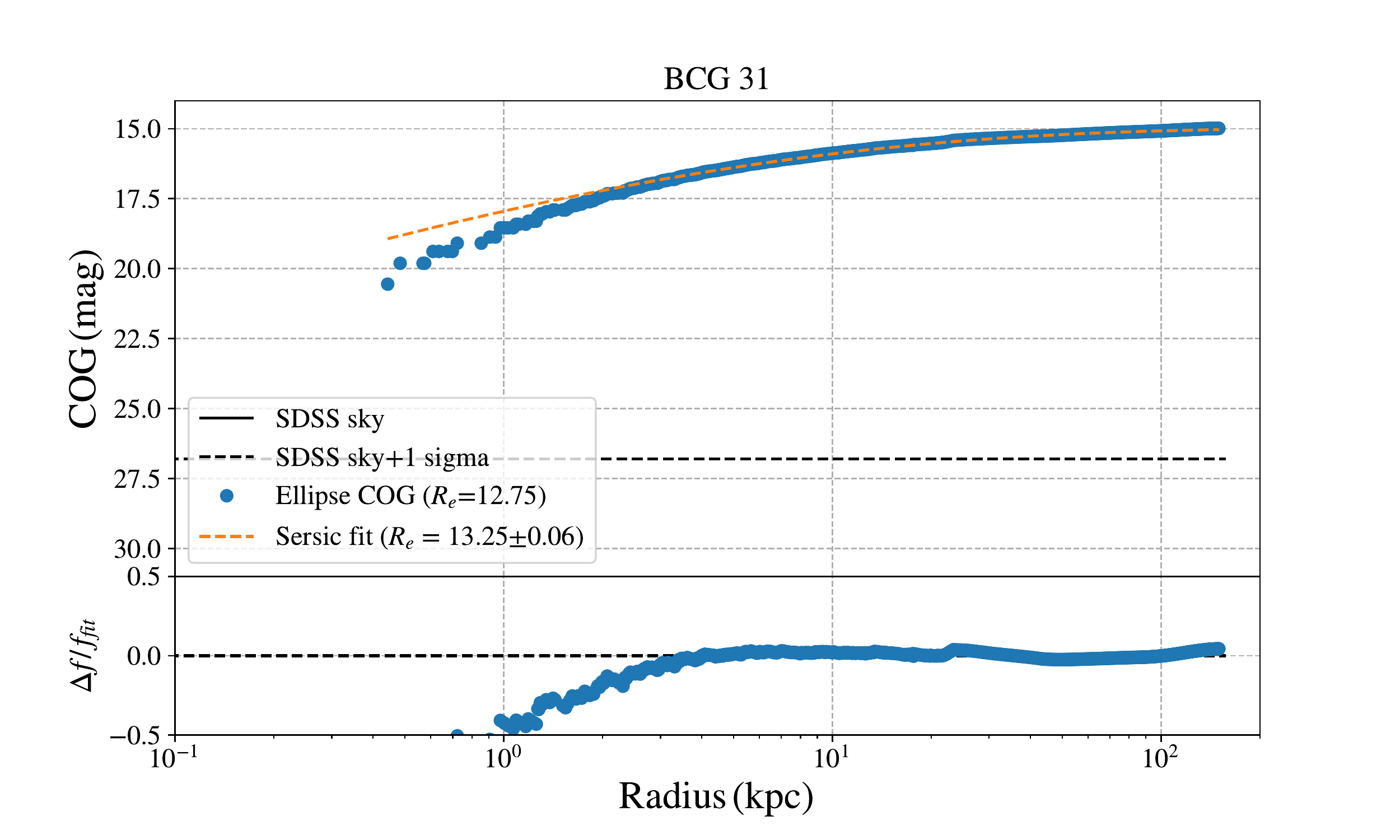}
\caption{\label{fig:31cog_eg.png}
An example (BCG no.~31) showing the curve of growth out to 150\,kpc from  {\tt Ellipse} (blue points), and a single S\'{e}rsic fitting to the profile (orange curve). 
The curve of growth, as well as the half-light radius, are presented as a function of the radius of the generalized ellipse profile in unit of kpc. The lower panel shows the difference between two profiles as $ ({\rm flux}_{\tt Ellipse}-{\rm flux}_{\tt Sersic} ) / {\rm flux}_{\tt Sersic}$. 
}
\end{figure}

Fig.~\ref{fig:31sb_eg.png} and Fig.~\ref{fig:31cog_eg.png} show the profiles of the 3 kinds of photometry for an example BCG, no.~31, in our Main sample. The profiles, as well as the half-light radii, are   functions of the radius of the generalized ellipse profile in unit of kpc \citep{peng02}. The radius or radial coordinate of the generalized ellipse profile is defined as:
\begin{equation}\label{form:gen_el}
r(x,y) = \left({|x-x_c|}^{C_0+2} + \left|\frac{y-y_c}{q}\right|^{C_0+2})^{\frac{ 1 }{ C_0+2 }}\right),
\end{equation}
where ($x_c,y_c$) is the center, $q$ is the axis ratio, and $C_0$ is the ``boxiness'' parameter ($C_0=0$ corresponds to a perfect ellipse; in running {\tt Ellipse} this is the value we adopt).

\begin{figure}
\centering
\includegraphics[width=0.4\textwidth]{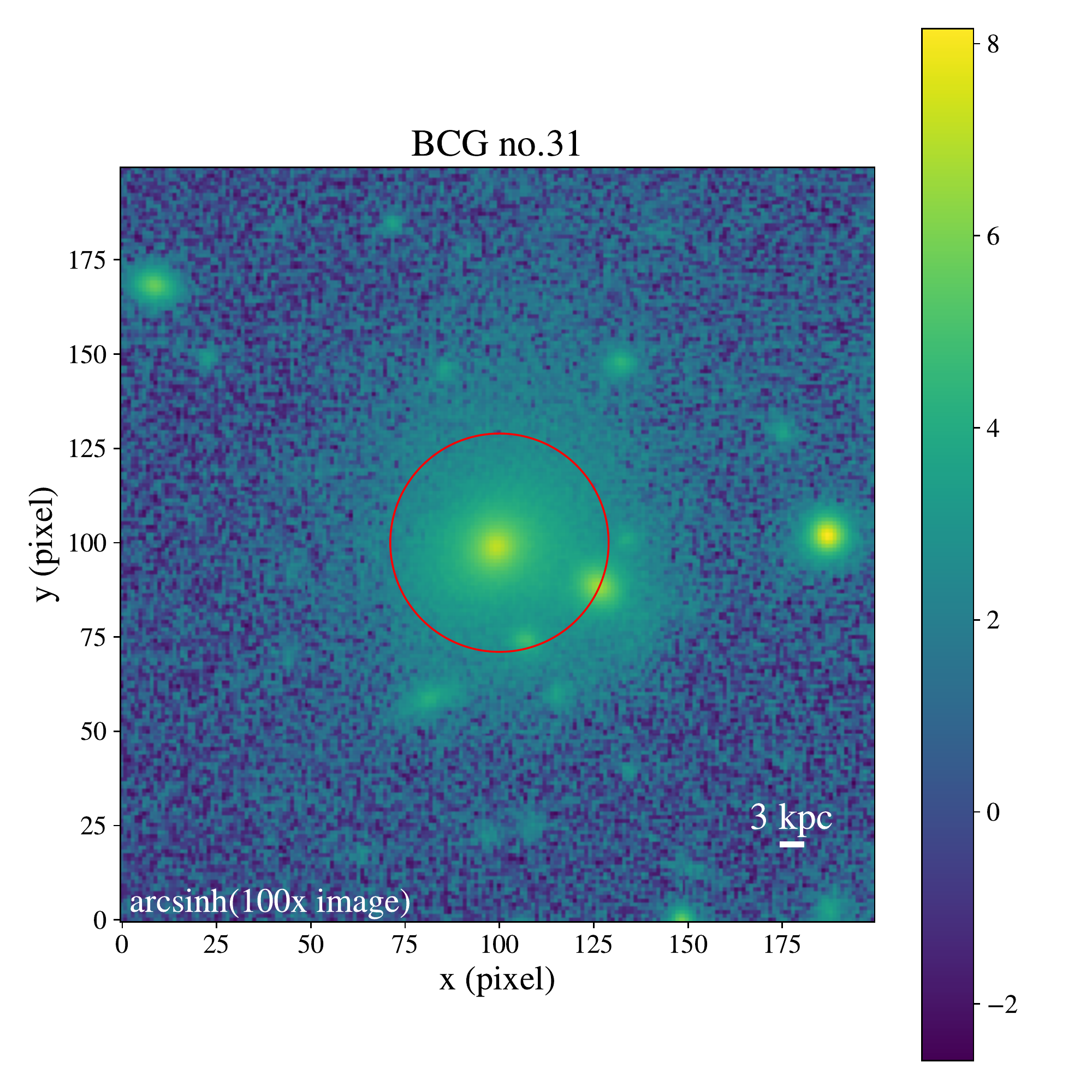}
\caption{\label{fig:31im.png}
The image of BCG no.~31, zoomed in to the central 200 pixels.  The red circle has a radius of 20\,kpc, and touches two satellites, which cause a dip in the surface brightness profile derived by {\tt Ellipse}.
}
\end{figure}

Most of the surface brightness profile of {\it{PyMorph}} and {\it{150\,kpc}} are consistent down to the sky level, but the {\it{150\,kpc}} are more sensitive to the light of cores, nearby neighbors, and other asymmetric structures. For example, there are 2 cores at around 20 kpc of BCG no.~31 that causes a ``bump'' (see Fig.~\ref{fig:31im.png}).

\begin{figure}[]
\plotone{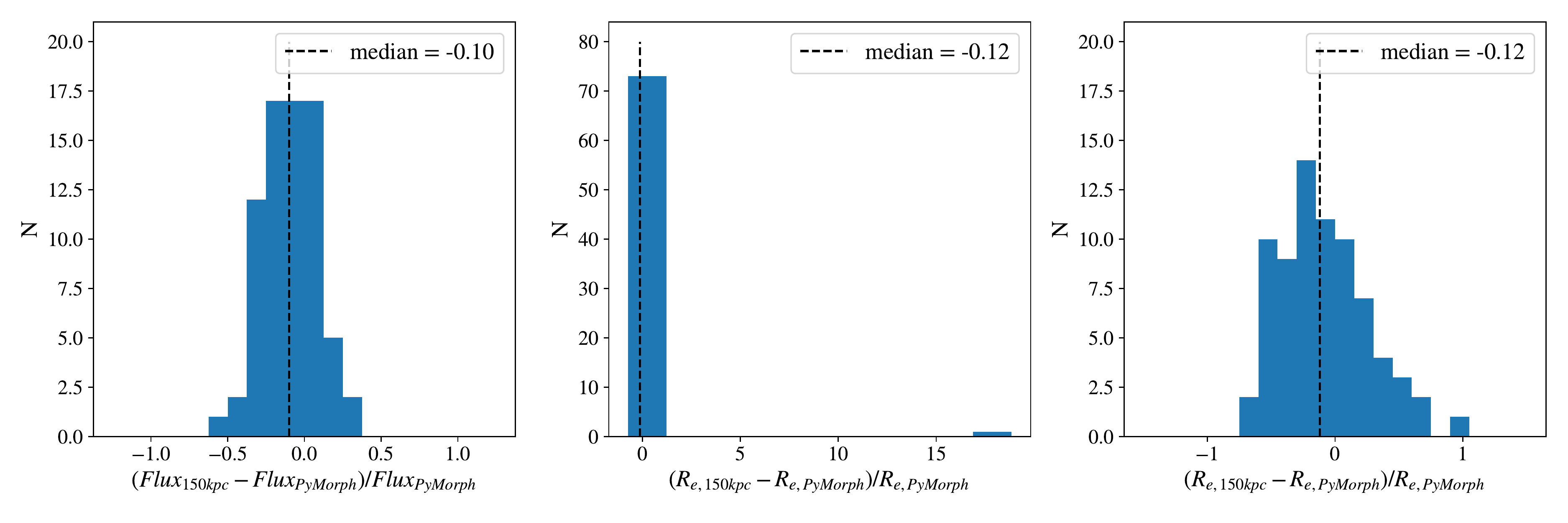}
\caption{
Comparison of total flux (left panel) and $R_e$ (middle panel) from {\it{150\,kpc}} and {\it{PyMorph}} of the UPenn sample of 74 BCGs.  
The panel on the right is a zoom-in to values around zero for $R_e$.
\label{fig:150PyMorph}}
\end{figure}

Below we provide pairwise comparison of the 3 kinds of photometric measurements.
First, in Fig.~\ref{fig:150PyMorph} we show the comparison of the total flux (left panel) and half-light radius  (middle panel) of {\it{PyMorph}} and {\it{150\,kpc}}   of the UPenn sample. 
For both quantities, the values based on {\it{150\,kpc}} are slightly smaller.  There is also an outlier to the distribution.
Upon inspection of the surface brightness profiles, we deem that  the values based on {\it{150\,kpc}} are incorrect (presumably due to its sensitivity to the residuals from bright neighbors).  The panel on the right shows the comparison based on whole UPenn sample, zoomed in to the range around zero.

\begin{figure}[]
\plotone{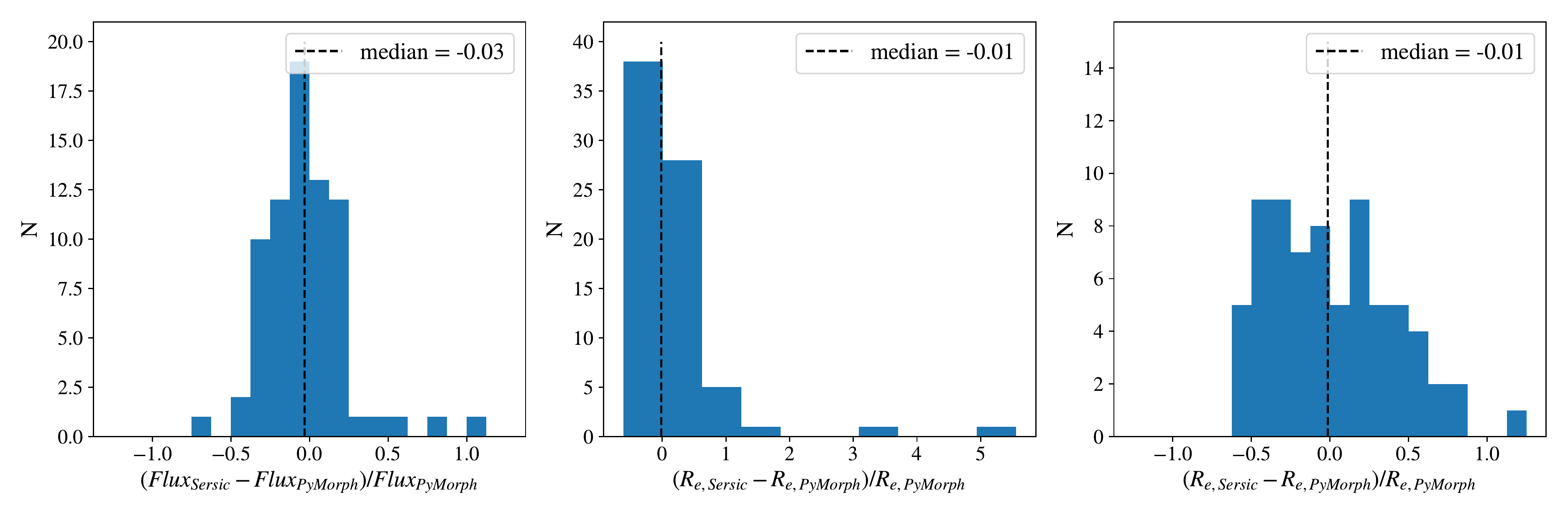}
\caption{
Comparison of total flux (left panel) and $R_e$ (middle panel) of {\it{Sersic}} and {\it{PyMorph}} of the UPenn sample.  
The panel on the right is a zoom-in to values around zero for $R_e$.
\label{fig:SerPyMorph}}
\end{figure}

Second, we compare the {\it{Sersic}} and {\it{PyMorph}} of the UPenn sample in Fig.~\ref{fig:SerPyMorph} (again, left panel for the total flux, and middle panel for $R_e$). There are   4 BCGs that have  $\Delta {\rm flux} > 50 \%$.  The panel on the right shows the comparison based on whole UPenn sample, zoomed in to the range around zero.

\begin{figure}[]
\plotone{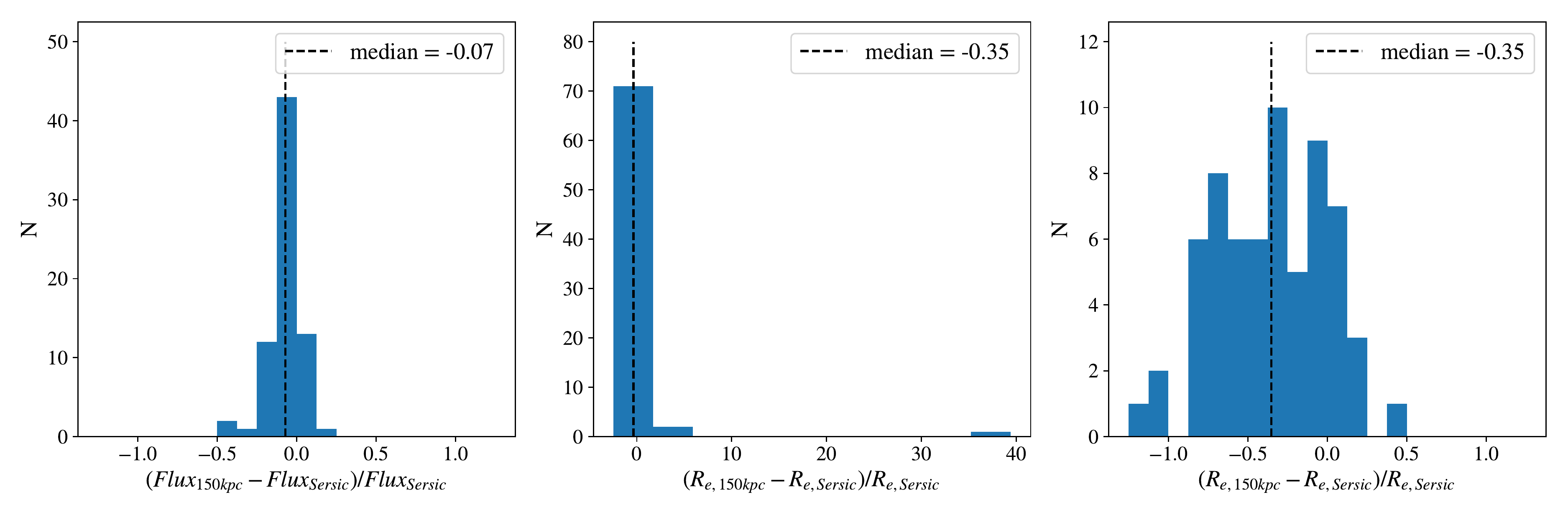}
\caption{
Comparison of total flux (left panel) and $R_e$ (middle panel) of {\it{150\,kpc}} and {\it{Sersic}} of the UPenn sample.  
The panel on the right is a zoom-in to values around zero for $R_e$.
\label{fig:150Ser}}
\end{figure}

The comparison between {\it{150\,kpc}} and {\it{Sersic}} is presented in Fig.~\ref{fig:150Ser}, which
 shows that the total flux of these two method are very consistent and the $R_e$ of {\it{Sersic}} is only larger by 7\%.

For the total flux, we use {\it{PyMorph}} and {\it{Sersic}} for the Main sample, and primarily {\it{Sersic}}  for the simulated sample (except for a few that {\it{150\,kpc}} is used). The offsets between these distribution are within $7\%$, although the spread can up to $50\%$. 
This is because {\it{PyMorph}} has a better background subtraction and uses a 2-component model, so we primarily use it. The flux of {\it{150\,kpc}} is only measured in a limited radius and is systematically smaller than the others, so it is the last choice among the three. The TNG sample does not have the {\it{PyMorph}} measurements, so {\it{Sersic}} is preferred.

To decide the maximum separation of cores, we use the median $R_e$ of {\it{PyMorph}} for the real data, and {\it{Sersic}} for the mock observations. The {\it{PyMorph}} median $R_e$ for the Upenn sample is 17.49 kpc, while that from {\it{Sersic}} is 17.88 kpc. Based on Fig.~\ref{fig:SerPyMorph}, we expect the two median values to be consistent within $1\%$. This is because {\it{Sersic}} and {\it{150kpc}} are based on empirical photometry that is more sensitive to the residuals of the bright neighbors. Also, the spread in $R_e$ is much larger than the total flux, so a fixed value (18 kpc) is better for direct comparisons between the observation and simulation.

\section{Notable objects}\label{appendix:note}

\begin{figure}
\centering
\includegraphics[width=0.7\textwidth]{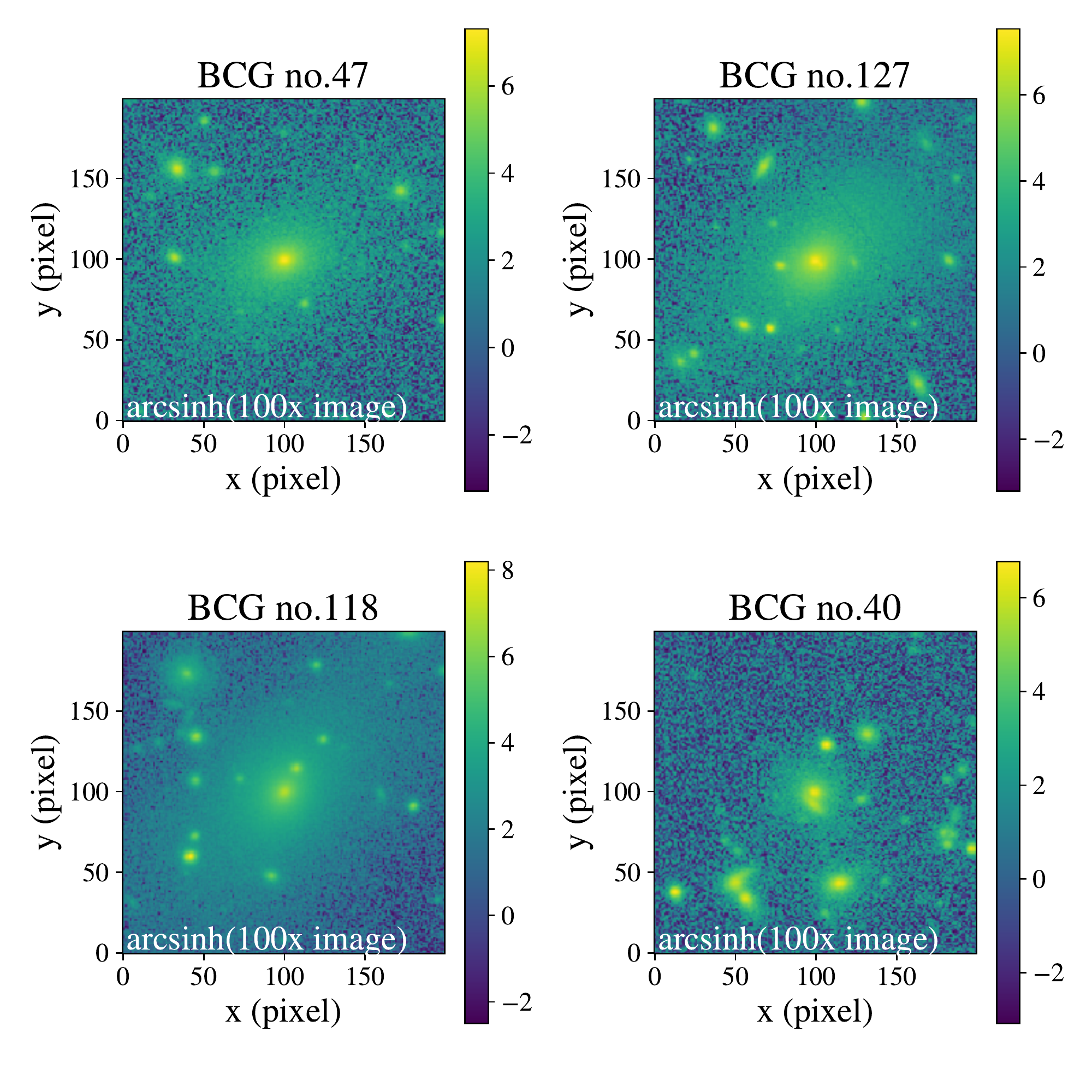}
\caption{\label{fig:im_note.png}
The SDSS $i$-band images of the 4 notable BCGs as noted in Section~\ref{appendix:note_BCG}, zoomed in to the central 200 pixels.
}
\end{figure}

In the Main sample, 
several BCGs pose challenges to our pipeline, and we have to apply special treatments for them
instead of the general procedure described in Sections~\ref{subsubsec:multi} and \ref{subsubsec:merg} (see Fig.~\ref{fig:im_note.png}).
In the simulated sample, for 2 simulated BCG we use the {\tt Ellipse} total flux within 150\,kpc instead of results from the single S\'{e}rsic fit.  We discuss these objects case by case in this Appendix.
We also show explicitly the BCGs (both real and simulated ones)  that are excluded in our analysis in Section~\ref{appendix:exclude}.

\subsection{Notable BCGs in the Main Sample}\label{appendix:note_BCG}

First, two of the BCGs (nos.~47 and 127) could have their brightest peak also detected as a core, because there are 2 nearly equally bright peaks at its center, and the SDSS pipeline sets the galaxy center in between the peaks.
 For these objects, we update the position of BCG to the brightest peak and apply our detection relative to this new position.

\begin{figure}
\centering
\includegraphics[width=0.7\textwidth]{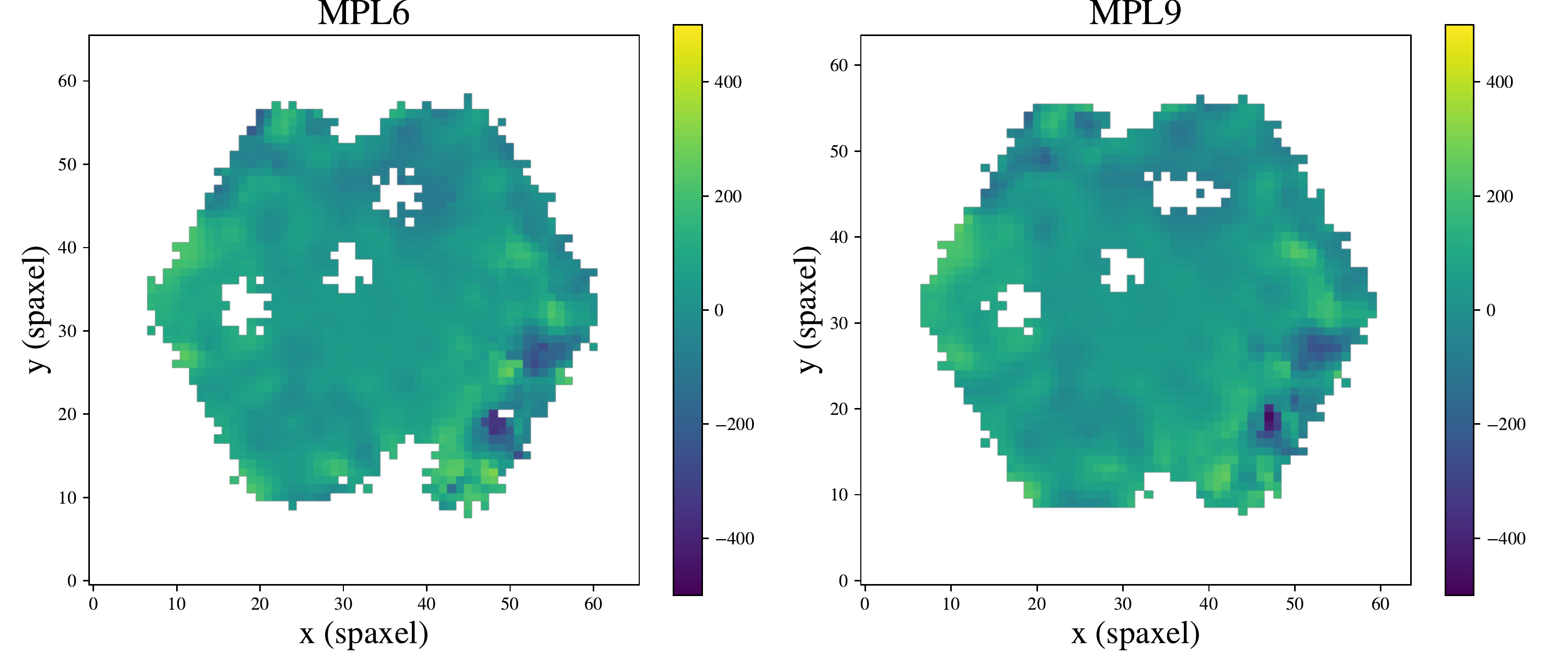}
\caption{\label{fig:118.png}
The MPL-6 and MPL-9 MaNGA stellar velocity map with ``DONOTUSE" mask applied, for BCG no.~118.
}
\end{figure}

Second, the spaxels of the core of BCG no.~118 are mostly masked out in DAP in MPL-9, but partially retained in MPL-6. Hence, we use the MPL-6 map for this source (See Fig.~\ref{fig:118.png}). 

Third, the IFU observation of BCG no.~40 is targeting a satellite instead of the BCG. This is because the target belongs to a MaNGA ancillary program investigating close pairs and mergers
\citep{wake17}, whose targets are selected from the NSA catalog with a velocity difference of less than 500\,km/s. Therefore, although we cannot measure its velocity difference relative to the BCG main body, it still satisfies our criteria.

Finally, recall that, we mainly use the 800$\times$800 pixel images and their  BACK$\_$SIZE\,$=160$ (1/5 image size) segmentation maps for our BCGs. The 1000$\times$1000 pixel image and the  BACK$\_$SIZE\,$=160$ (1/3 image size) set is used in a few cases (4 out of 89 in the Main sample) when the segmentation and {\tt Ellipse} models of the two versions are different, and we prefer the 1000$\times$1000 pixel version. The 800$\times$800 image and the  BACK$\_$SIZE\,$=160$ (1/3 image size) is used for BCG no.~68, however, because the usual settings cannot mask out a bright elongated galaxy in its image.

\subsection{Notable BCGs in TNG300}\label{appendix:note_TNG_BCG}

\begin{figure}
\centering
\includegraphics[width=0.7\textwidth]{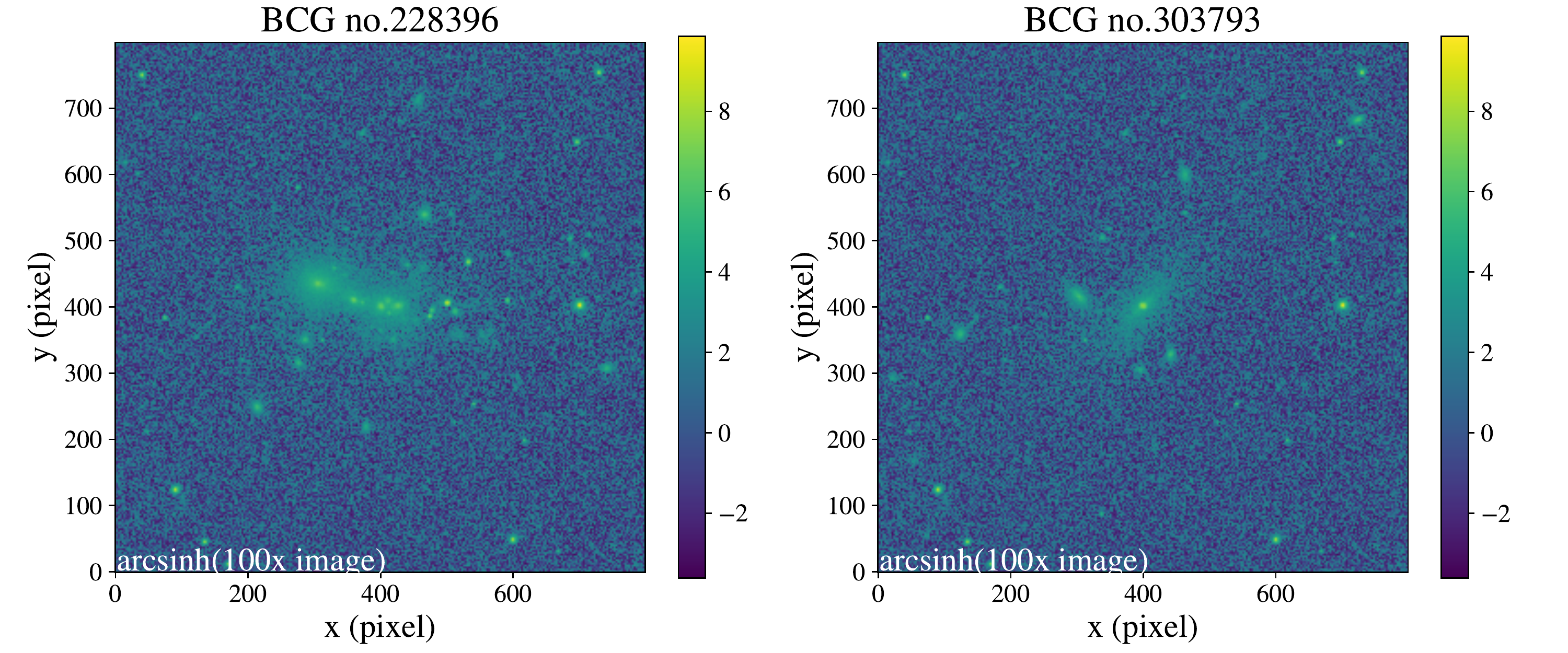}
\caption{\label{fig:TNG_note}
The images of the 2 notable BCGs in TNG300.
}
\end{figure}

There are 2 objects in the TNG sample that the single S\'{e}rsic fit fails and we use the {\tt Ellipse} total flux within 150\,kpc instead (Fig.~\ref{fig:TNG_note}).

The BCG of halo 228396 consists of 4 blending bright objects and this complex morphology causes a single S\'{e}rsic fit to fail. We still use the {\tt Ellipse} photometry because,  even though the curve of growth of the primary object is strongly affected by the other objects, they are all considered as the components of the BCG.

The BCG of halo 303793 has an unusual morphology that has a bright core in the center and a faint and very elongated outskirt. Since the light is very concentrated and the S\'{e}rsic fit fails, we use the flux within 150\,kpc.

\subsection{Changed BCGs}
\label{appendix:changed}

\begin{figure}
\centering
\plottwo{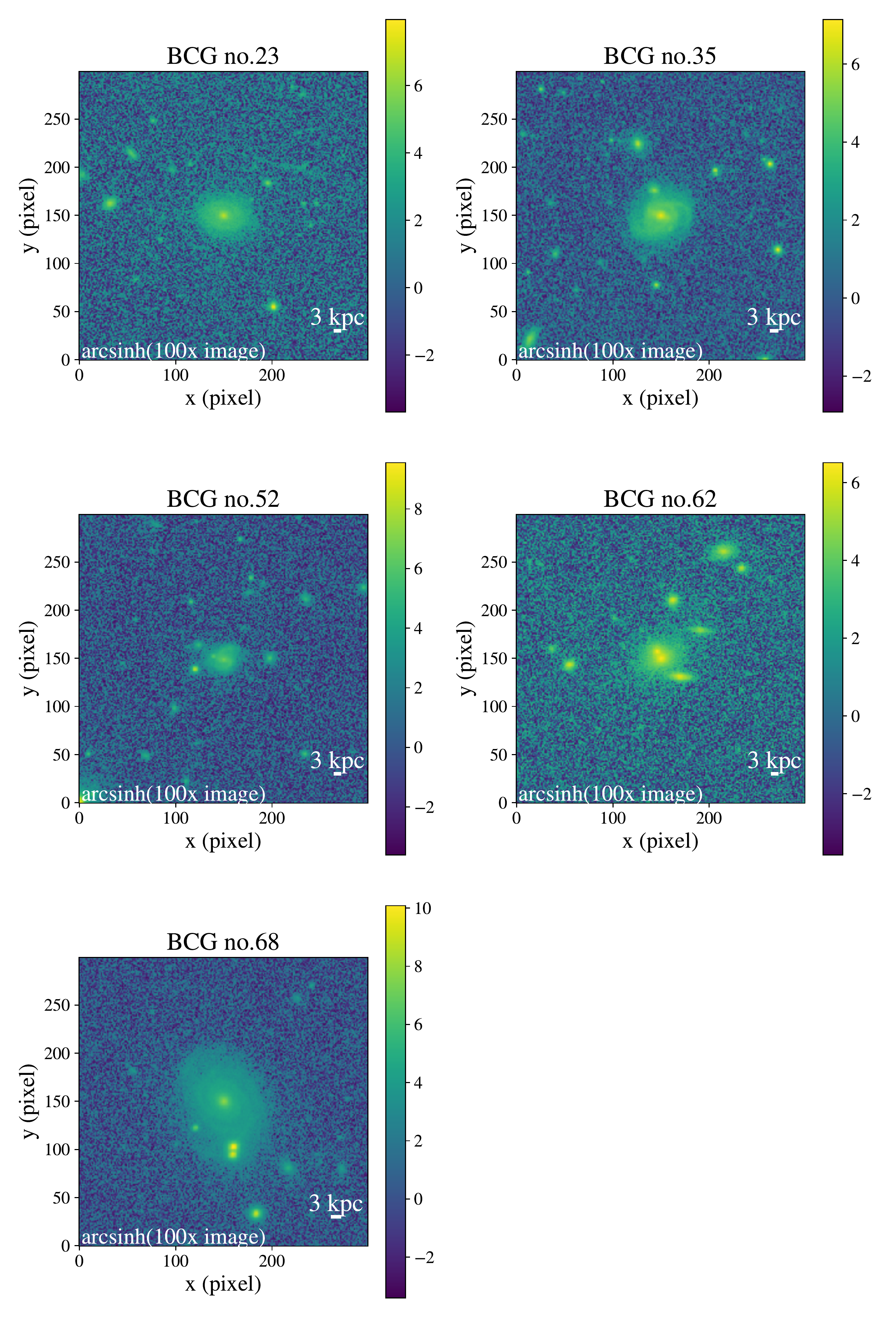}{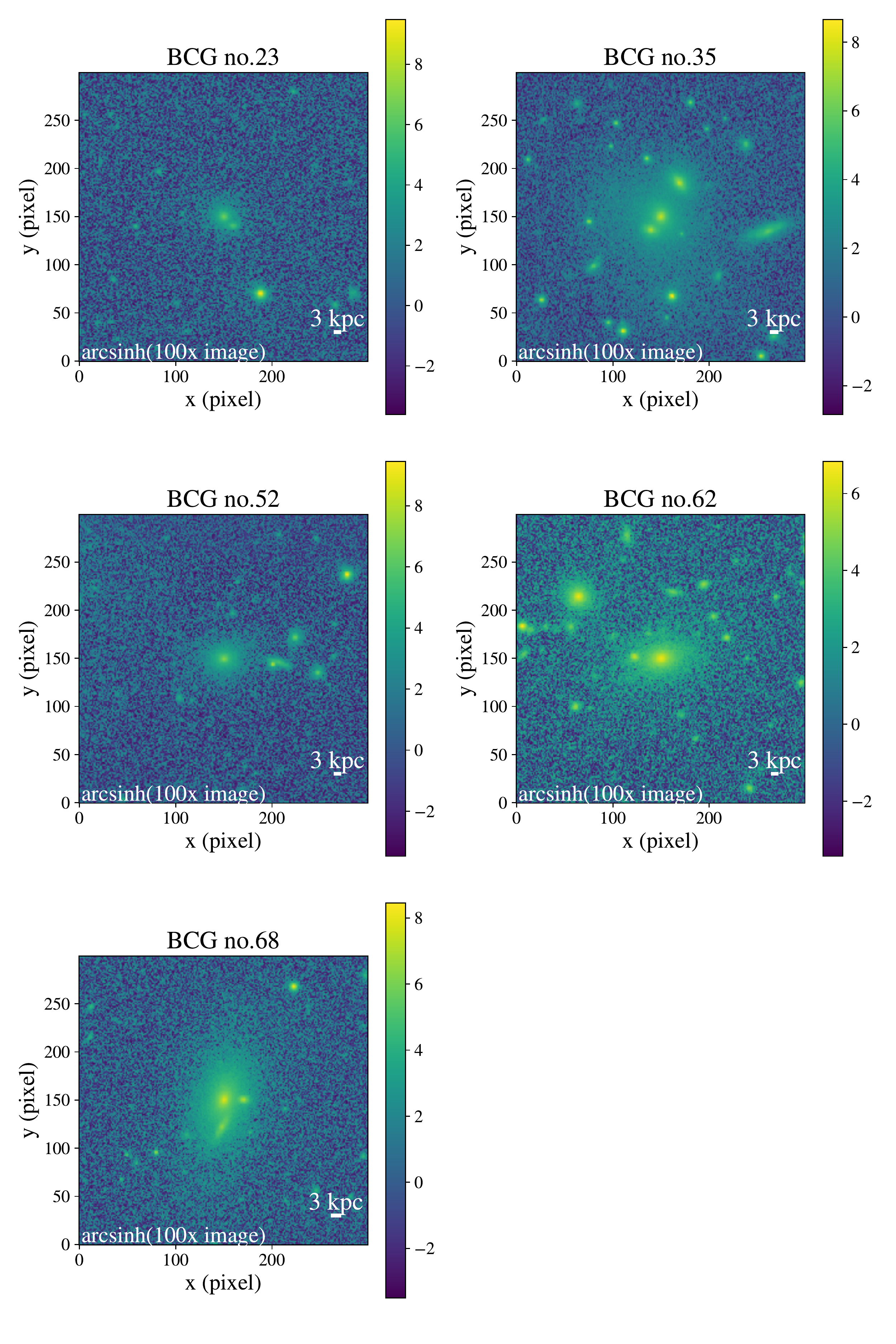}

\caption{\label{fig:changed_bcg}
{\it Left:} The SDSS images of the 5 Y07 BCGs that are changed during our BCG identification process, zoomed in to the central 300 pixels.  It can be noticed that there is a core in BCG no.~62; it has a velocity offset of $-286$\,km/s.  Adopting this galaxy as the BCG would change our $f_{\rm mc}$ to $0.13\pm 0.04$.
{\it Right:} The images of the 5  BCGs identified by us, zoomed in to the central 300 pixels.
}
\end{figure}

\begin{figure}
\epsscale{0.8}
\plottwo{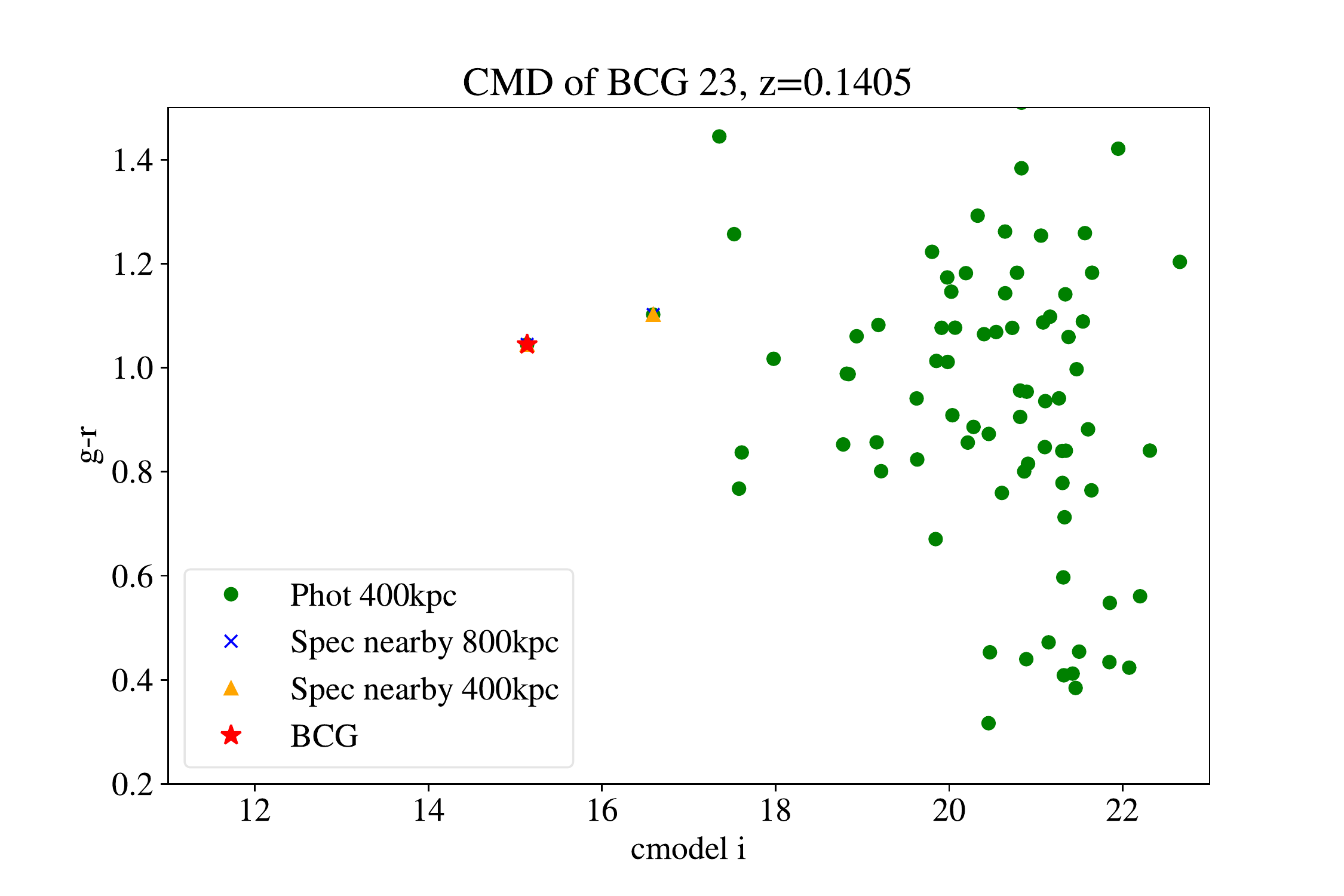}{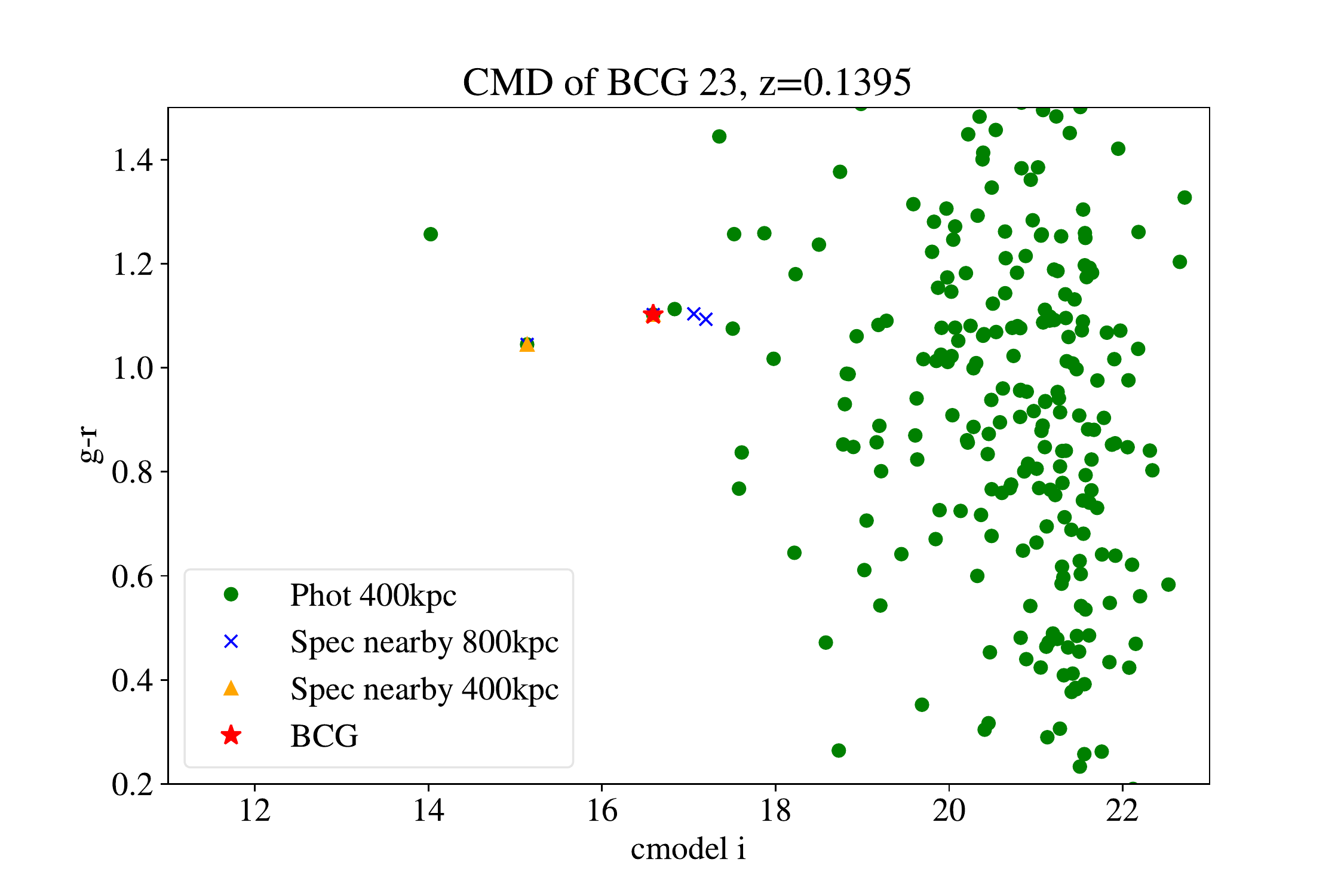}
\plottwo{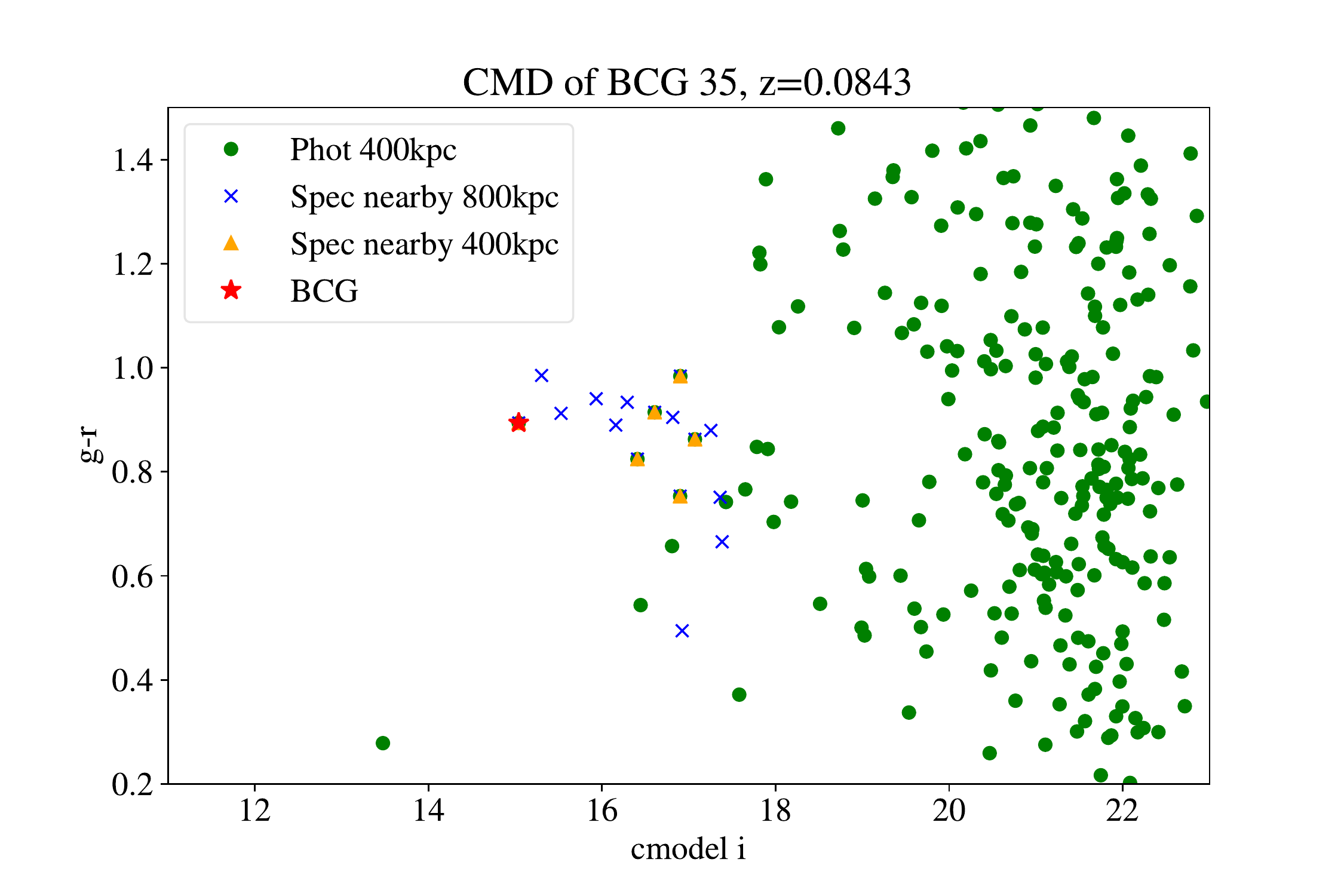}{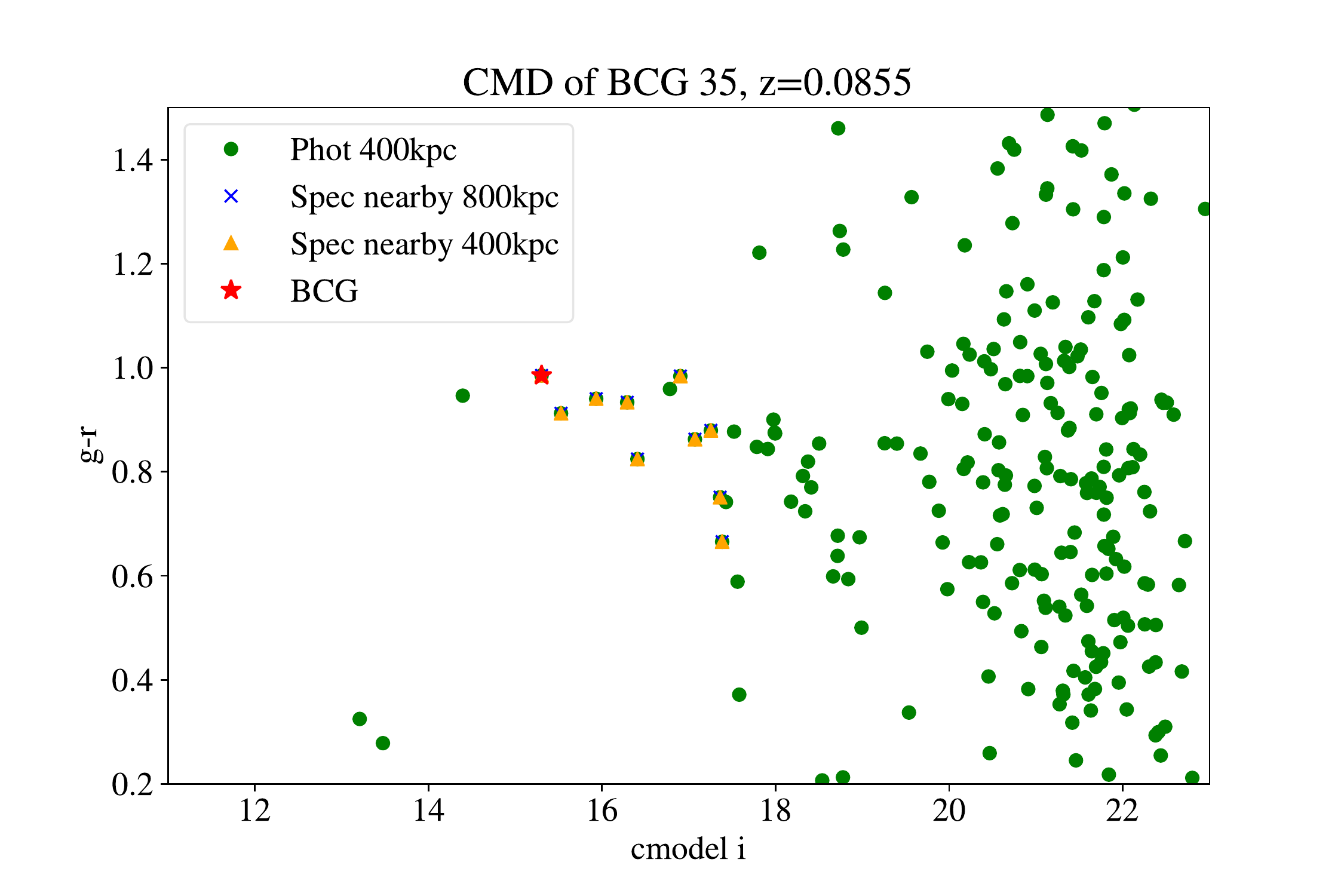}
\plottwo{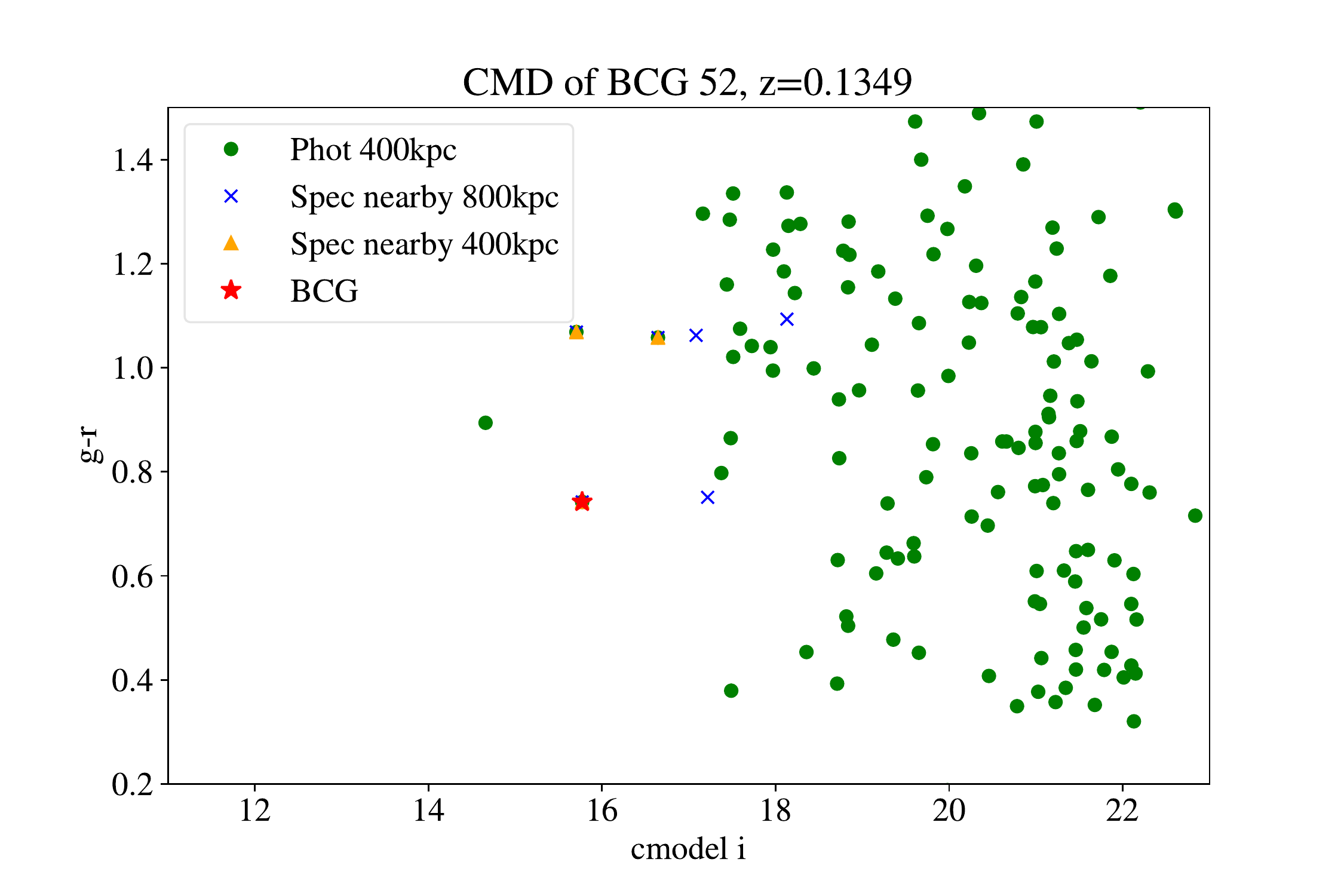}{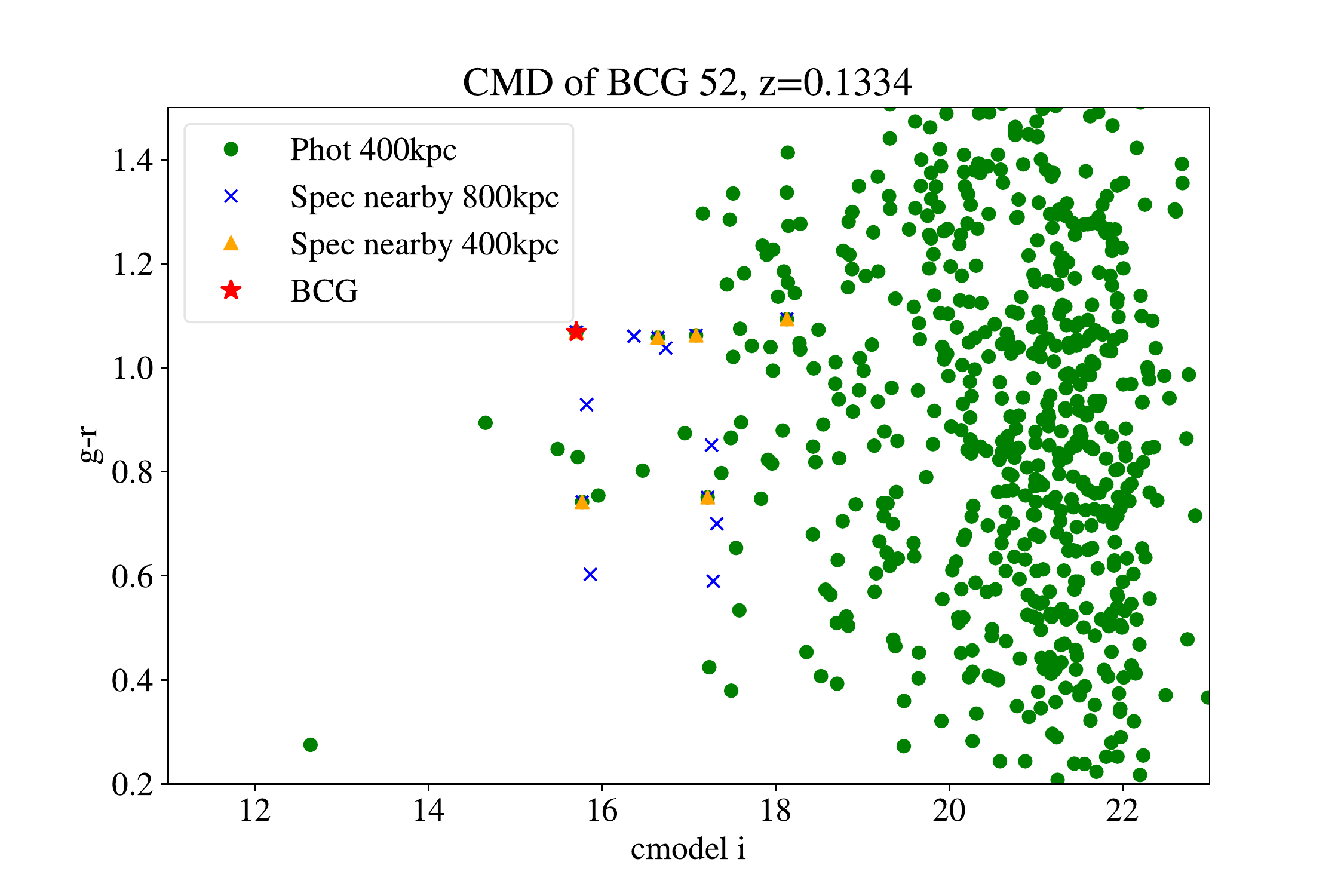}
\caption{
{\it Left:} Color-magnitude diagrams (CMDs) of 3 Y07 BCGs (nos.~23, 35, and 52) that are changed in our BGC identification process. The color is extinction corrected model $g-r$ magnitude, and the X-axis is the cmodel (de Vaucouleurs$+$Exponential) $i$-band magnitude. The red star represents the BCG. The green dots are the SDSS galaxies within a 400 kpc radius. The blue crosses are the SDSS galaxies with spectroscopy within 800 kpc radius and redshift offset\,$<0.01$. The yellow triangles are SDSS galaxies with spectroscopy within a 400 kpc radius. {\it Right:} The CMDs showing the location of our chosen BCGs.
}
\label{fig:changed_cmd}
\end{figure}

\begin{figure}
\epsscale{0.8}
\plottwo{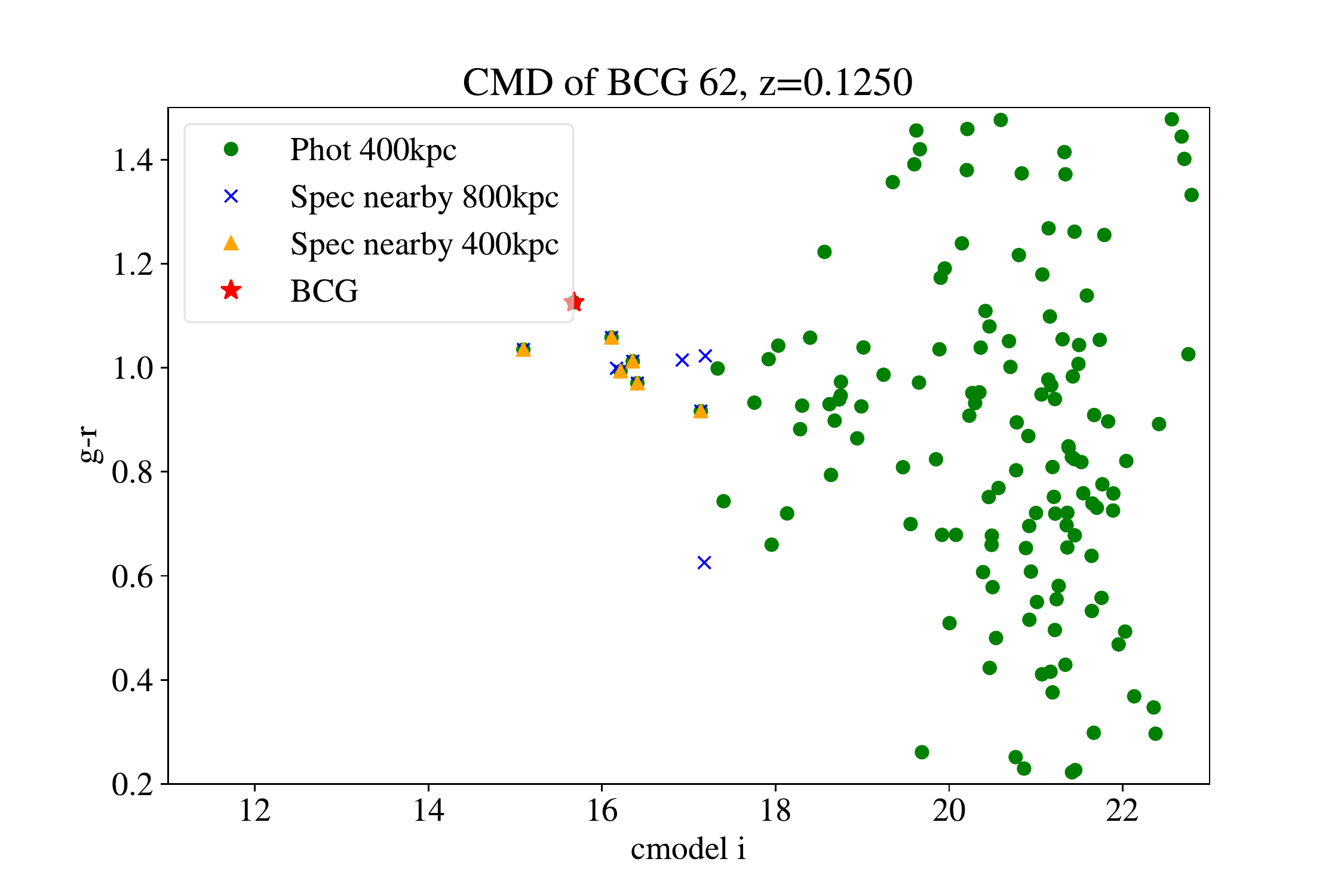}{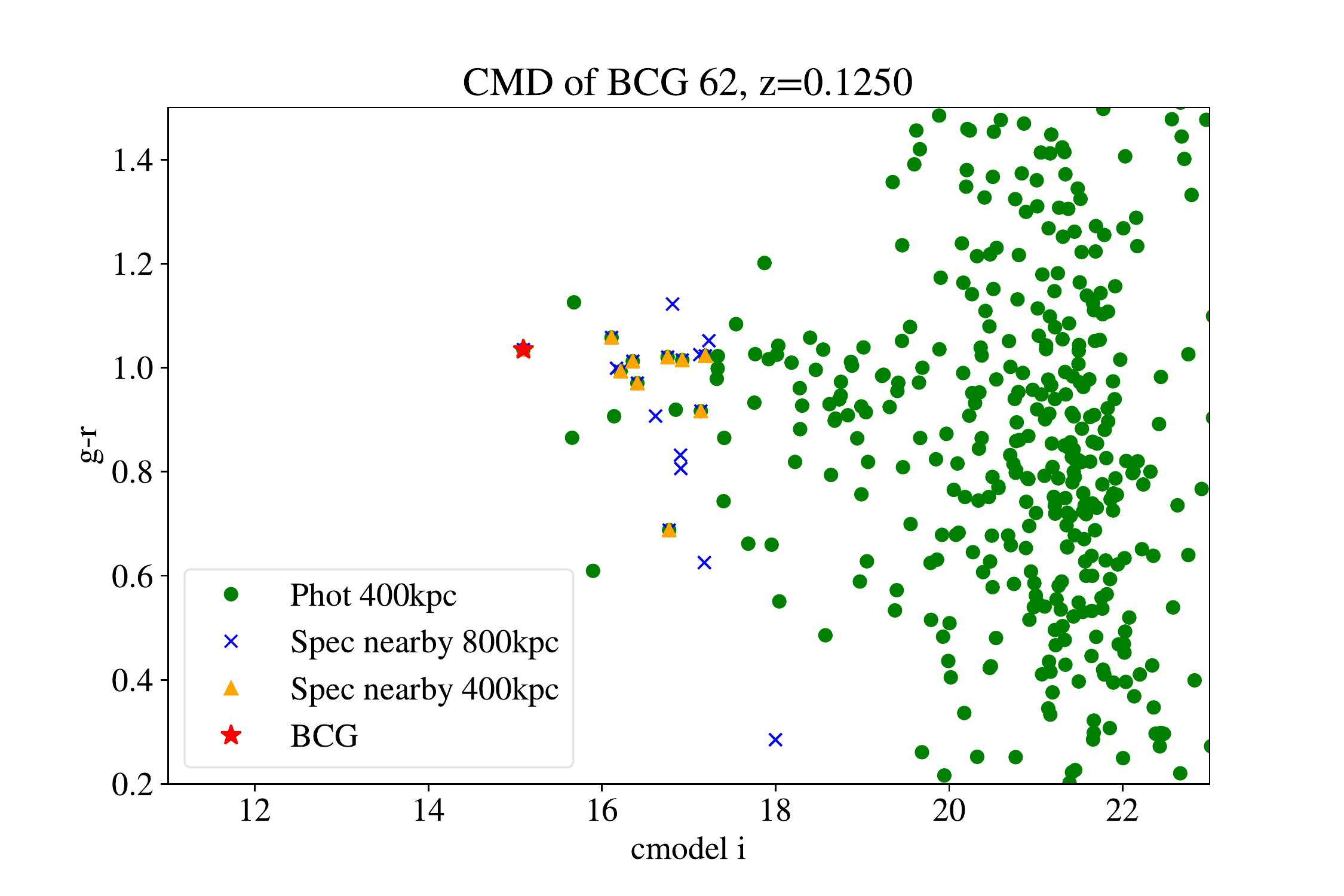}
\plottwo{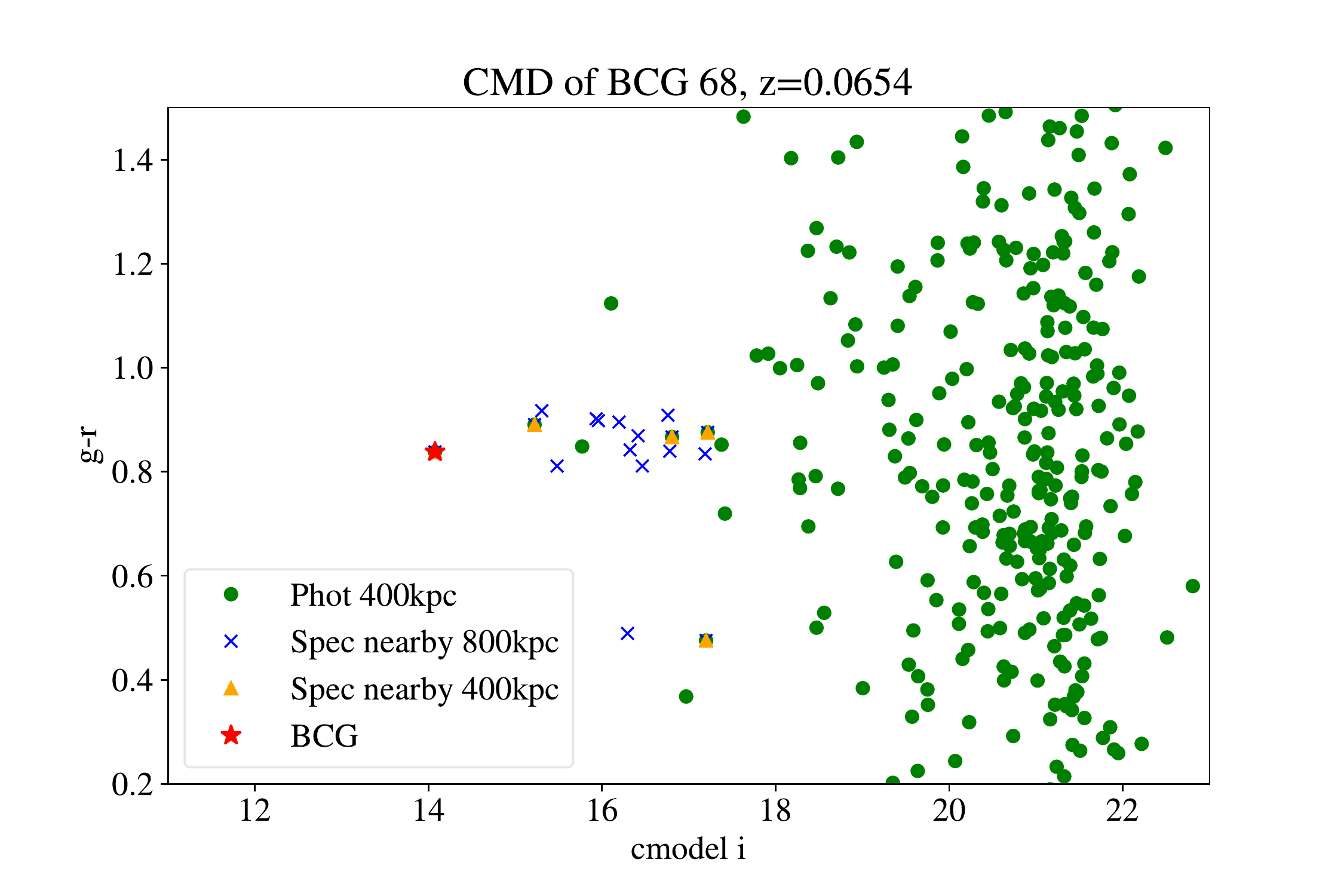}{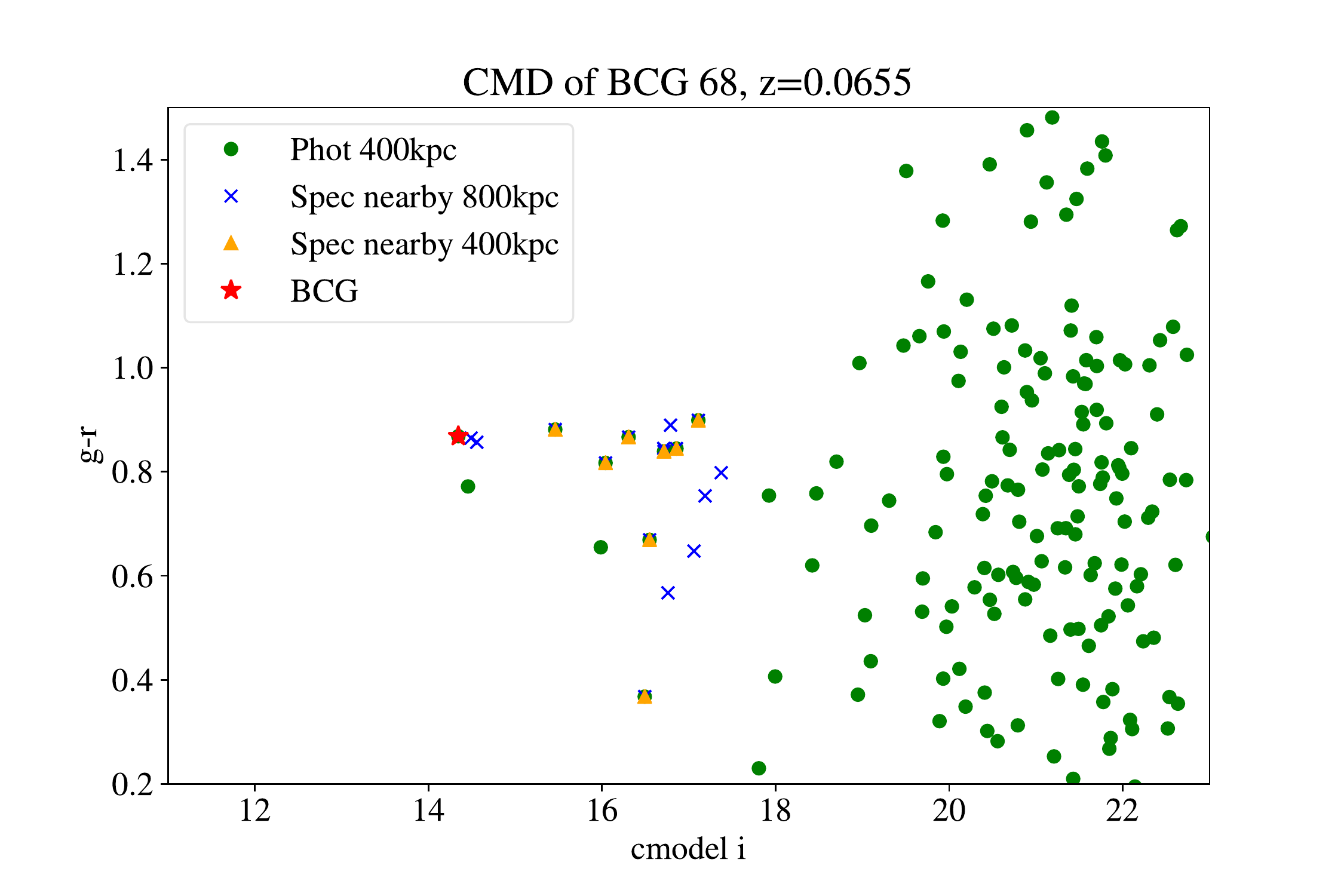}
\caption{
Same as Fig.~\ref{fig:changed_cmd}, but for BCGs nos.~62 and 68.
}
\label{fig:changed_cmd2}
\end{figure}

In the BCG identification process, we have changed 5 BCGs originally identified by Y07 to another galaxy. The SDSS images of the original BCGs in Y07 and the new BCGs are shown in Fig.~\ref{fig:changed_bcg}. The color-magnitude diagrams of these clusters are shown in Figs.~\ref{fig:changed_cmd} and \ref{fig:changed_cmd2}. BCGs nos.~23, 35, 52, 68 are changed because they are spiral galaxies. BCG no.~62 is not the most luminous galaxy within a projected distance of 800\,kpc from the cluster center, so we select the truly most luminous one as the BCG\footnote{The reason for this is that additional redshifts from SDSS-III \citep{eisenstein11} became available after the construction of the Y07 catalog.  We note that this only occurs to this particular cluster; for all other clusters, the new redshifts from SDSS-III do not play a role in the choice of BCGs.}, where the luminosity is taken from the NSA catalog. 
Among the changed galaxies, nos.~52, 62, 68 are in our main sample (please see Section~\ref{appendix:BCG_id} below for more details).
As can be seen from Fig.~\ref{fig:changed_bcg}, the original BCG of cluster no.~62 has a core; using MaNGA velocity map we have measured a velocity offset of $-286$\,km/s.  Had we adopted it as the BCG, our multiple-core frequency would increase slightly from $0.11\pm 0.04$ to $0.13\pm 0.04$.

\subsection{Excluded BCGs}\label{appendix:exclude}

\subsubsection{Excluded Objects in the MPL-9 sample at the BCG Identification Stage}\label{appendix:BCG_id}

\begin{figure}[]
\centering
\includegraphics[width=0.7\textwidth]{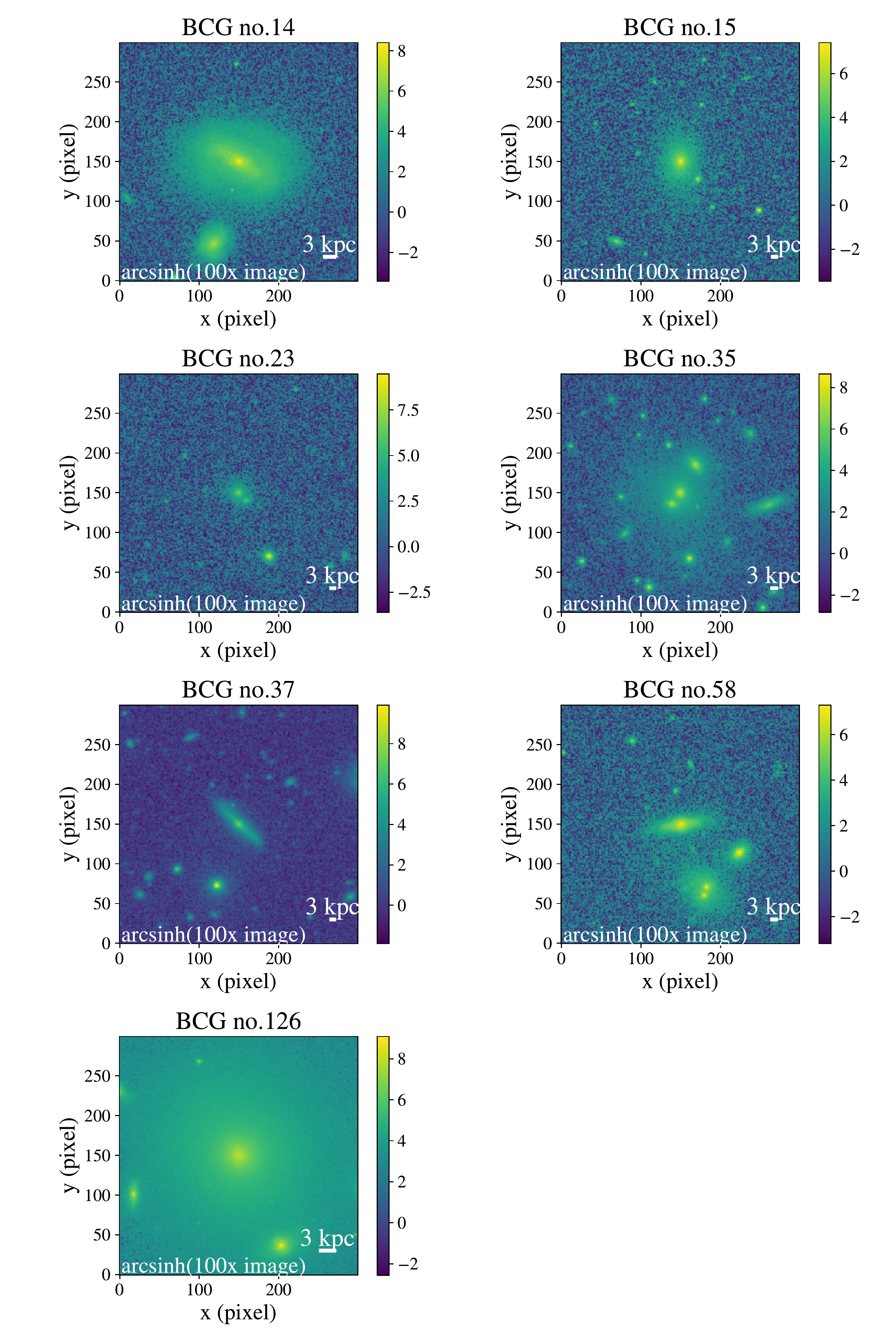}
\caption{
The SDSS images of the 7 BCGs removed during our BCG identification process, zoomed in to the central 300 pixels.  Note that for nos.~23 and 35 it is the BCG that we have chosen, not the one from the Y07 catalog.
\label{fig:removed}
}
\end{figure}

\begin{figure}
\epsscale{1.1}
\plottwo{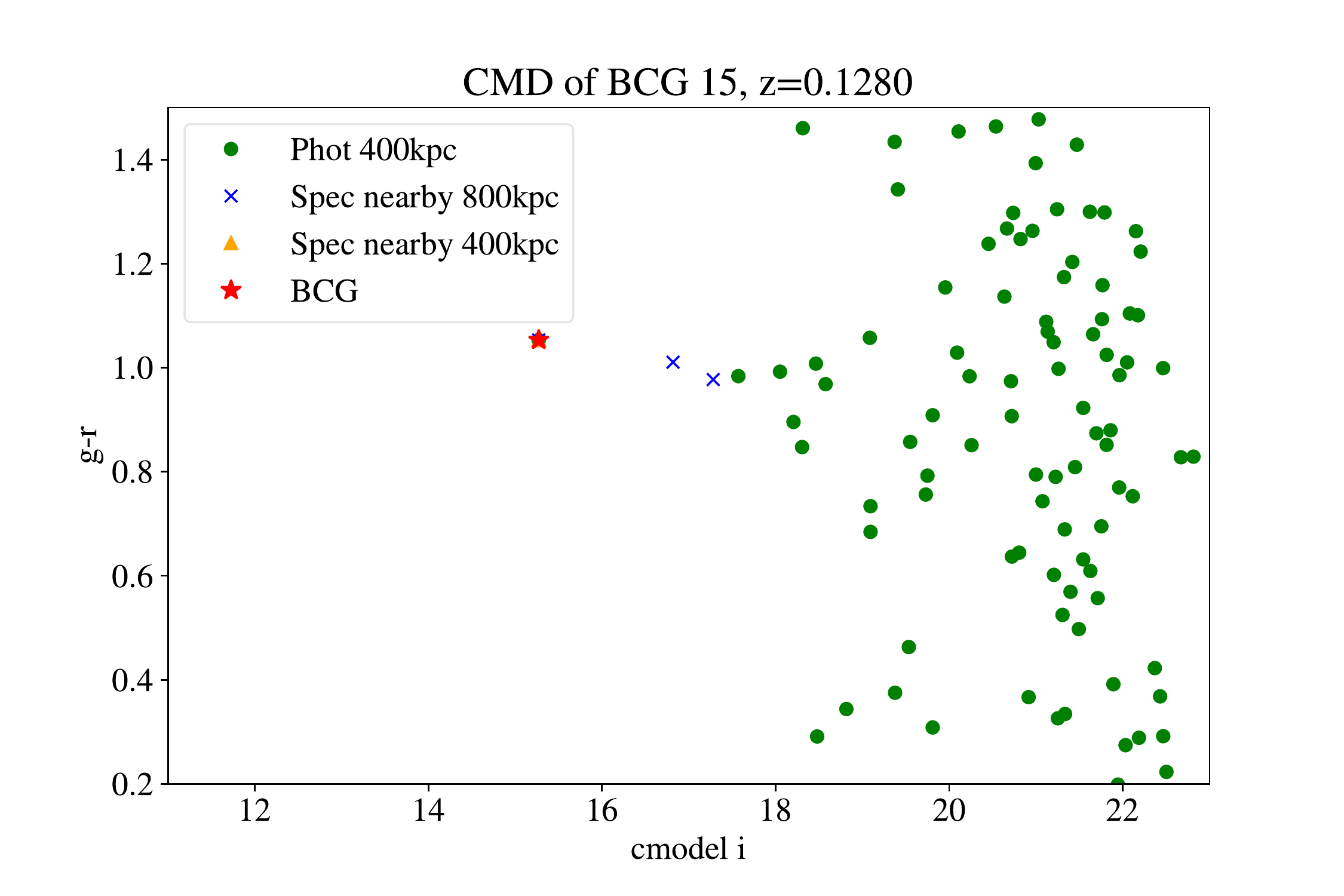}{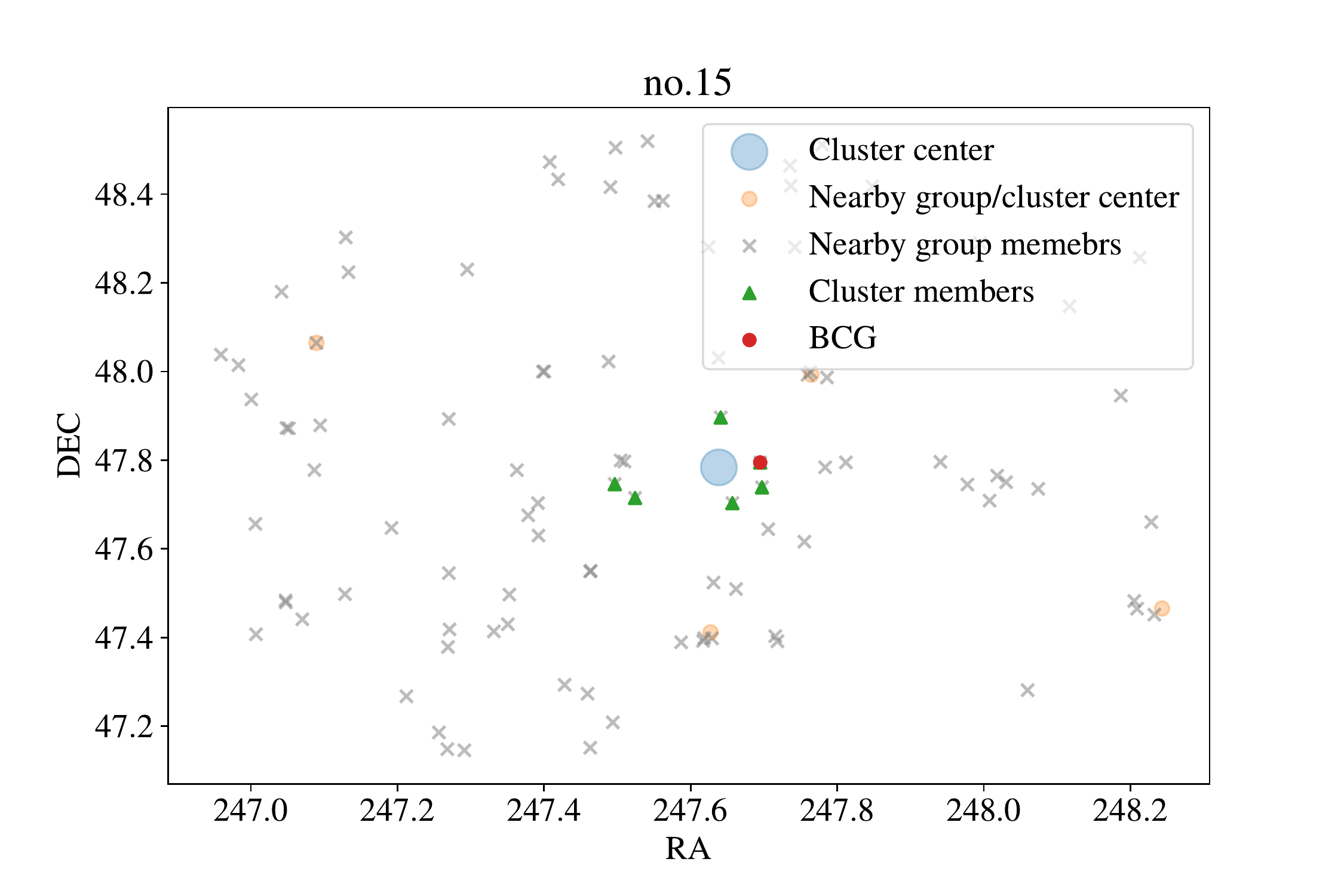}
\caption{
{\it Left:} 
The color-magnitude diagram (CMD) of BCG no.~15. The color is extinction corrected model $g-r$ magnitude, while the X-axis is the cmodel (de Vaucouleurs$+$Exponential) $i$-band magnitude. The red star represents the BCG. The green dots are the SDSS galaxies within a 400\,kpc radius. The blue crosses are the SDSS galaxies with spectroscopy within an 800\,kpc radius and redshift offset\,$<0.01$. The is no SDSS galaxies with spectroscopy within a 400\,kpc radius (yellow triangle).
{\it Right:} The spatial distribution member galaxies of the cluster that hosts BCG no.~15. The red dot represents the BCG identified by the Y07 catalog. The blue circle is the center of the cluster. 
The green triangles are the spectroscopically confirmed cluster members. 
The gray crosses (yellow circles) are the other galaxies (clusters) in the Y07 catalog that are within 0.75 degree radius and ${\rm redshift\  offset}<0.1$. 
\label{fig:15cmd}
}
\end{figure}

7 objects are excluded in the BCG identification process (Fig.~\ref{fig:removed}). BCG no.~15 has only few neighbors that can be seen in the SDSS image, and a few spectroscopically confirmed members (Fig.~\ref{fig:15cmd}, right panel) in the cluster catalog. Also, its red sequence is not obvious (left panel). BCGs nos.~14, 37, 58 have a spiral morphology, and we could not find other BCG candidates for their host clusters. 
 BCGs nos.~23 and 35 are chosen based on our selection criteria  (see Section~\ref{appendix:changed}), but they are not observed by MaNGA. The BCG of the Coma cluster (no.~126) is also excluded not only because it is very nearby and resolved, but also because it is observed by the Coma ancillary program  and the  DAP product does not include the maps of the BCG.

\begin{figure}[]
\epsscale{1.15}
\plottwo{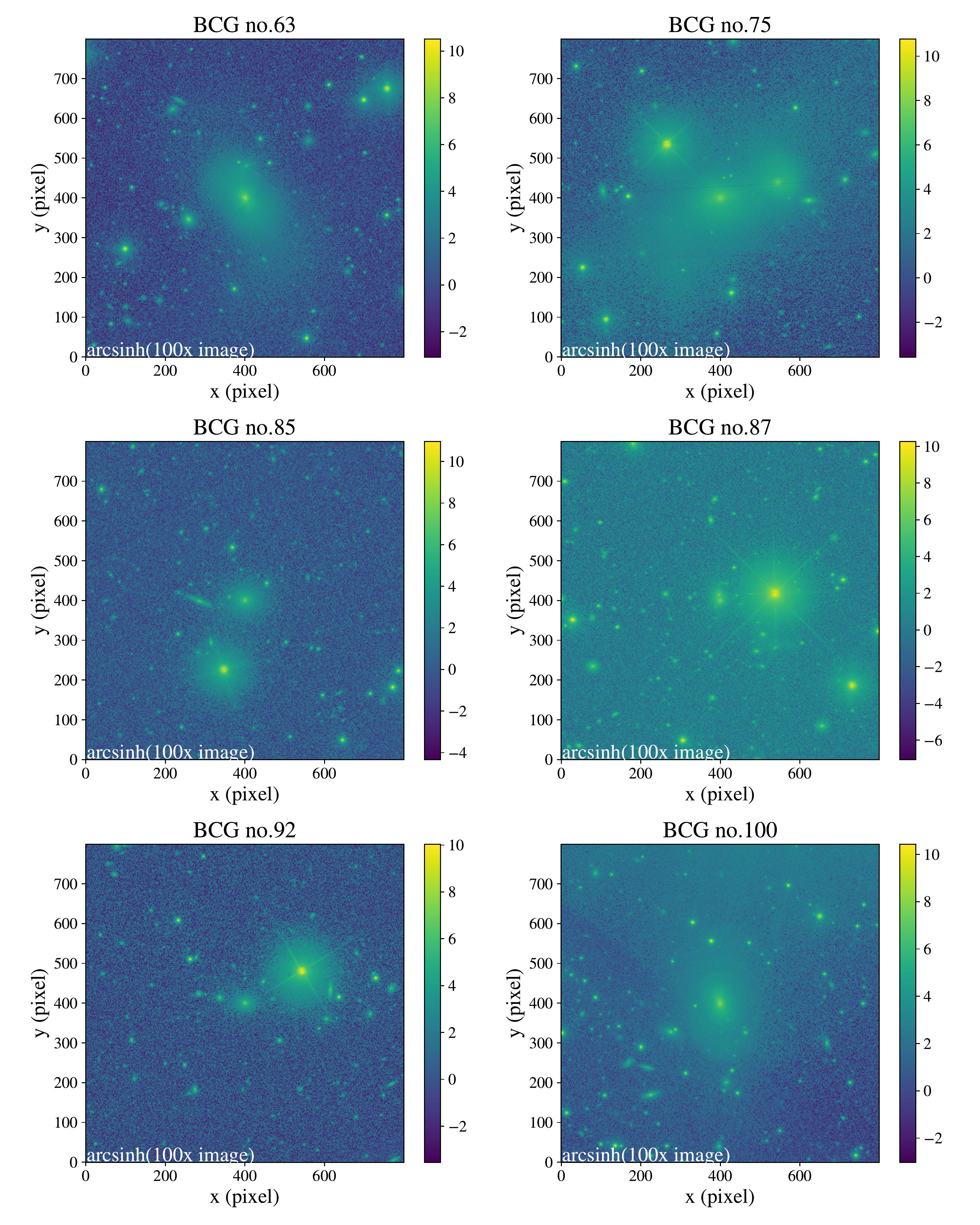}{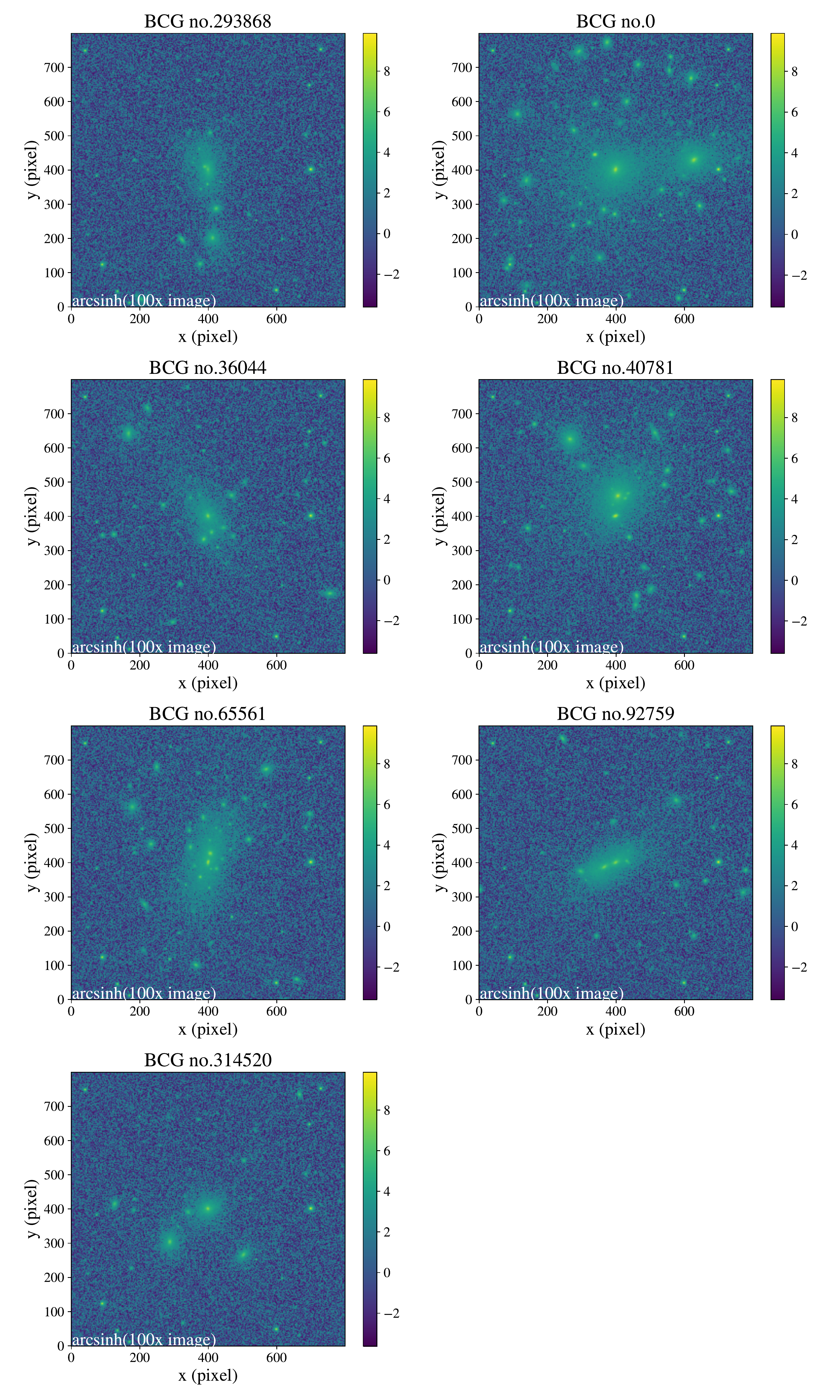}
\caption{
{\it Left:} The SDSS $i$-band images of the 6 BCGs that are excluded due to image quality issues.
{\it Right:} The images of the 7 simulated BCGs that are excluded.
\label{fig:tng_maj2}}
\end{figure}

\subsubsection{Excluded Objects in the MPL-9 sample  due to Image Quality Issues}\label{appendix:bad_im}

6 BCGs are excluded due to the image quality issues. As shown in Fig.~\ref{fig:tng_maj2} (left panel), BCG nos.~63, 75, and 100 have artificial features across the BCG. BCG nos.~75, 85, 87, 92 have bright stars close to the BCG. These can cause problems in background subtraction and affect the resulting {\tt Ellipse} models.

\subsubsection{Excluded BCGs in TNG300}\label{appendix:exclude_TNG_BCG}

7 BCGs are excluded in the simulated sample, because their morphologies are too complex and we cannot obtain a reliable total flux (Fig.~\ref{fig:tng_maj2}, right panel). We emphasize that {\it no such problems occur for 
the real BCGs in our Main sample}. However, mergers might be related to the complex morphology, so we present 2 multiple-core frequencies in the main text, one including and the other excluding these objects (Section~\ref{subsec:TNG_results}). 

The BCG of halo 293868 has many small neighbors touching it and its {\tt Ellipse}  surface brightness profile is non-monotonic and is unreliable, forcing us to exclude it. 
This object (and its host halo) is the one that we remove in Section~\ref{sec:tngsample}.

The BCG of halo 0 has a bright neighbor in the field of view and the outskirt light of the neighbor affects the outskirt of the curve-of-growth. 
Also, the  mass of this halo is  $1.2\times 10^{15} h^{-1}\,M_{\odot}$, far exceeding the massive end of our Main sample.

The BCG of halo 36044 also has many small neighbors touching it and the blending affects the curve of growth at the outskirt. Its $R_e$ of the single S\'{e}rsic fit is larger than $500''$, which is unreliable large.

The BCG of halo 40781 also has many small neighbors touching it and the masking causes some dips in the {\tt Ellipse} curve of growth and in turn, the single S\'{e}rsic fit failed. 

The BCG of halo 65561 has a complex morphology that not only has many small neighbors touching it, and the {\it center part consists of 2 cores with comparable brightness}. This leads to a non-monotonic surface brightness profile and unreasonable curve of growth, hence the single S\'{e}rsic fit fails. 

The BCG of halo 92759 has bright neighbors touching it and the blending affects the curve of growth. Its $R_e$ of the single S\'{e}rsic fit is larger than $700''$, which is unreasonably large.

The BCG of halo 314520 has a bright neighbor in the field of view and the outskirt light of the neighbor affects the outer parts of the curve of growth. Its $R_e$ of the single S\'{e}rsic fit is larger than $100''$.

\clearpage

\section{Tables of the BCGs and the detected cores}
\label{appendix:tab}

\small
\setlength{\tabcolsep}{2pt}

\begin{longtable}{|c|c|c|c|c|c|c|c|c|c|}
\caption{The 128 BCGs in the MPL-9 sample. The columns are the ID in this work, RA, DEC, $z$ from the Y07 catalog (for the changed BCGs mentioned in 
          Section B.3, from SDSS or MaNGA), halo mass, MaNGA plate-ifu ID, and 3 sample selection flags. 
The 115 BCGs described in Section.~\ref{sec:mangabcg} have the ``Selected'' flag set to 1.
``phot'' denotes the source of the photometry: 0 for {\it S\'{e}rsic}, 1 for M16, and 2 for F19 (see Section~\ref{appendix:Re_const}).
 ``IFU18kpc'' set to 1 means the IFU coverage is at least 18\,kpc. The 79 BCGs in the Main sample can be selected by setting selected\,$=1$ and IFU\_18kpc\,$=1$. \label{tab:BCG}}\\
\hline
  \multicolumn{1}{|c|}{ID} &
  \multicolumn{1}{c|}{RA} &
  \multicolumn{1}{c|}{Dec} &
  \multicolumn{1}{c|}{$z$ (Y07)} &
  \multicolumn{1}{c|}{$\log(M_{180m})$} &
  \multicolumn{1}{c|}{plateifu} &
  \multicolumn{1}{c|}{selected} &
  \multicolumn{1}{c|}{phot} &
  \multicolumn{1}{c|}{IFU18kpc} \\
\hline
  1 & 213.834862 & 52.3459366 & 0.0745125 & 14.00 & 8591-3704 & 1 & 1 & 0\\
  2 & 195.088192 & 26.7887007 & 0.1460274 & 14.00 & 11009-6104 & 1 & 1 & 1\\
  3 & 119.023042 & 33.7445348 & 0.0742991 & 14.01 & 8977-3703 & 1 & 1 & 0\\
  4 & 126.502387 & 40.9811098 & 0.0575645 & 14.01 & 10496-6104 & 1 & 0 & 0\\
  5 & 249.101963 & 28.6177833 & 0.1444254 & 14.02 & 11823-12704 & 1 & 1 & 1\\
  6 & 131.301993 & 29.3062335 & 0.0999816 & 14.02 & 10499-6101 & 1 & 1 & 1\\
  7 & 132.787323 & 27.355888 & 0.1198074 & 14.02 & 9506-6103 & 1 & 2 & 1\\
  8 & 248.02191 & 13.6474084 & 0.0522497 & 14.02 & 8609-9102 & 1 & 0 & 0\\
  9 & 177.546947 & 53.7225048 & 0.060313 & 14.02 & 11872-12701 & 1 & 1 & 0\\
  10 & 127.528158 & 45.3517744 & 0.1482441 & 14.03 & 8725-6104 & 1 & 2 & 1\\
  11 & 240.348548 & 26.1161536 & 0.0875589 & 14.03 & 9089-6103 & 1 & 0 & 0\\
  12 & 118.11363 & 19.5401717 & 0.1155163 & 14.03 & 9497-6101 & 1 & 2 & 1\\
  13 & 222.810125 & 32.3782013 & 0.0883005 & 14.03 & 9002-3703 & 1 & 2 & 0\\
  14 & 235.475817 & 28.1340056 & 0.03322 & 14.03 & 9888-12701 & 0 & 2 & 0\\
  15 & 247.694963 & 47.7948292 & 0.1279589 & 14.03 & 8483-6104 & 0 & 0 & 0\\
  16 & 212.955985 & 52.8167901 & 0.076489 & 14.04 & 8591-6102 & 1 & 1 & 0\\
  17 & 168.743724 & 53.6250028 & 0.1048493 & 14.04 & 9000-9101 & 1 & 1 & 1\\
  18 & 181.827567 & 46.7275499 & 0.1014518 & 14.04 & 8261-3702 & 1 & 0 & 0\\
  19 & 258.84568 & 57.4112548 & 0.0273026 & 14.05 & 8625-12704 & 1 & 2 & 0\\
  20 & 229.429418 & 27.8953518 & 0.1195061 & 14.05 & 9891-12701 & 1 & 2 & 1\\
  21 & 148.962806 & 1.57813456 & 0.100966 & 14.05 & 10845-3701 & 1 & 1 & 0\\
  22 & 213.970379 & 50.323853 & 0.0738916 & 14.06 & 9864-12702 & 1 & 2 & 1\\
  23 & 163.03258 & 44.77343 & 0.1395 & 14.06 & N/A & 0 & 0 & 0\\
  24 & 119.617129 & 37.7866182 & 0.042836 & 14.07 & 9181-12702 & 1 & 2 & 0\\
  25 & 122.867045 & 43.6383289 & 0.1429993 & 14.08 & 10213-3701 & 1 & 1 & 1\\
  26 & 121.449567 & 25.2566188 & 0.1408135 & 14.08 & 9503-3703 & 1 & 0 & 1\\
  27 & 48.5733376 & $-0.6096748$ & 0.1152619 & 14.08 & 8081-3701 & 1 & 2 & 0\\
  28 & 215.964753 & 40.258839 & 0.0821963 & 14.08 & 8335-6103 & 1 & 1 & 0\\
  29 & 139.94524 & 33.7497418 & 0.0229126 & 14.08 & 10505-6102 & 1 & 0 & 0\\
  30 & 246.426331 & 43.9317681 & 0.1331387 & 14.09 & 8555-3702 & 1 & 1 & 1\\
  31 & 122.615469 & 40.4306353 & 0.0982483 & 14.09 & 9486-6103 & 1 & 2 & 1\\
  32 & 206.209062 & 52.7760145 & 0.1398773 & 14.09 & 9884-6102 & 1 & 1 & 1\\
  33 & 225.682826 & 53.046861 & 0.1338617 & 14.09 & 8593-3701 & 1 & 0 & 1\\
  34 & 228.451854 & 28.0329749 & 0.114432 & 14.10 & 9891-3701 & 1 & 2 & 0\\
  35 & 122.18108 & 14.7892 & 0.08554 & 14.10 & N/A & 0 & 0 & 0\\
  36 & 234.16229 & 25.9068083 & 0.0946412 & 14.11 & 9889-3703 & 1 & 0 & 0\\
  37 & 46.49724 & $-0.16648$ & 0.10705 & 14.11 & 9194-9101 & 0 & 0 & 1\\
  38 & 122.560974 & 20.2072517 & 0.1247102 & 14.11 & 9490-9102 & 1 & 1 & 1\\
  39 & 225.62102 & 52.7339826 & 0.133075 & 14.12 & 8593-3704 & 1 & 1 & 1\\
  40 & 244.683117 & 25.9474202 & 0.14518 & 14.12 & 9046-6102 & 1 & 0 & 1\\
  41 & 137.141134 & 16.0465505 & 0.0719147 & 14.12 & 8248-6101 & 1 & 1 & 0\\
  42 & 157.650761 & 35.9166146 & 0.1235215 & 14.12 & 8943-3704 & 1 & 1 & 1\\
  43 & 121.603933 & 17.4177001 & 0.1041438 & 14.13 & 10497-12702 & 1 & 1 & 1\\
  44 & 247.724345 & 24.5620962 & 0.0632161 & 14.13 & 9892-3701 & 1 & 0 & 0\\
  45 & 146.523062 & 34.6242592 & 0.1342496 & 14.13 & 10838-12702 & 1 & 0 & 1\\
  46 & 322.115223 & 0.18204297 & 0.1382348 & 14.13 & 8616-3702 & 1 & 1 & 1\\
  47 & 168.406003 & 47.4871527 & 0.1120715 & 14.14 & 10509-12704 & 1 & 0 & 1\\
  48 & 137.44131 & 41.9558691 & 0.1400345 & 14.14 & 8247-9102 & 1 & 1 & 1\\
  49 & 238.299753 & 27.5557304 & 0.1474349 & 14.14 & 9888-6104 & 1 & 1 & 1\\
  50 & 60.0838005 & $-5.5852549$ & 0.1308573 & 14.14 & 8728-3703 & 1 & 2 & 1\\
  51 & 241.840776 & 28.8516181 & 0.126337 & 14.15 & 9030-9101 & 1 & 0 & 1\\
  52 & 175.54471 & 55.45825 & 0.13339 & 14.15 & 8995-6103 & 1 & 1 & 1\\
  53 & 125.388172 & 55.1520274 & 0.0796762 & 14.17 & 10494-12704 & 1 & 0 & 1\\
  54 & 227.041415 & 29.2222 & 0.1109796 & 14.17 & 9891-9101 & 1 & 2 & 1\\
  55 & 204.112347 & 54.8983344 & 0.1067703 & 14.17 & 11020-12702 & 1 & 1 & 1\\
  56 & 127.105745 & 24.6230533 & 0.0883995 & 14.18 & 8939-6104 & 1 & 2 & 0\\
  57 & 117.48113 & 29.4201944 & 0.0623623 & 14.18 & 8146-12704 & 1 & 0 & 1\\
  58 & 157.921418 & 36.0233668 & 0.0852559 & 14.19 & 8943-9102 & 0 & 2 & 1\\
  59 & 157.723276 & 41.2211227 & 0.092116 & 14.19 & 8455-12703 & 1 & 1 & 1\\
  60 & 183.997375 & 35.7173765 & 0.1333452 & 14.19 & 8554-6103 & 1 & 2 & 1\\
  61 & 131.63527 & 29.5987137 & 0.0701255 & 14.20 & 10499-12702 & 1 & 0 & 1\\
  62 & 209.18658 & 44.90331 & 0.12504 & 14.21 & 8328-3703 & 1 & 0 & 1\\
  63 & 252.565435 & 23.5798277 & 0.0360614 & 14.21 & 11979-12701 & 0 & 0 & 0\\
  64 & 181.368271 & 51.4799899 & 0.0853738 & 14.21 & 10508-6102 & 1 & 1 & 0\\
  65 & 255.471297 & 35.0510989 & 0.108797 & 14.21 & 8614-12701 & 1 & 0 & 1\\
  66 & 240.832659 & 25.453721 & 0.0895342 & 14.24 & 9092-12704 & 1 & 0 & 1\\
  67 & 312.909237 & $-0.0558668$ & 0.107683 & 14.25 & 9191-6103 & 1 & 0 & 1\\
  68 & 205.6749428 & 26.23977838 & 0.065362 & 14.25 & 8983-12703 & 1 & 2 & 1\\
  69 & 118.36082 & 29.3594463 & 0.0605648 & 14.26 & 8937-12705 & 1 & 2 & 0\\
  70 & 254.933112 & 32.6153255 & 0.1013 & 14.26 & 9883-9101 & 1 & 2 & 1\\
  71 & 241.877644 & 23.2363 & 0.0894408 & 14.26 & 9087-6102 & 1 & 0 & 0\\
  72 & 246.172233 & 25.3282368 & 0.1284911 & 14.27 & 9048-3703 & 1 & 2 & 1\\
  73 & 232.542819 & 29.0083925 & 0.0841925 & 14.27 & 9042-3702 & 1 & 2 & 0\\
  74 & 169.669505 & 45.5527444 & 0.1126036 & 14.27 & 8466-6104 & 1 & 2 & 1\\
  75 & 220.178483 & 3.46542146 & 0.0273424 & 14.27 & 11835-9101 & 0 & 0 & 0\\
  76 & 129.372907 & 43.8693671 & 0.1354883 & 14.28 & 11745-9102 & 1 & 0 & 1\\
  77 & 219.437785 & 48.5900153 & 0.1227145 & 14.30 & 11011-9102 & 1 & 1 & 1\\
  78 & 242.80765 & 36.9734312 & 0.0674232 & 14.30 & 11944-9102 & 1 & 0 & 0\\
  79 & 191.698030 & 54.887602 & 0.08506 & 14.30 & 11863-6101 & 1 & 1 & 0\\
  80 & 141.886587 & 1.76024544 & 0.1485534 & 14.32 & 10513-9102 & 1 & 0 & 1\\
  81 & 239.195267 & 25.857048 & 0.0741077 & 14.32 & 9092-12702 & 1 & 0 & 1\\
  82 & 167.245462 & 50.3484349 & 0.1156362 & 14.33 & 9001-12701 & 1 & 0 & 1\\
  83 & 242.992215 & 29.8385381 & 0.050068 & 14.33 & 9028-3704 & 1 & 2 & 0\\
  84 & 248.398468 & 25.8187461 & 0.144223 & 14.33 & 9049-12703 & 1 & 2 & 1\\
  85 & 238.373566 & 27.3898901 & 0.0909815 & 14.35 & 9888-9102 & 0 & 2 & 1\\
  86 & 117.456003 & 34.8839132 & 0.1311572 & 14.35 & 8717-1901 & 1 & 1 & 0\\
  87 & 160.966477 & 1.06169372 & 0.1159049 & 14.36 & 10837-3704 & 0 & 0 & 0\\
  88 & 223.139803 & 50.9228509 & 0.1310365 & 14.37 & 9865-12703 & 1 & 1 & 1\\
  89 & 124.129955 & 43.7284448 & 0.1423332 & 14.39 & 10213-12701 & 1 & 1 & 1\\
  90 & 230.903167 & 28.64307479 & 0.084046 & 14.39 & 9043-3704 & 1 & 1 & 0\\
  91 & 233.333131 & 31.2120472 & 0.0674265 & 14.40 & 9890-6104 & 1 & 1 & 0\\
  92 & 214.844448 & 37.8724736 & 0.1361261 & 14.40 & 8337-3702 & 0 & 0 & 1\\
  93 & 187.44772 & 36.6821227 & 0.144796 & 14.40 & 8981-12701 & 1 & 0 & 1\\
  94 & 59.3550829 & $-5.4206679$ & 0.0651988 & 14.40 & 8728-3704 & 1 & 2 & 0\\
  95 & 258.119888 & 64.0608367 & 0.073438 & 14.41 & 11983-12704 & 1 & 0 & 1\\
  96 & 119.918918 & 54.00637 & 0.1032 & 14.42 & 8716-3702 & 1 & 0 & 0\\
  97 & 167.096253 & 44.150282 & 0.0587373 & 14.42 & 8258-3703 & 1 & 1 & 0\\
  98 & 112.169414 & 41.4730136 & 0.1190837 & 14.44 & 8131-3703 & 1 & 2 & 0\\
  99 & 176.224235 & 51.2670667 & 0.1293536 & 14.44 & 8989-12704 & 1 & 2 & 1\\
  100 & 116.577525 & 18.3686844 & 0.0526085 & 14.44 & 9492-9101 & 0 & 2 & 0\\
  101 & 118.354928 & 34.2757163 & 0.1396946 & 14.45 & 9484-12702 & 1 & 1 & 1\\
  102 & 133.518892 & 29.0535413 & 0.0843655 & 14.46 & 10499-12703 & 1 & 0 & 1\\
  103 & 245.362216 & 42.7612901 & 0.1355748 & 14.46 & 8555-12701 & 1 & 0 & 1\\
  104 & 244.157663 & 42.4487743 & 0.1382145 & 14.51 & 8600-9101 & 1 & 0 & 1\\
  105 & 133.652535 & 0.64257394 & 0.1069627 & 14.52 & 10839-9102 & 1 & 0 & 1\\
  106 & 259.14868 & 27.7789882 & 0.1195438 & 14.52 & 9085-6102 & 1 & 2 & 1\\
  107 & 205.734769 & 55.6039491 & 0.0665768 & 14.54 & 11020-12704 & 1 & 1 & 1\\
  108 & 255.677051 & 34.0599931 & 0.09891 & 14.55 & 8613-12705 & 1 & 0 & 1\\
  109 & 52.6672592 & $-6.973111$ & 0.1443454 & 14.57 & 9189-12702 & 1 & 0 & 1\\
  110 & 205.454711 & 26.3734822 & 0.075452 & 14.59 & 8983-12704 & 1 & 2 & 1\\
  111 & 255.638101 & 33.5166395 & 0.0863756 & 14.59 & 8613-6102 & 1 & 1 & 0\\
  112 & 238.238718 & 27.6633004 & 0.082813 & 14.60 & 9888-12703 & 1 & 1 & 1\\
  113 & 124.516049 & 54.6190831 & 0.1174246 & 14.61 & 10494-12702 & 1 & 0 & 1\\
  114 & 122.535493 & 35.2752678 & 0.0839909 & 14.62 & 10214-12704 & 1 & 0 & 1\\
  115 & 231.030854 & 29.8889698 & 0.1134961 & 14.62 & 9044-12705 & 1 & 1 & 1\\
  116 & 146.478197 & 43.0467329 & 0.0730291 & 14.64 & 8461-12701 & 1 & 2 & 1\\
  117 & 239.71939 & 26.4386225 & 0.0873412 & 14.67 & 9094-9101 & 1 & 0 & 1\\
  118 & 245.129711 & 29.8910243 & 0.0960166 & 14.68 & 9025-9101 & 1 & 2 & 1\\
  119 & 157.934731 & 35.0413911 & 0.1204611 & 14.70 & 8943-12705 & 1 & 2 & 1\\
  120 & 247.436995 & 40.8116548 & 0.0293382 & 14.75 & 11942-12705 & 1 & 0 & 0\\
  121 & 231.100286 & 30.0060316 & 0.1174951 & 14.76 & 9044-12703 & 1 & 2 & 1\\
  122 & 322.612308 & $-0.0067582$ & 0.1374996 & 14.78 & 8616-12703 & 1 & 2 & 1\\
  123 & 321.443159 & 0.93108017 & 0.1350092 & 14.78 & 8615-12704 & 1 & 1 & 1\\
  124 & 177.217964 & 51.5522256 & 0.1325608 & 14.83 & 8991-6102 & 1 & 1 & 1\\
  125 & 127.024478 & 44.7667566 & 0.1449603 & 14.85 & 11745-12701 & 1 & 0 & 1\\
  126 & 194.89879 & 27.9592634 & 0.0239087 & 14.87 & 8480-12701 & 0 & 0 & 0\\
  127 & 126.371086 & 47.1335691 & 0.1290448 & 15.01 & 8725-12704 & 1 & 1 & 1\\
  128 & 239.583348 & 27.2334134 & 0.0908067 & 15.09 & 9094-12702 & 1 & 0 & 1\\
\hline\end{longtable}


\clearpage

\small
\setlength{\tabcolsep}{2pt}
\begin{longtable}{|c|c|c|c|c|c|c|c|}
\caption{The 15 cores detected in the Main sample with flux ratio $\mathcal{F}_{\rm core}\ge 0.05$. The columns are BCG ID, source of the systemic velocity, mean and velocity velocity offset in the core region (in unit of km/s), R.A.~ \& Dec.~(both in degrees), flux ratio, selection method (vis\,$=1$ denotes cores identified through visual inspection). \label{tab:ob_core}}\\
\hline
  \multicolumn{1}{|c|}{Number} &
  \multicolumn{1}{c|}{sys\_v} &
  \multicolumn{1}{c|}{v\_off\_mean} &
  \multicolumn{1}{c|}{v\_off\_median} &
  \multicolumn{1}{c|}{core\_ra\_im} &
  \multicolumn{1}{c|}{core\_dec\_im} &
  \multicolumn{1}{c|}{flux\_ratio} &
  \multicolumn{1}{c|}{vis} \\
\hline
  6 & DAP & 365.36 & 389.71 & 131.30121 & 29.308043 & 0.182 & 1\\
  26 & DAP & $-276.83$ & $-286.22$ & 121.4511 & 25.255268 & 0.060 & 0\\
  40 & DAP & 99.74 & 95.60 & 244.68321 & 25.946503 & 0.110 & 1\\
  53 & DAP & $-125.05$ & $-110.32$ & 125.38854 & 55.153046 & 0.045 & 1\\
  57 & DAP & 339.90 & 340.83 & 117.48328 & 29.417328 & 0.053 & 0\\
  57 & DAP & 149.73 & 64.86 & 117.48128 & 29.419292 & 0.011 & 1\\
  68 & DAP & 75.07 & 70.78 & 205.67506 & 26.23668 & 0.063 & 0\\
  68 & DAP & 204.02 & 229.98 & 205.67241 & 26.239765 & 0.054 & 0\\
  72 & DAP & 335.53 & 267.27 & 246.17114 & 25.329369 & 0.115 & 1\\
  77 & DAP & 50.21 & 47.83 & 219.43756 & 48.590576 & 0.005 & 1\\
  103 & corrected & $-138.94$ & $-92.70$ & 245.36404 & 42.76216 & 0.075 & 0\\
  103 & corrected & $-212.12$ & $-191.30$ & 245.36154 & 42.76228 & 0.395 & 1\\
  104 & DAP & 257.89 & 237.99 & 244.15991 & 42.448227 & 0.051 & 1\\
  123 & DAP & 477.52 & 453.52 & 321.44464 & 0.9318187 & 0.147 & 1\\
  125 & DAP & 185.18 & 141.89 & 127.02314 & 44.76726 & 0.0199 & 1\\
\hline\end{longtable}

\begin{longtable}{|r|r|r|r|r|r|}
\caption{The cores of the 225 BCGs in TNG300. The columns are the sublaho ID, their  position with respect to the BCG in $h^{-1}\,$ckpc, flux ratio, and the visual inspection flag (where 1 denotes identification via visual inspection; c.f.~Fig.~\ref{fig:tng_maj}), host halo mass (in $h^{-1}M_\odot$).  No.~65561 is an excluded BCG with bad photometry (with 2 cores).\label{tab:TNG_BCG}}\\
\hline
  \multicolumn{1}{|c|}{ID} &
  \multicolumn{1}{c|}{core\_$\Delta$x} &
  \multicolumn{1}{c|}{core\_$\Delta$y} &
  \multicolumn{1}{c|}{flux\_ratio} &
  \multicolumn{1}{c|}{visual} &
  \multicolumn{1}{c|}{$\log M_{200m}$} \\
\hline
 22739 & 11.5535 & $-6.9935$ & 0.202 & 1 & 14.832\\
  65561 & 1.9560 & 1.9410 & 0.002 & 1 & 14.594\\
  65561 & 3.4750 & 15.0028 & 0.172 & 1 & 14.594\\
  143445 & 6.8048 & $-9.0130$ & 0.050 & 1 & 14.288\\
  154823 & 17.4887 & $-5.7579$ & 0.101 & 0 & 14.351\\
  179992 & 0.6054 & 6.5247 & 0.093 & 1 & 14.307\\
  197011 & 16.3293 & 9.3378 & 0.106 & 0 & 14.262\\
  200512 & 7.6236 & 1.3997 & 0.117 & 1 & 14.260\\
  216339 & 8.3862 & $-11.3124$ & 0.220 & 1 & 14.217\\
  228396 & 14.1833 & $-0.7329$ & 0.149 & 0 & 14.211\\
  237651 & $-8.9273$ & 4.0888 & 0.194 & 1 & 14.053\\
  265310 & $-1.1209$ & $-4.5295$ & 0.002 & 1 & 14.154\\
  267972 & $-11.2348$ & 2.3378 & 0.183 & 0 & 14.112\\
  267972 & $-15.1513$ & 4.1710 & 0.126 & 0 & 14.112\\
  269979 & 8.1404 & $-14.0914$ & 0.116 & 1 & 14.125\\
  292034 & $-3.1866$ & $-16.8615$ & 0.134 & 0 & 14.091\\
  296390 & $-7.1637$ & 10.8193 & 0.115 & 0 & 14.046\\
  308262 & $-14.6869$ & 12.4339 & 0.116 & 0 & 14.029\\
  308262 & $-13.2060$ & 14.6144 & 0.114 & 0 & 14.029\\
  311982 & 10.5523 & 3.9004 & 0.112 & 0 & 14.047\\
  319481 & $-2.0510$ & $-5.1047$ & 0.003 & 1 & 14.010\\
  336616 & $-3.5629$ & 3.3941 & 0.010 & 1 & 14.007\\
  339186 & $-0.8054$ & 5.3912 & 0.003 & 1 & 14.018\\
\hline\end{longtable}

\bibliographystyle{aasjournal}



\end{CJK*}
\end{document}